\documentclass[%
 aip,
 jcp,
 amsmath,amssymb,
 reprint,%
]{revtex4-2}

\usepackage{graphicx}
\usepackage{dcolumn}
\usepackage{bm}
\usepackage{xcolor}
\usepackage{soul}
\usepackage{amsmath}
\usepackage{mathrsfs}
\usepackage{bbm}
\usepackage{dsfont}
\usepackage{cancel}
\usepackage[FIGBOTCAP]{subfigure}
\usepackage{multirow}

   \usepackage{longtable}
\usepackage{cancel}
\usepackage{soul}
\usepackage[utf8]{inputenc}
\usepackage[T1]{fontenc}
\usepackage{mathptmx}
\DeclareUnicodeCharacter{00A0}{ }

\begin{document}


\title{Chiral molecule candidates for trapped ion spectroscopy by \textit{ab initio} calculations: from state preparation to parity violation
}

\author{Arie Landau}
\affiliation{Schulich Faculty of Chemistry, The Helen Diller Quantum Center and the Solid State Institute, Technion-Israel Institute of Technology, Haifa, 3200003, Israel}
\affiliation{The Institute of Advanced Studies in Theoretical Chemistry, Technion-Israel Institute of Technology, Haifa 3200003, Israel}

\author{Eduardus}
\affiliation{Van Swinderen Institute for Particle Physics and Gravity (VSI),
University of Groningen, Groningen, The Netherlands}

\author{Doron Behar}
\author{Eliana Ruth Wallach}
\affiliation{Schulich Faculty of Chemistry, The Helen Diller Quantum Center and the Solid State Institute, Technion-Israel Institute of Technology, Haifa, 3200003, Israel}
\affiliation{Physics Department, Technion-Israel Institute of Technology, Haifa 3200003, Israel}

\author{Lukáš F. Pašteka}
\affiliation{Van Swinderen Institute for Particle Physics and Gravity (VSI),
University of Groningen, Groningen, The Netherlands}
\affiliation{Department of Physical and Theoretical Chemistry, Faculty of Natural Sciences, Comenius University, Mlynsk\'a dolina, 84215 Bratislava, Slovakia}
\author{Shirin Faraji}
\affiliation{Zernike Institute for Advanced Materials, Faculty of Science and Engineering, University
of Groningen, Nijenborgh 4, 9747AG Groningen The Netherlands}
\author{Anastasia Borschevsky}
\affiliation{Van Swinderen Institute for Particle Physics and Gravity (VSI),
University of Groningen, Groningen, The Netherlands}

\author{Yuval Shagam}
\thanks{Author to whom correspondence should be addressed: yush@technion.ac.il}

\affiliation{Schulich Faculty of Chemistry, The Helen Diller Quantum Center and the Solid State Institute, Technion-Israel Institute of Technology, Haifa, 3200003, Israel}

\begin{abstract}
Parity  non-conservation (PNC) due to the weak interaction is predicted to give rise to enantiomer dependent vibrational constants  in chiral molecules, but the phenomenon has so far eluded experimental observation.
The enhanced sensitivity of molecules to physics beyond the Standard Model (BSM), has led to substantial advances in molecular precision spectroscopy, and these may be applied to PNC searches as well.
Specifically, trapped molecular ion experiments leverage the universality of trapping charged particles to optimize the molecular ion species studied toward BSM searches, but in searches for PNC only a few chiral molecular ion candidates  have been proposed so far.
Importantly, viable candidates need to be internally cold and their internal state populations should be detectable with high quantum efficiency.
To this end, we focus on molecular ions that can be created by near threshold resonant two-photon ionization and detected via state-selective photo-dissociation.
Such candidates need to be stable in both charged and neutral chiral versions to be amenable to these methods.
Here, we present a collection of suitable chiral molecular ion candidates we have found, including CHDBrI$^+$ and CHCaBrI$^+$, that fulfill these conditions according to our \textit{ab initio} calculations.
We find that organo-metallic species have a low ionization energy as neutrals and relatively high dissociation thresholds.
Finally, we compute the magnitude of the PNC values for vibrational transitions for some of these candidates.
An experimental demonstration of state preparation and readout for these candidates will be an important milestone toward measuring PNC in chiral molecules for the first time.
\end{abstract}

\maketitle

\section{\label{sec:intro}{Introduction}}
Following the observation of parity non-conservation (PNC) in $\beta$ decay\cite{wu1957} and atomic spectroscopy\cite{BOUCHIAT1982358,Wood1997}, the symmetry between the two mirror configurations of a chiral molecule was also predicted to be broken by the weak interaction.\cite{YAMAGATA1966495,Quack2008}
In most chiral molecules, the weak interaction is supposed to make one enantiomer slightly more energetically stable.
Consequently the effect alters the tunneling dynamics between enantiomers \cite{Quack2008,quack2022,Hund1927}, and creates enantiomer specific vibrational transition frequencies.\cite{LETOKHOV1975275} 
This effect may also pertain to the mystery of how life originated with a specific handedness with many examples of a single naturally occurring enantiomer.\cite{YAMAGATA1966495,Senami2019,Quack2002,Glavin2019}
Observation of PNC in molecules, which has so far not been achieved in the laboratory, will improve our understanding of such phenomena and is a fundamentally important question in chemistry. 

The majority of efforts to measure PNC in molecules have focused on neutral species,\cite{Daussy1999,Satterthwaite2021, Albert2017, Stoeffler2010, Cournol2019,Fiechter2021,Figgen2009} despite the advantages associated with charged molecules. However, recent advances in precision spectroscopy with trapped molecular ions that leverage long coherence times encourage the development of this platform.\cite{Cairncross2017,Zhou2020,roussy2022new,Hanneke2016,Fan2021} Moreover, the generality of ion trapping allows the selection of a molecular ion with optimized sensitivity toward the question at hand. This capability has been exploited in searches for CP (charge and parity) violation \cite{Cairncross2017,roussy2022new}, searches for Dark Matter \cite{Roussy2021} and precision rotational spectroscopy\cite{Chou2020} for example. Furthermore, methods for production of molecular ions that are cold internally and externally have been demonstrated.\cite{Tong2010,roussy2022new} Finally, chiral molecules with particularly asymmetric electronic wavefunctions, such as radical states, are predicted to have enhanced PNC by reduced cancellation of PNC contributions.\cite{Kuroda2022,Senami2019} Such radicals have improved lifetimes as radical ions due to suppression of reactions between the molecules. 

The advantages associated with trapped ions  have led to a few proposals to utilize chiral molecular ions to observe PNC.\cite{Quack2008,Stohner2004,Gottselig2004,Segal2017,Senami2019} However, the scarcity of existing theoretical modeling for chiral molecular ions makes it challenging to choose a suitable candidate for such an experiment. One challenge is that molecular ions are often prepared through ionization of neutrals and thus have a low dissociation threshold. The susceptibility to predissociation may shorten the interrogation times of vibrationally excited states.

Here, we conduct an \textit{ab initio} study in search of suitable chiral molecular ions that can be prepared through state-selective near threshold photo-ionization (STPI) and have a sufficiently high dissociation threshold to facilitate vibrational precision spectroscopy with long interrogation times. We propose STPI as a pathway to create internally cold, chiral molecular ions as has been applied to a few diatomic species successfully.\cite{Tong2010,Loh2012,Cairncross2017,Schmidt2020} Using STPI on a cold neutral molecular beam only allows molecules in certain quantum states to enter the trapped ion ensemble.
This step is important, since the complexity of such polyatomic molecules makes them particularly difficult to cool as the hope for a cycling transition diminishes when fewer symmetries exist.\cite{BergerChiralLaserCooling2016,Kozyryev2016,Kozyryev2017,Mitra2022}

In a quantum projection noise limited statistical uncertainty on the measured frequency is given by 
\begin{equation}
    \delta f = \frac{1}{C \tau \sqrt{N}}
    \label{eq:qpn}
\end{equation}
where $C$ is the contrast, $\tau$ is the coherence time and N is the total number molecules measured. Many  details may cause $C$, $\tau$ and $N$ to vary and technical noise may cause the observed experimental uncertainty to be far above this limit. The quality of the state preparation and state resolved detection pertains to the contrast attained $C$. The quantum efficiency of the detection process relates to $N$. Finally the molecular ion stability is crucial to estimate since it may limit $\tau$, eliminating one of the most important advantages of using charged, chiral molecules. The estimated statistical uncertainty $\delta f$ compared to the vibrational PNC shift governs the experiment feasibility along with consideration of sources of systematic uncertainty.

Four separate stages are envisioned for the future experiment:
\begin{enumerate}
\item Generation and cooling of neutral molecules in a supersonic expansion, or other generic beam method.
\item Ionization of the molecules via 1+1{$^\prime$}  -- two-photon resonant ionization.
\item Ion trapping and spectroscopy experiment.
\item State-selective photo-dissociation spectroscopy  with time of flight mass spectrometry for detection.
\end{enumerate}

This paper focuses on finding suitable chiral molecular ion candidates for stages 2 and 4. A plan for step 3 is discussed in Ref.~\citenum{erez2022}, but many other alternatives could be applied. Step 1 is straightforward for some of the candidates discussed in this work, and more involved for others. {Since the scheme addresses the same molecule in neutral and ionized forms, we refer to the neutral molecule and its companion charged molecule interchangeably.} However, the precision spectroscopy stage is focused on the cation version of the molecule.

To limit the molecule complexity, we study chiral molecules with up to 5-atoms in a tetrahedral structure. Four-atom chiral molecules are not considered here as these are less likely to be chiral in both the neutral and charged configuration simultaneously, often having low transition state barriers between enantiomers.~\cite{Changala2017, Quack2008}

We find several suitable candidates with various spin multiplicities of the form CHXBrI$^+$ where X $\in \{ \mathrm{D,Ca}\}$ and isotopically chiral CHX$^{79}$Br$^{81}$Br$^+$ where X $\in \{ \mathrm{D,Ca,Yb}\}$. All of these candidates are chiral in both neutral and charged forms. We discuss how our results indicate that these molecules fit stages 2 and 4. Additionally, we compute the magnitude of the expected PNC for some of these candidates, here and in Ref.~\citenum{eduardus2023}. We show for example that candidates where $\mathrm{X}\in \{$Li, Na$\}$ fit most, but not all of the requirements, and thus are not suitable candidates. Despite not explicitly investigating CHYbBrI$^+$, we can infer that it is also amenable to the experiment by considering  similar molecules such as CHYbBr$_2^+$ examined in the study.

\section{Scheme description}
\label{schemeoutline}

\begin{figure*}[t]
    \centering
    \includegraphics[width = 15cm]{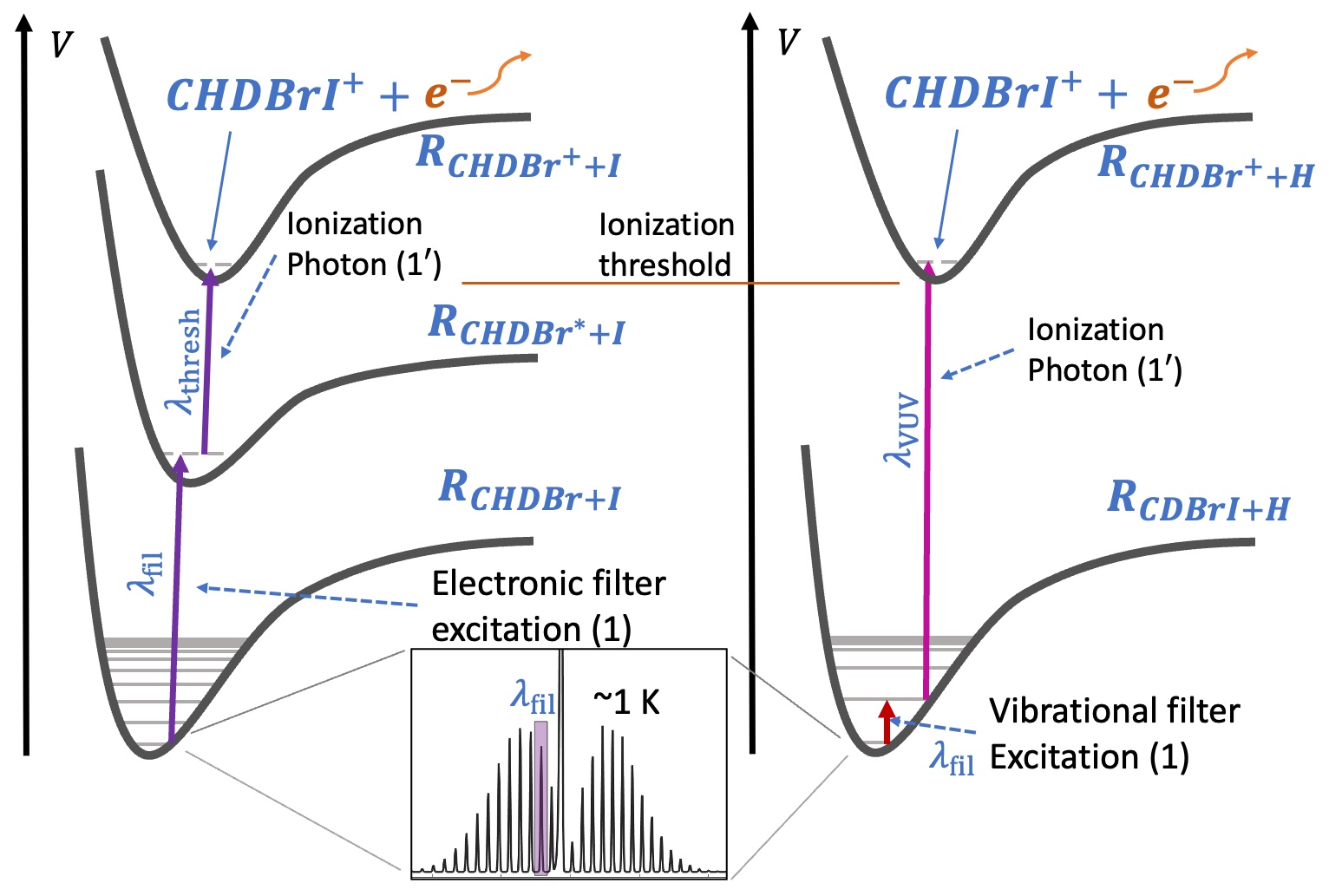}
    \caption{Resonant two-color near threshold ionization procedure schematically outlined for CHDBrI$^+$ (STPI). The first photon excitation is resonant with a transition that ideally excites a single rotational state of the molecule. This can be a ro-vibronic transition as shown on the left or a ro-vibrational transition as shown on the right. The selective filtering excitation is followed by a second (vacuum) ultra-violet photon that ionizes the molecule directly or through an autoionizing state such that the molecule's final quantum state  suffers minimal change. }
    \label{fig:REMPI}
\end{figure*}

Polyatomic molecules have many degrees of freedom, making  a general method to internally cool them challenging to find.
Most cold polyatomic molecule experiments resort to generic collision-based cooling methods such as supersonic expansion \cite{morse1996supersonic} or buffer gas cooled beams \cite{Hutzler2012}, but these methods lower the internal temperature of molecules to a few kelvins only.
Recent advances in laser cooling schemes suggest a promising pathway toward cold polyatomic molecules.\cite{BergerChiralLaserCooling2016,Kozyryev2016,Kozyryev2017,vilas2022magneto,zhu2022functionalizing}
In fact, the Franck-Condon factors for some isotopically chiral molecules have been found to be favorable for laser cooling,\cite{BergerChiralLaserCooling2016} which places cold and controlled neutral chiral molecules within reach.
For molecular ions that originate as neutrals there is an alternative avenue by which they may be prepared, populating a low number of states through a filtering step discussed here.

The proposed scheme begins with a cold beam of the candidate molecules in neutral form. 
When the neutral molecules enter the ion trap they are ionized by a two-color resonant process.
The first photon frequency is resonant with a specific resolved molecular transition. Therefore, it excites only molecules in certain quantum states.
The second photon ionizes the molecules that populate this intermediate state, either directly to the continuum near the ionization threshold or through an autoionizing Rydberg state.
The molecular ion can only minimally heat up internally due to the low energy at which the electron is ejected or the selectivity of the autoionization process.

We discuss two avenues to achieve the state selectivity. The first is to use a ro-vibronic (ro-vibrational and electronic) excitation for the first photon, followed by ionization by a second photon (Fig.~\ref{fig:REMPI} left).
This places both of these photon energies in the UV (ultraviolet) or Deep UV (DUV) range.
The main risk with this method is that many organic molecules are vulnerable  to dissociation in electronically excited states and this may occur before the second photon completes the ionization process.
A second way of achieving cold molecular ions is to use a vibrational excitation in the mid-IR along the C$-$H stretch where the molecule is stable as the selective filtering step, followed by a second photon in the DUV or VUV (vacuum UV) to ionize the molecule (Fig.~\ref{fig:REMPI} right).
This avoids the pre-dissociation of the molecule in the intermediate state since the process is energetically forbidden and the state density is significantly lower. 

Once the molecules are ionized and internally cold, they are still moving at hundreds of meters per second due to the neutral beam velocity.
The ensemble can be decelerated by applying a uniform electric field to the molecular ion cloud.\cite{roussy2022new,Zhou2020}
Subsequently, the ions are trapped with the same electrodes and separated from the rest of the neutral molecules.
The ionization process serves as a selective step for specific quantum states since the molecules that are not excited by the first photon remain neutral and will not be decelerated and trapped, leading to a cold sample of molecular ions.
This method is utilized for the HfF$^+$ experiment \cite{Zhou2020,roussy2022new} following state-selective 1+1{$^\prime$} ionization  and produces 40k ions in fewer than 4 quantum rotational states with a translational temperature of $\sim2$~K. 
This same method is also used to create state-selected N$_2^+$ with 2+1{$^\prime$}  scheme, where the ions are subsequently trapped by sympathetic cooling with laser cooled atomic ions.\cite{Tong2010}
Throughout the manuscript we refer to this state-selective near threshold photoionization step as STPI.

While the first photon transitions may be challenging to fully resolve for  complex polyatomic molecules, any  partial selectivity that arises in this step is likely to produce molecular ions that are very cold.
Since the spectral width of nanosecond pulsed dye lasers is on the order of GHz, this tool is excellent for addressing individual rotational states, but the hyperfine structure will not be resolved.
Population in hyperfine states is unlikely to harm initial forms of our precision measurement and can be overcome by depletion or tailored pulses and polarization selections.\cite{Lee2022,Leibscher2022}
Another advantage of trapped molecular ions is that there is sufficient time to perform multiple depletion processes, even when the available power is low.

Our scheme requires that chiral molecular ion candidates must be stable both as neutrals and as ions. 
For the 5-atom molecules we consider, removal of an electron from { such} a small chiral molecule often makes the molecular ion very weakly bound, with a tendency to dissociate, particularly when the molecule is vibrationally excited (Section~\ref{sec:STPD}).
In fact, as we will show, many candidates that are amenable to STPI dissociate far too easily to support vibrational spectroscopy. We therefore search for candidates with sufficiently high dissociation thresholds (>1.2~eV).
We find that adding a metal atom to a halogen substituted methane substantially increases the dissociation threshold while simultaneously reducing the ionization threshold of the neutral version of the species.
For example, we find that for CHCaBrI$^+$ the most favorable dissociation channel is I + CHCaBr$^+$ at an energy of   2.55 eV  as compared to CHDBrI$^+$ that dissociates to  I + CHDBr$^+$ at 1.29 eV (Section~\ref{sec:STPD}). Simultaneously, the ionization threshold of CHCaBrI is more than 4~eV lower than non-metallic counterparts (Section~\ref{sec:STPI}).

To detect the internal state of our chiral molecular ions we turn to photodissociation, due to the relatively low dissociation threshold.
In an ion trap, state-selective photodissociation (SPD) serves as a highly efficient method for probing internal states of the molecular ions by detecting the fragments via mass spectrometry.\cite{Ni2014,Zhou2020,Karr2012,carollo2018two}
An alternative we pursue involves the detection of the internal state of the molecule using single-photon photo-dissociation.
Upon dissociation, any internal rotational state of the molecule will be translated into a different kinetic energy of the photo-fragments. Detection of the photofragments may be performed by coupling a velocity map imaging setup to an ion trap for example. Our initial design is promising, but details will be reported elsewhere. In this work we check that the wavelengths associated with the dissociation processes would be feasible by calculating the dissociation energies for the various dissociation channels (Section~\ref{sec:STPD}).

Both STPI (Section~\ref{sec:STPI}) and  SPD (Section~\ref{sec:STPD}) need to be developed experimentally, but this papers aims to rule out many candidates that would not fit these schemes and suggest some that do fit it based on {\it ab initio} calculations.

{
\section{Computational Details}
\label{comp}

\label{method}

The computed properties presented in the next sections are, the vertical and adiabatic ionization energies and several excitation energies (EEs) of the neutral systems, in addition, the dissociation channel energies, isotopic vibrational modes, rotational constants, transition state energies of the cation systems. 
Below are the details for all of these computations.

The energy differences we are interested in are, the ionization potentials, adiabatic  and vertical, the low-energy dissociations, the activation energy between the S and R enantiomers, and the low-level EEs of the neutral systems.  
All the electronic configurations considered herein have either a closed-shell structure with a singlet multiplicity or a radical configuration with a doublet multiplicity.

Radical systems are frequently studied using single reference methods.
In particular, density functional theory (DFT) is often used in determining the geometrical structures and energy differences of reactions that include radicals.~\cite{zador2009reaction,qi2005hydrogen,zador2009reaction,luo2018kinetic,xu2019theoretical,xiao2020experimental,lisovskaya2021pulse} Sometimes, coupled-cluster with singles, doubles and perturbative triples [CCSD(T)] is used to benchmark the proper exchange-correlation functional for geometry optimization.~\cite{xu2019theoretical}
Configuration interaction and M\o{}ller Plesset perturbation theory from second to forth order can also be used to optimize structures.~\cite{goddard1989computational,rauk1994carboxyl} 
Notice that in Ref.~\citenum{qi2005hydrogen} it was shown that DFT performs significantly worse than the unrestricted second order M\o{}ller Plesset perturbation method in both geometry optimization and interaction energy calculations for radicals.
However, all the above mentioned methods represent single-reference approach, in which the radicals' treatment is based on an unrestricted wave-function (Hartree Fock or DFT).
The unrestricted approximation for the radical wave-functions suffer, for example, from spin-contamination.
Note that only in some studies the very important $\langle S^2 \rangle$ expectation value, which gauges the spin-contamination of the wave-function, is reported.~\cite{rauk1994carboxyl}

Therefore, in this study, we examine the following scheme for describing the closed-shell configurations as well as the radicals: the geometry is calculated at the second-order M\o{}ller Plesset (MP2) level while the energies are obtained at the CCSD(T) level using the same basis sets (i.e., CCSD(T)//MP2).
However, since these open-shell radical doublet states calculations are based on the unrestricted Hartree-Fock wave-function, with the well-known problematic behaviour that manifests in spin-contamination,  we carefully monitor this approximation.
We perform, in addition, equation-of-motion CCSD (EOM-CCSD) calculations to verify the validity of the CCSD(T)//MP2 scheme.
Indeed, we find that for some radicals the MP2 geometries are inconsistent with the EOM-CCSD ones (see Table~\ref{tbl:geom_cmpr} in the SI).  
On the other hand, EOM-CCSD for ionization potential (EOM-IP-CCSD) or electron affinity (EOM-EA-CCSD), depending on the molecule under investigation, is a proper approach for describing radicals.
It is a single-reference but multi-configuration approach that operates in Fock space, in which the reference and the target states are treated in a balanced fashion and as a result it provides a well-defined spin state.\cite{Krylov:EOM} 
Notice that EOM-CCSD analytic gradients are available within the Q-Chem program.~\cite{pieniazek2008charge}

The eight low-lying EEs are calculated for the neutral molecules. For the molecules with a singlet state we use EOM-CCSD for excitation energies (EOM-EE-CCSD) and for the system with a doublet neutral state we use EOM-EA-CCSD. 
As discussed above, the EOM-CCSD energies are calculated at the EOM-CCSD geometry using the same basis set, which is suitable for doublet states, whereas for the closed-shell singlet states we use the CCSD energies at the MP2 geometry (in a few cases we used CCSD geometry since MP2 did not converge, the difference between the two is negligible, as shown before in Ref.~\citenum{helgaker2014molecular}).

The harmonic frequencies of the optimized structures are calculated using EOM-IP-CCSD for doublet state cations and CCSD for singlets. 
The isotopic effect is considered explicitly for these calculations. In systems with two hydrogens we use the masses of hydrogen and deuterium, and in the Br$_2$ case we use the different Br masses, $^{79}$Br and $^{81}$Br.
In addition, we calculate these frequencies using the DFT $\omega$B97M-V functional. 
The agreement between the two methods suggests that this functional is suitable for the frequency analysis for the systems at hand.
Moreover, the $\omega$B97M-V $\langle S^2\rangle$ expectation values is found in the  $0.75-0.756$ range for all the reported doublet states.
Therefore, we employ  $\omega$B97M-V to calculate the transition states between the S and R enantiomers in order to evaluate if the enantiomer states are time invariant.
All the non-relativistic calculations are done with the Q-Chem electronic-structure package.~\cite{feng2019implementation}

We consider two basis sets, of  triple-zeta (TZ) and quadruple-zeta (QZ) quality.
The frozen-core scheme is used within the coupled-cluster based calculations. 
The Dunning correlation consistent cc-pVXZ~\cite{dunning1989gaussian,woon1993gaussian} basis sets are used for all the light atoms (H, Li, C, O, Na and Ca) within the XZ sets, where X=T or Q.
For EEs we use the augmented version of these basis sets,~\cite{dunning1992augccpvxz} aug-TZ and aug-QZ. 
For the heavier atoms (Cl, Br and I) we use these basis sets paired to small-core pseudopotentials, i.e., cc-pVXZ-PP.~\cite{peterson2003pp1,peterson2003pp2}
For Yb we use def2-XZVPP paired to a small-core pseudopotential.~\cite{gulde2012error}

These pseudopotentials were optimized to provide an accurate description of the Pauli repulsion of the cores, their Coulomb and exchange effects on the valence space, and scalar-relativistic corrections.~\cite{peterson2003pp1,peterson2003pp2}
Therefore, we also examined the  spin–orbit effect by taking the difference between the fully relativistic calculations, using the Dirac-Coulomb Hamiltonian including the  Gaunt interaction, and the spin-free version,~\cite{Dyall1994spinFree} which provides results without spin-orbit coupling for the four-component Hamiltonian.
The spin-orbit contributions to the adiabatic ionization potential (AIP) and the vertical ionization potential (VIP)  of the nine molecules were calculated using the dyall.cv2z basis set within EOM-IP-CCSD.
The eight low-lying EEs of  the neutral molecules ({$^2$CHCaBr$_2$} and {$^2$CHCaBrI})  with  doublet multiplicity were calculated via EOM-EA-CCSD and the dyall.av2z basis set, except for Ca for which we used dyall.v2z. 
For neutral molecules with  singlet multiplicity we used EOM-EE-CCSD/dyall.v2z, these include: {$^1$CH$_2$Br$_2$} (four excited states), {$^1$CHLiBr$_2$} (four excited states), {$^1$CHLiBrI} (four excited states), {$^1$CHNaBr$_2$} (five excited states) and {$^1$CHNaBrI} (two excited states).
All in all we calculated the spin–orbit effects for ten VIPs and AIPs, twenty-four EEs using EOM-EA-CCSD and nineteen EEs using EOM-EE-CCSD.
The maximal spin–orbit contribution among the molecules reduces the AIPs by 0.02 eV, for the VIPs by 0.08 eV, for the EEs using EOM-EA-CCSD by 0.003 eV, and for EOM-EE-CCSD EEs by 0.05 eV. 
The average shifts are at least 4-fold smaller and with identical sign for each molecule.
We conclude that the spin–orbit effect is negligible and cancels out since we investigated the energy difference between very similar chemical systems. Therefore, we report the values obtained by Q-Chem.~\cite{feng2019implementation}
All the relativistic calculations are performed using the  relativistic electronic structure package DIRAC22.~\cite{code2020dirac}

Finally, we evaluate the stability of the selected cations by calculating their low lying dissociation channel energies.
For fragments with singlet multiplicity we use CCSD and for doublet fragments we use EOM-IP-CCSD or EOM-EA-CCSD using the QZ basis set and the Stuttgart/Cologne pseudopotentials.
The spin-orbit corrections in this case are not negligible and are added to the non-relativistic dissociation energies.
The  spin–orbit effect was calculated only for the lowest dissociation channel of each system using  the dyall.cv2z basis set with the same methodology. 
The average  spin–orbit effect for the  eleven dissociation energies is -0.15 eV, with the maximum value -0.27 eV, constituting less then 10\% of the calculated values that neglect relativistic effects.


Benchmark calculations are presented in the Supplemental Information Section~\ref{banchmark}.
In summary, the presented values and geometries were obtained using the following schemes:
EOM-CCSD/QZ is used for the ionization energies (Section~\ref{sec:STPI}), EOM-CCSD/aug-QZ for the excitation energies (Section~\ref{sec:STPI}).
The cation's vibrational frequencies  using EOM-CCSD/TZ and $\omega$B97M-V/TZ  (Section~\ref{sec:structure}).
Transition states and their vibrational analysis using $\omega$B97M-V/TZ  (Section~\ref{sec:structure}).
Dissociation energies using EOM-CCSD/QZ, in addition, spin–orbit corrections are added via EOM-CCSD/dyall.cv2z  (Section~\ref{sec:STPD}).
}

\section{State-Selective Photoionization (STPI)}

\label{sec:STPI}

State-selective ionization combined with control  of the excess energy in the process can be achieved using  a resonant 1+1{$^\prime$}  process, where 1 and 1{$^\prime$}  denote the number of photons involved in each of the two stages in the process and the "prime" tag indicates that the two stages use different photon energies. This notation is common for such REMPI (Resonance enhanced multi-photon ionization) methods.
The advantage of a 1+1{$^\prime$}  process is that the power in each stage can be carefully tuned to avoid the competing 1+1 process. 
This is in contrast to the 2+1{$^\prime$} and 3+1{$^\prime$}  processes, where the simultaneous interaction of multiple photons in the first stage usually necessitates focusing the laser beam, which precludes fine control of the first photon beam power since the ionization volume and power are coupled.
To approximate the amenability of our candidates to 1+1{$^\prime$}  resonant photo-ionization we calculate the ionization threshold of the molecule as well as electronically excited states for the resonant transition of the first photon.

The ionization threshold has an adiabatic and vertical component and we calculate both for each of the initial set of candidates: CHXBrI where X $\in \{ \mathrm{D,Li,Na,Ca}\}$ and CHX$^{79}$Br$^{81}$Br where X $\in \{ \mathrm{D,Li,Na,Ca,Yb}\}$, (Table~\ref{tbl:vip_aip}).
Between the AIP and the VIP and we expect to find resonant autoionizing states that are often used to control the emitted electron energy and quantum states while maintaining a large coupling to the continuum. 
Controlled ionization is also possible by choosing a near threshold energy for the 1{$^\prime$}  photon.

 \begin{table}[]
 	\caption{Vertical ionization potentials (VIPs) and adiabatic ionization potentials (AIPs) in eV calculated at the EOMCCSD/QZ level. }
   \label{tbl:vip_aip}
   \def\arraystretch{1.5}
 \begin{center}
   \begin{tabular}{l|c|c||l|c|c}
    \hline
    \hline
   Molecule         &    VIP    &  AIP    &    Molecule         &    VIP    &  AIP    \\    
    
\hline                                                        
  {CHD$^{79}$Br$^{81}$Br}   &  10.68   &   10.22   & {CHDBrI }      &   9.98   &    9.64   \\
 {CHLi$^{79}$Br$^{81}$Br}   &   8.27   &    7.17   & {CHLiBrI}      &   8.12   &    7.05   \\
 {CHNa$^{79}$Br$^{81}$Br}   &   7.85   &    6.91   & {CHNaBrI}      &   7.74   &    6.92   \\
 {CHCa$^{79}$Br$^{81}$Br}   &   5.56   &    5.38   & {CHCaBrI}      &   5.53   &    5.35   \\    
 {CHYb$^{79}$Br$^{81}$Br}   &   5.60   &    5.43   \\ 
  \hline
  \hline
   \end{tabular}  \\
   \end{center}
 \end{table}

To estimate the resonant, filtering, transition energy, we calculate the energies of the first eight vertical electronic excitations. For each candidate these are listed in Table~\ref{tbl:bribr_ees}. The first photon (1) frequency would need to be resonant with one of these states. Naturally, there are many specific rovibrational transitions associated with each state to choose from, but these energies provide a rough estimate of where these transitions with maximal Frank-Condon factors are expected.

\begin{figure}[b]

\subfigure[]{
   \includegraphics[width=1\columnwidth, angle=0,scale=1,
draft=false,clip=true,keepaspectratio=true]{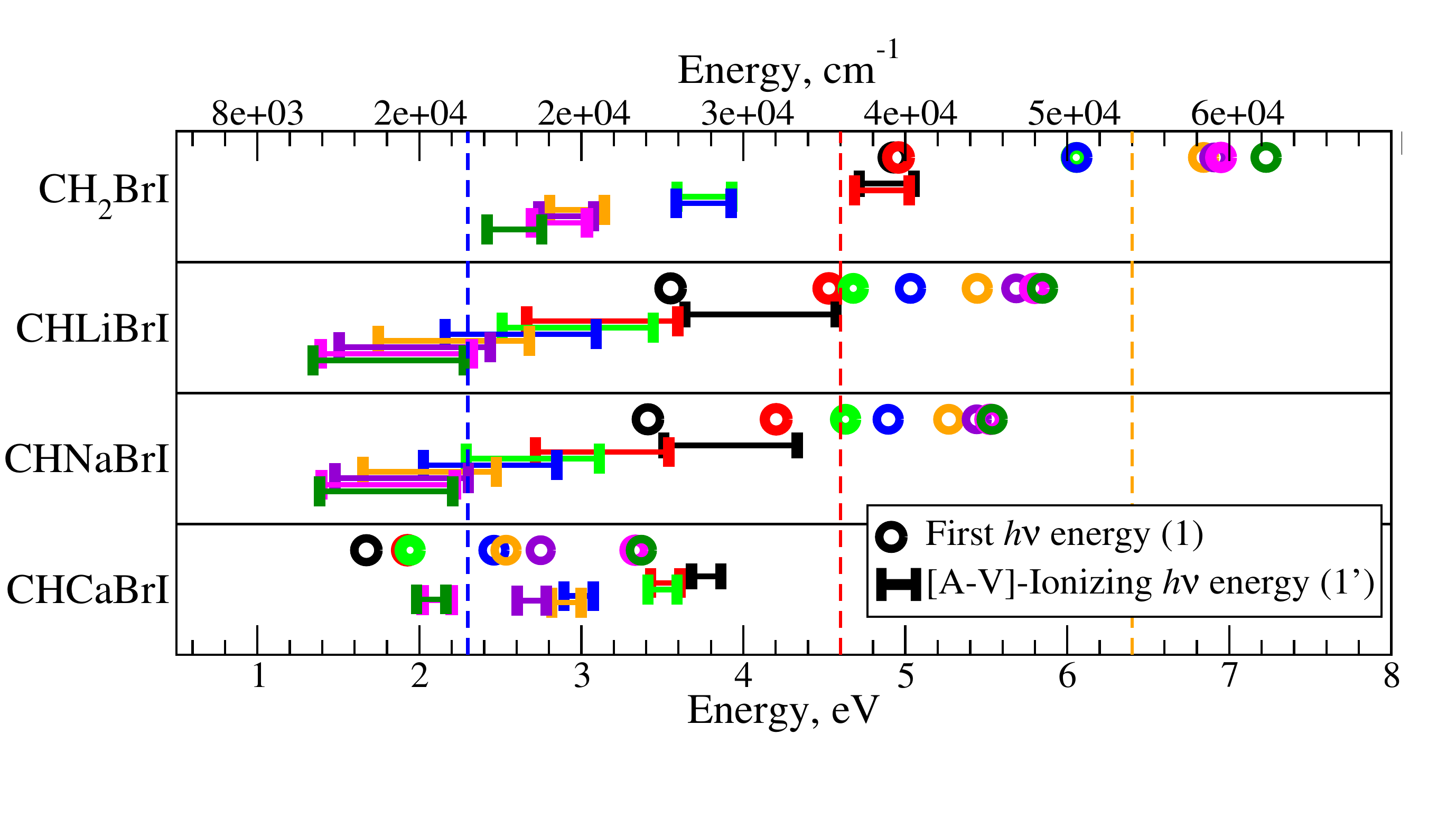}
    \label{ieee_bri.a}
}
\subfigure[]{
   \includegraphics[width=1\columnwidth, angle=0,scale=1.,
draft=false,clip=true,keepaspectratio=true]{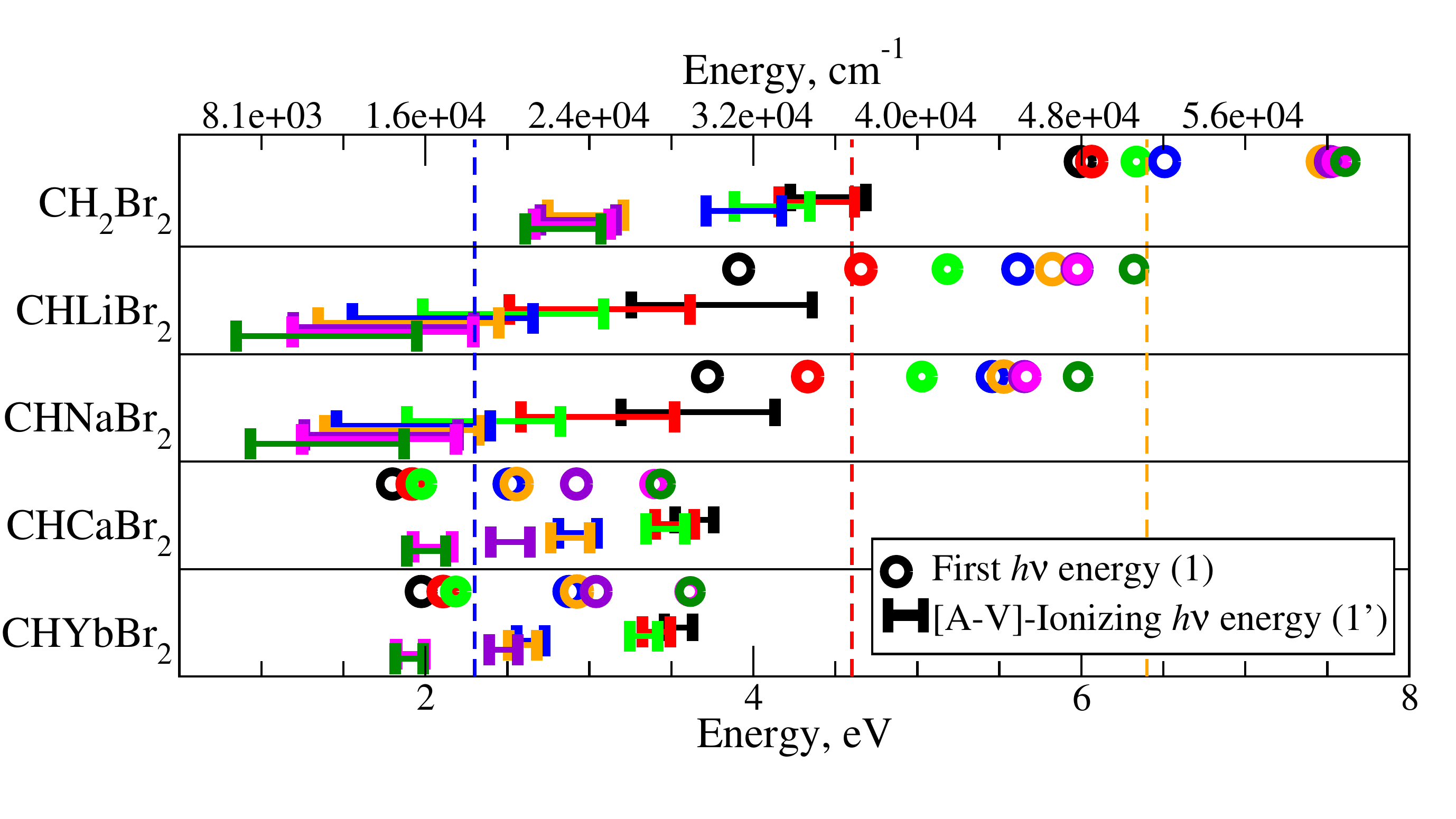}
    \label{ieee_bri.b}
}
\caption[]{Photon energies for 1+1{$^\prime$}  REMPI ionization of CHXBrI where X $\in \{ \mathrm{D,Li,Ca,Na}\}$ (a) and CHX$^{79}$Br$^{81}$Br where X $\in \{ \mathrm{D,Li,Ca,Na,Yb}\}$ (b) are shown. The electronic excitation energies  that approximately represent the first photon energy are depicted by the circles. After the electronic excitation, the remaining energy to the AIP and the VIP are depicted by the left and right edges of the errorbar, which corresponds to the second photon energy range. The electronic excitation symbol and remaining energy for ionization markings are color-coded per excitation. The vertical, dashed lines denote the maximal photon energy reachable with tunable dye laser directly (blue), frequency doubled (red), and frequency tripled (yellow). For brevity Br$_2$  and H$_2$ are used, which corresponds to $^{79}$Br$^{81}$Br and HD.}
 \label{ieee_bri}

\end{figure}

Once excited to the intermediate state, we can use the AIP and VIP to estimate the energy of the 1{$^\prime$}  photon that is needed to ionize the molecules by subtracting the respective electronic state energy.
Figure~\ref{ieee_bri} shows the predicted photon energies needed in the 1+1{$^\prime$}  process for each candidate.
The energies of the electronically excited states are depicted by circles (first photon energy (1)) and the ionizing photon energy (1{$^\prime$}) is depicted by an error bar that stretches between the AIP and the VIP.
The color-coding of the markers of the 1 and 1{$^\prime$}  photons corresponds to different electronic excitations (Table~\ref{tbl:bribr_ees}). Excitations one through eight are plotted rising order  and colored as black, red, light green, blue,  orange, purple, magenta, dark green respectively. The excitation energy and first photon energy are equal, which serves as a key to help distinguish between the different intermediate states.

\begin{table*}[]
 	\caption{The eight lowest EEs calculated via EOM-EE-CCSD/aug-QZ for singlet states and via EOM-EA-CCSD/aug-QZ for doublets, where for the singlets we use the MP2/QZ geometry and for doublets EOM-EA-CCSD/QZ. For brevity Br$_2$  and H$_2$ are used, which corresponds to $^{79}$Br$^{81}$Br and HD.}
        \label{tbl:bribr_ees}
   \def\arraystretch{1.5}
 \begin{center}
   \begin{tabular}{cccc|ccccc}
    \hline
    \hline
  $^1$CH$_2$BrI    &   $^1$CHLiBrI   &   $^1$CHNaBrI      &  $^2$CHCaBrI  & $^1$CH$_2$Br$_2$   &    $^1$CHLiBr$_2$    &   $^1$CHNaBr$_2$        &   $^2$CHCaBr$_2$   &     $^2$CHYbBr$_2$   \\
   4.93    &   4.55    &   3.41   &  1.67    &  5.99    &    3.91    &   3.72    &   1.80   &    1.97    \\
   4.96    &   4.53    &   4.20   &  1.93    &  6.06    &    4.66    &   4.33    &   1.92   &    2.11    \\
   6.05    &   4.68    &   4.63   &  1.94    &  6.33    &    5.18    &   5.03    &   1.97   &    2.18    \\
   6.06    &   5.03    &   4.89   &  2.46    &  6.51    &    5.61    &   5.46    &   2.51   &    2.87    \\
   6.84    &   5.44    &   5.27   &  2.53    &  7.47    &    5.82    &   5.53    &   2.56   &    2.92    \\
   6.91    &   5.68    &   5.44   &  2.75    &  7.52    &    5.97    &   5.65    &   2.92   &    3.04    \\
   6.95    &   5.80    &   5.52   &  3.33    &  7.56    &    5.98    &   5.66    &   3.39   &    3.61    \\
   7.23    &   5.85    &   5.54   &  3.37    &  7.61    &    6.32    &   5.98    &   3.43   &    3.62    \\
  \hline
  \hline
   \end{tabular}  \\
   \end{center}
 \end{table*}

To understand if it is feasible to realize the wavelengths  in the lab we  compare the energies of the photons to the tuning range of a dye laser pumped by a 532 nm source. The choice of dye lasers originates in their broad tunability over the range and relatively narrow linewidth that commonly allows resolving of individual rotational lines. 
The {maximal lasing frequencies} for tunable dye lasers are depicted by vertical dashed lines including their frequency conversion add-ons such as doubling and tripling.

In the second approach discussed to achieve the filtering step, the molecule would be excited along a specific rovibrational transition (Fig.~\ref{fig:REMPI} right).
The vibrational excitation needs to be adequately high to ensure that the intermediate state is not thermally populated.
The relevant energies for the second transition are the AIPs and VIPs (Table~\ref{tbl:vip_aip}) minus the vibrational excitation. 
This means that the second photon would need to be in the VUV range for most of the molecules except for the Ca and Yb substituted candidates, where the  transition is reachable with a tripled dye laser.

A clear trend emerges from Table~\ref{tbl:bribr_ees} and Fig.~\ref{ieee_bri}. Substituting the deuterium with an  alkali atom reduces the energies of the excited electronic states by more than 1 eV. 
Replacing deuterium with Ca or Yb further reduces this energy since these candidates are now radicals in the neutral state.
Furthermore, these substitutions cause the ionization energy to dramatically drop (see Table~\ref{tbl:vip_aip}).
This makes this subset of candidates particularly appealing from a state preparation perspective.

For the scheme to work one of the important steps is for the intermediate transition to be resolved. This naturally becomes challenging with the growth of the molecular complexity and certainly the molecules discussed here are far from simple. However, any sort of partially resolved transitions should already assist in reducing the target molecular ion internal temperature in the process. Moreover, for rovibrational transitions, these tend to be resolved for similar molecules.\cite{Roberts2022}

\section{Cation properties} 
\label{sec:stability}
In this section we consider the candidates in charged form to test their amenability for precision spectroscopy in search of PNC. We discuss the binding energy of the candidates (Section~\ref{sec:STPD}) as well as the structure of the molecules (Section~\ref{sec:structure}).

\subsection{Dissociation}
\label{sec:STPD}

Since the candidate cations are produced by shedding an electron as we describe in Section~\ref{sec:STPI}, we suspect that their binding energy might be relatively low. For the chiral molecular ion to be a viable candidate, we must verify that it will not fragment too easily and certainly not when undergoing vibrational excitation. {This is because one of the main avenues to search for PNC is through vibrational spectroscopy and comparison between enantiomers.\cite{LETOKHOV1975275} If the molecular ion unintentionally dissociates during the spectroscopy stage the precision of the measurement will be substantially reduced, which lowers its appeal as a candidate.}

Here we calculate the dissociation threshold for the various dissociation channels for the candidates. 
The resulting thresholds for the lowest energy dissociative channel are given in Table~\ref{tbl:dis.soc}. 
The dissociation channels were calculated by taking the difference between the cation energy and the sum of the fragment energies.
EOM-CCSD/QZ//EOM-CCSD/QZ (IP, ionization potential, or EA, electron attachment) is used for systems with doublet multiplicity and CCSD/QZ//MP2/QZ for the singlet state molecules. The relativistic spin-orbit coupling correction is added on top of the non-relativistic results. This correction is obtained by taking the difference between results obtained using the full relativistic Hamiltonian and its spin-free version; the same approach but a smaller basis set was used for this calculation (see Section~\ref{comp} for details). 
For CHLiBr$_2^+$ and CHNaBr$_2^+$ we find two channels with similar dissociation energies and both are listed. 
For $^1$CHYbBr$_2^+$, the EOM-CCSD calculation of $^2$CHYbBr$^+$ did not converge, thus we present the CCSD(T) results instead.  
The spin multiplicity of the system which is either singlet or doublet is noted in a superscript before the molecular formula. 
  
Additional, higher dissociation channel energies are listed in Table~\ref{tbl:dis.mult}.

 \begin{table}[b]
        \caption{Lowest Dissociation energies in eV of the cations in their ground states, doublet states calculated using the EOM-CCSD/QZ at the EOM-CCSD/QZ geometry and singlets using CCSD/QZ at the MP2/QZ geometry (values marked with an `$*$' are calculated at the CCSD(T)/QZ//MP2/QZ level instead of EOM-CCSD/QZ). Spin-orbit corrections are obtained using the same approach with the dyall.cv2z basis set. For brevity Br$_2$  and H$_2$ are used, which corresponds to $^{79}$Br$^{81}$Br and HD.}
   \label{tbl:dis.soc}
   \def\arraystretch{1.5}
 \begin{center}
   \begin{tabular}{c|c|c||c}
    \hline
    \hline
        Dissociation Channel &          EOMCC     &  Spin-Orbit  &  Total \\
   \hline
{{$^2$CH$_2$BrI$^+  \rightarrow$}  \text{$^1$CH$_2$Br$^+$ + $^2$I}}  &   1.50 &  -0.21  &  1.29  \\
   \hline
{{$^2${CHLiBrI$^+   \rightarrow$}  \text{$^1$CHLiBr$^+$ + $^2$I}}}   &   2.24 &  -0.27  &  1.97  \\
   \hline
{$^2${CHNaBrI$^+    \rightarrow$}  \text{$^1$CHNaBr$^+$ + $^2$I}}    &   2.54 &  -0.28  &  2.26  \\
   \hline
{{$^1$CHCaBrI$^+    \rightarrow$}  \text{$^2$CHCaBr$^+$ + $^2$I}}    &   2.82 &  -0.27  &  2.55  \\
   \hline
   \hline
{{$^2$CH$_2$Br$_2^+ \rightarrow$}  \text{$^1$CH$_2$Br$^+$ + $^2$Br}} &   1.43 &  -0.14  &  1.29  \\
   \hline
{{$^2${CHLiBr$_2^+  \rightarrow$}  \text{$^1$CLiBrH$^+$ + $^2$Br}}}  &   2.72 &  -0.16  &  2.56  \\
{{$^2${CHLiBr$_2^+  \rightarrow$}  \text{$^2$CLiBr$^+$ + $^1$HBr}}}  &   2.68 &  -0.04  &  2.62  \\
   \hline
{$^2${CHNaBr$_2^+   \rightarrow$}  \text{$^2$CNaBr$^+$ + $^1$HBr}}   &   2.79 &  -0.04  &  2.75  \\
{$^2${CHNaBr$_2^+   \rightarrow$}  \text{$^1$CHNaBr$^+$ + $^2$Br}}   &   2.96 &  -0.15  &  2.81  \\
   \hline
{$^1${CHCaBr$_2^+   \rightarrow$}  \text{$^2$CHCaBr$^+$ + $^2$Br}}   &   3.25 &  -0.14  &  3.11  \\
   \hline
{{$^1$CHYbBr$_2^+   \rightarrow$}  \text{$^2$CHYbBr$^+$ + $^2$Br}}   & * 3.26 &  -0.17  &  3.09  \\
  \hline
  \hline
   \end{tabular}  \\
   \end{center}
 \end{table}

To estimate  the minimal acceptable dissociation threshold  we can consider, for example, the worst case scenario of a vibrational excitation along the C$-$H stretch mode where the photon energy is roughly $0.4$~eV (explicit calculations in Section~\ref{sec:structure}). 
Thus the energy of the first vibrational excitation for the C$-$H stretch is  $\sim0.4$~eV above the minimum electronic energy of CHDBrI$^+$.
In addition, the presence of the excitation laser may dissociate {the} molecule in the first excited vibrational state if it couples between that state and the continuum. 
This would cause a loss of population during the spectroscopy. 
However, this loss is  manageable since in a Ramsey spectroscopy experiment this dissociation would be limited to the $\frac{\pi}{2}$ pulses,~\cite{erez2022} but  could also occur from black body radiation. We would nonetheless prefer that the dissociation threshold is higher than 0.8~eV or about double the highest vibrational mode energy.
Alternatively, using a lower energy vibrational mode is also possible. 
{For a dissociation channel such as {{CH$_2$BrI$^+  \rightarrow$}  \text{CH$_2$Br$^+$ + I}} the zero point energies (ZPEs) of CH$_2$BrI$^+$ and the CH$_2$Br$^+$ fragment are similar since the Iodine is associated with the lowest energy vibrational modes.} 
{Therefore,} we neglect the contribution of the {ZPEs to the reduction of the dissociation threshold of}  the molecules {since their difference is small relative to the threshold}. 
The energies for the vibrational modes for the {cation-}candidates are shown in Table~\ref{tbl:catvib}. 
For the metal substituted candidates we can use the same cutoff of 0.8 eV as the molecular properties and dissociation channels are similar to the other systems.
The vibrational energies were calculated at the coupled-cluster level as well as using density functional theory with the  $\omega$B97M-V functional. 
The agreement between the two approaches suggests that  $\omega$B97M-V is also suitable for calculating the energies and frequencies of transition states between the two enantiomers.   

 \begin{table*}[]
        \caption{Vibrational transition frequencies of the cations in cm$^{-1}$. EOM-IP-CCSD/TZ (EOMCC) for doublet state cations and CCSD/TZ (CCSD) for  singlet. In addition, $\omega$B97M-V/TZ calculations are presented  as DFT. Herein the isotopic effect is considered explicitly using the mass of deuterium and using the different Br masses, i.e., Br$_2$=$^{79}$Br$^{81}$Br. The left column represents the mode number. {Vibrational scaling factors of 0.942-0.946 for  coupled cluster and 0.946-0.949 for  $\omega$B97M-V should be applied\cite{Irikura2005,NIST2022} as verified by our comparison of the computed vibrational transition frequencies for neutral CHDBrI and CHDBr$_2$ to experimental results in Table~\ref{freq_exprt} of Section~\ref{banchmark}.}}
        \label{tbl:catvib}
   \def\arraystretch{1.5}
 \begin{center}
   \begin{tabular}{c|cc|cc|cc|cc|cc}
    \hline
    \hline
  \# & \multicolumn{2}{c|}{$^2$CHDBrI$^+$} & \multicolumn{2}{c|}{$^2$CHDBr$_2^+$} & \multicolumn{2}{c|}{$^1$CHCaBrI$^+$} & \multicolumn{2}{c|}{$^1$CHCaBr$_2^+$} & \multicolumn{2}{c}{$^1$CHYbBr$_2^+$} \\
     &  EOMCC  & DFT   &  EOMCC  & DFT   &  CCSD  & DFT   &  CCSD  & DFT   &  CCSD  & DFT    \\
     \hline
  1  & 151.2  &  155.9  &  190.9  &  193.8  &  129.3  & 132.0  &  149.4  & 147.5  & 113.5  &  110.6  \\
  2  &   523  &  535    &    561  &  551    &  145.9  & 144.7  &  150.4  & 150.3  & 118.8  &  118.8  \\
  3  &   616  &  620    &    642  &  649    &    205  &  205   &    218  & 216    & 182.8  &  180.2  \\
  4  &   716  &  714    &    760  &  755    &    428  &  442   &    462  & 470    &   445  &  436    \\
  5  &   794  &  791    &    771  &  796    &    479  &  500   &    489  & 509    &   484  &  479    \\
  6  &  1104  & 1097    &   1057  & 1118    &    511  &  520   &    519  & 525    &   489  &  488    \\
  7  &  1284  & 1267    &   1295  & 1276    &    912  &  894   &    956  & 932    &   953  &  938    \\
  8  &  2362  & 2349    &   2365  & 2353    &   1095  & 1091   &   1116  & 1110   &  1111  & 1105    \\
  9  &  3225  & 3212    &   3228  & 3212    &   3202  & 3182   &   3208  & 3191   &  3206  & 3189    \\
  \hline
  \hline
   \end{tabular}  \\
   \end{center}
 \end{table*}

All the molecules CHXBrI$^+$ where X $\in \{ \mathrm{D,Li,Ca,Na}\}$ and  CHXBr$^+_2$ where X $\in \{ \mathrm{D,Li,Ca,Na,Yb}\}$ have dissociation thresholds above 0.8~eV. The higher the threshold the less susceptible the molecule will be to spontaneous dissociation by absorbing a photon from the black body.
This is in contrast to the other chiral molecular ions we have computed such as CHCl$_2$F$^+$, CHBrClF$^+$, CHBr$_2$F$^+$ and CHIBrCl$^+$ (Table~\ref{tbl:dis.nogood.2}), which range from instability with a negative energy for dissociation to weakly bound molecular ions with thresholds below the 0.8~eV cutoff.
These molecules would likely not support spectroscopy of the C$-$H stretch transition and other lower energy modes  without compromising the molecule lifetime.
Notably, we find that all the metal substituted molecules are very stable with a high dissociation threshold. This high stability combined with the predicted convenient wavelengths for STPI (Section~\ref{sec:STPI}), makes them particularly appealing for precision spectroscopy.

 \begin{table}[]
        \caption{The lowest CCSD(T) dissociation energies at the MP2 geometries using the TZ and QZ basis sets for the unsuitable candidates. In this case CHIBrCl$^+$ is close to the stability cutoff, but notice that inclusion of the Spin-Orbit contribution will reduce its threshold further. 
        }
   \label{tbl:dis.nogood.2}
   \def\arraystretch{1.5}
 \begin{center}
   \begin{tabular}{c|c|c||c|c|c}
    \hline
    \hline
          \text{Molecule}          &  TZ        &  QZ       & \text{Molecule}                  &  TZ        &     QZ  \\
 \hline
\multicolumn{3}{l||}{$^2${CHCl$_2$F$^+ \rightarrow$}}       & \multicolumn{3}{l}{$^2${CHBrClF$^+ \rightarrow$}}       \\
  \text{$^2$CCl$_2^+$ + $^1$HF  }  &    -0.06  &   -0.11  &  \text{$^1$CClHF$^+$ + $^2$Br }  &    0.42   &  0.53  \\
  \text{$^2$CClHF$^+$ + $^2$Cl }   &    -0.17  &   -0.09  &  \\
\hline
      \multicolumn{3}{l||}{$^2${CHBr$_2$F$^+ \rightarrow$}} & \multicolumn{3}{l}{$^2${CHIBrCl$^+ \rightarrow$}}       \\
 \text{$^1$CBrHF$^+$ + $^2$Br }  &     0.01  &    0.10    &   \text{$^1$CHBrCl$^+$ + $^2$I } &    0.62   &  0.74   \\
 \hline
  \hline
   \end{tabular}  \\
   \end{center}
 \end{table}

The dissociation energies for the candidates (Table~\ref{tbl:dis.nogood.2}) are computed at the CCSD(T) at the MP2 geometries.  Since these values are well below our 0.8 eV cutoff and these candidates are unlikely to be suitable, we did not perform the EOM-CCSD calculations as done for the candidates in Table~\ref{tbl:dis.soc}.  

With the molecule stability established, we turn to discuss  dissociation for state-selective detection. 
Using dissociation for state-selective detection is common in trapped molecular ion experiments due to the low dissociation thresholds.\cite{Ni2014,Zhou2020,Karr2012,carollo2018two} 
Since the fragments have a different mass than the parent molecular ion, they are straightforward to distinguish with high quantum efficiency.
The rise of ion traps that are coupled to mass spectrometers~\cite{Schowalter2012, Schmid2017} facilitates a high quantum efficiency avenue to detect these fragments and thus measure the internal state distribution of the parent molecular ion.

It is challenging to predict the dominant dissociation channel for these complicated 5-atom molecules that have a large density of states originating in multiple channels above the dissociation threshold. 
However, if we consider the different dissociation channels, we see that the wavelengths for a 2 photon resonant dissociation process should be feasible.
For example, for CHDBrI$^+$ the dissociation energy is 1.29 eV, which is comparable to commercially available diode lasers in the near infrared, which range from 0.8~eV to 1.8~eV. 
These lasers can be combined with a selective excitation step realized with microwaves or a second diode laser to create the state-selective 1+1{$^\prime$}  dissociation process, similarly to the STPI described in Section~\ref{sec:STPI}. 
The 1{$^\prime$}  photon can also be tuned to the most favorable dissociation process for detection, which is not necessarily immediately above the dissociation threshold. 
In Table~\ref{tbl:dis.mult}, we show additional  dissociation channel thresholds for the promising candidates to help guide the experiment. 
In this case the spin-orbit contribution is not included; however we expect a similar shift of the dissociation energies to the shifts noted in Table~\ref{tbl:dis.soc}. 
Thus, these values represent an estimate for experimental studies; a more refined computation will be necessary in the future.

 \begin{table*}[]
 	\caption{Multiple dissociation channel energies are listed in eV. The doublet states are calculated at the EOM-CCSD/QZ level and singlet states at CCSD/QZ. Values marked with a $*$ are calculated at the CCSD(T)/QZ level.}
   \label{tbl:dis.mult}
      \def\arraystretch{1.5}
 \begin{center}
   \begin{tabular}{c|c|c|c|c|c}
    \hline
    \hline
    Dissociation Channel           & EOMCC & Dissociation Channel                 &   EOMCC & Dissociation Channel       &   EOMCC    \\
   \hline
\multicolumn{2}{l|}{{$^2$CH$_2$Br$_2^+   \rightarrow$}}       &  \multicolumn{2}{l|}{$^2${CHLiBr$_2^+ \rightarrow$}} &  \multicolumn{2}{l}{$^2${CHNaBr$_2^+   \rightarrow$}}   \\
              \text{$^1$CH$_2$Br$^+$ + $^2$Br }   &   1.43  &   \text{$^2$CLiBr$^+$ + $^1$HBr}      &   2.67      &    \text{$^1$CHNaBr$^+$ + $^2$Br}      &    2.96     \\
              \text{$^2$CHBr$^+$ + $^1$HBr }      &   2.48  &   \text{$^1$CHLiBr$^+$ + $^2$Br}      &    2.71     &    \text{$^2$CNaBr$^+$ + $^1$HBr}      &    2.79     \\
                                                  &          &   \text{$^2$CHBr$^+$ + $^1$LiBr}      &  *3.51      &    \text{$^2$CHBr$^+$ + $^1$NaBr}      &   *3.56     \\
   \hline
  \multicolumn{2}{l|}{$^2${CH$_2$BrI$^+  \rightarrow$}} &   \multicolumn{2}{l|}{$^2${CHLiBrI$^+   \rightarrow$}}     &      \multicolumn{2}{l}{$^2${CHNaBrI$^+   \rightarrow$}}     \\
           \text{$^1$CH$_2$Br$^+$ + $^2$I} &   1.50 &          \text{$^1$CHLiBr$^+$ + $^2$I }  &    2.24 &              \text{$^1$CHNaBr$^+$ + $^2$I }  &   2.54   \\
           \text{$^1$CH$_2$I$^+$ + $^2$Br} &   1.76 &          \text{$^1$CHLiI$^+$ + $^2$Br }  &    2.69 &              \text{$^1$CHNaI$^+$ + $^2$Br }  &   3.01   \\
           \text{$^2$CHI$^+$ + $^1$HBr  }  &   2.74 &          \text{$^2$CLiI$^+$ + $^1$HBr }  &    2.68 &              \text{$^2$CNaI$^+$ + $^1$HBr }  &   2.90   \\
           \text{$^2$CHBr$^+$ + $^1$HI  }  &   3.12 &          \text{$^2$CLiBr$^+$ + $^1$HI }  &    2.77 &              \text{$^2$CNaBr$^+$ + $^1$HI }  &   2.94   \\
  \hline
\multicolumn{2}{l|}{{$^1$CHCaBrI$^+   \rightarrow$}}  &
\multicolumn{2}{l|}{$^1${CHCaBr$_2^+   \rightarrow$}} &  \multicolumn{2}{l}{{$^1$CHYbBr$_2^+   \rightarrow$}}    \\  
 \text{$^2$CHCaBr$^+$ + $^2$I}        &         2.82 &
  \text{$^2$CHCaBr$^+$ + $^2$Br}      &      3.25   &     \text{$^2$CHYbBr$^+$ + $^2$Br}      &      *3.26     \\
  \text{$^2$CHCaI$^+$ + $^2$Br}        &         3.27  &
  \text{$^1$CCaBr$^+$ + $^1$HBr}      &      4.42   &     
 \text{$^1$CHBr$_2^+$ +  $^1$Yb}     &       4.70     \\
 & &
  \text{$^1$CBr$_2$ +  $^1$CaH$^+$}   &      4.81   &     \text{$^1$CBr$_2$ +  $^1$YbH$^+$}   &       4.73      \\
  & &
  \text{$^1$CHBr$_2^+$ +  $^1$Ca}     &      4.96   &     \text{$^1$CYbBr$^+$ + $^1$HBr}      &       4.44  \\
\hline
  \hline
   \end{tabular}  \\
   \end{center}
 \end{table*}

A second avenue we consider for detection is dissociation by a single photon process.
In this approach  the fingerprints of the internal state of the molecular ion will  appear in the kinetic energies and angular distributions of the photo-fragments. 
The parent molecule's internal state will affect the kinetic energy of the photofragments due to energy conservation since a single photon is absorbed.
We aim to probe two rotational states in this manner.\cite{erez2022} For distinguishing between vibrational states, this method may not be as straight forward since the energy difference may remain within one of the photofragments.
While measurement of the kinetic energy of photo-fragments has been achieved when dissociating trapped molecular ions,\cite{Zhou2020} the energy resolution needed to resolve individual rotational states of the parent molecule would need to be  below 10 m/s for the candidates considered. 
An ion trap that is optimized for detection of photofragment energies by coupling to a velocity map imaging detector should be able to reach the required resolutions.
Currently, we are building such an apparatus. The ion trajectory simulations, which will be reported elsewhere, indicate that this is feasible.

As indicated in Table~\ref{tbl:dis.soc} for the single photon dissociation process, realizing the required wavelengths is also possible with a single photon for a wide range of diode and pulsed dye lasers.

\subsection{Cation geometries}
\label{sec:structure}

The geometry of the molecular cation is one of the most important properties  that must be verified to support a measurement of PNC through comparison of vibrational transition frequencies. 
For many 4-atom candidates such as hydrogen peroxide, the chiral geometry does not survive the removal of an electron, where the cation has a planar structure.~\cite{Changala2017}
However, even the neutral form of hydrogen peroxide does not support time-invariant chiral states since it has a relatively low transition state between enantiomers relative to twist mode energy splitting.
The eigenstates of such a system are the symmetric and anti-symmetric superposition states of the S and R molecule configurations.\cite{Quack2008}

\begin{figure}[]

   \begin{tabular}{|c|c|}
\hline
   \includegraphics[width=0.45\columnwidth, angle=0,
draft=false,clip=true]{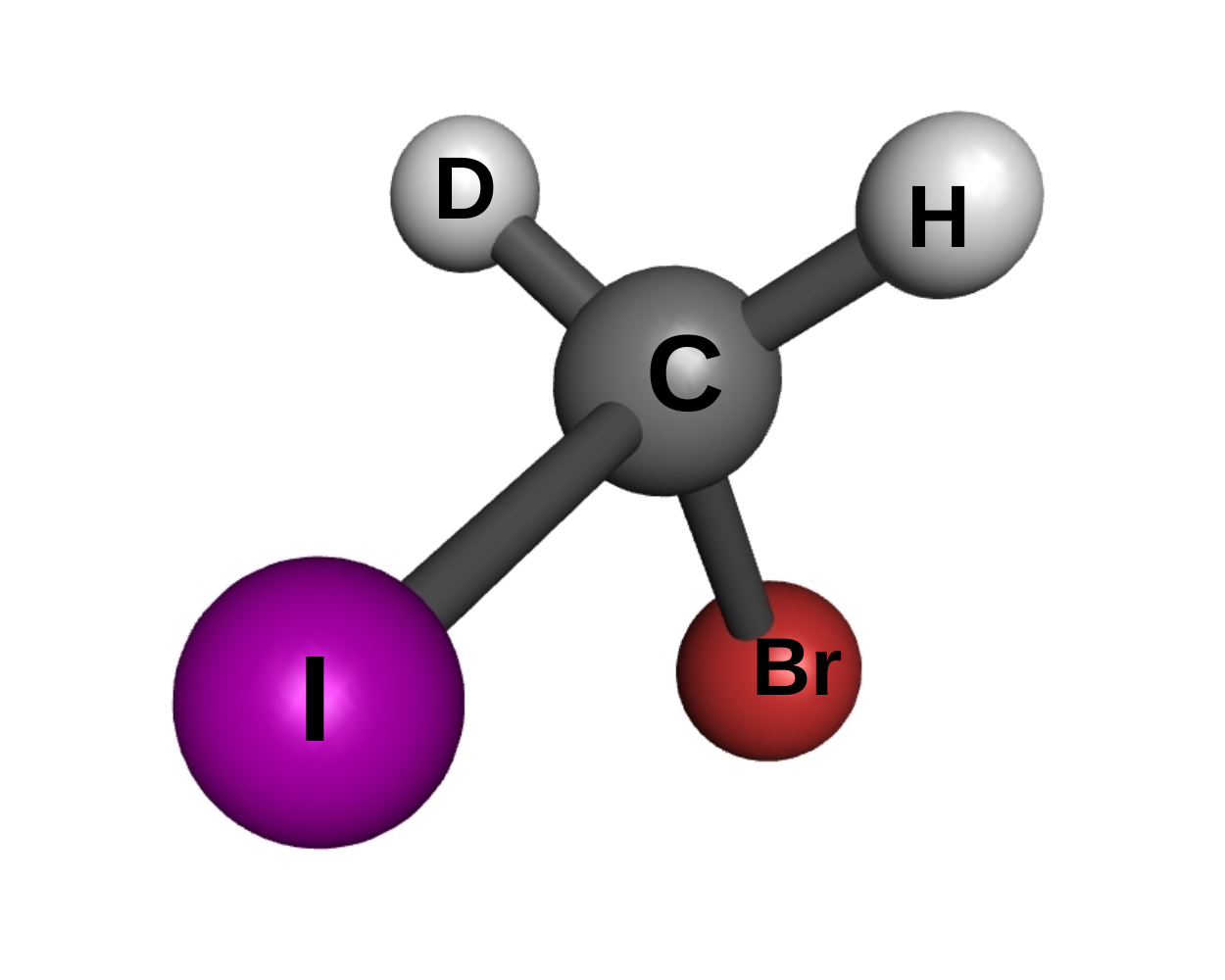}
&
   \includegraphics[width=0.45\columnwidth, angle=0,
draft=false,clip=true]{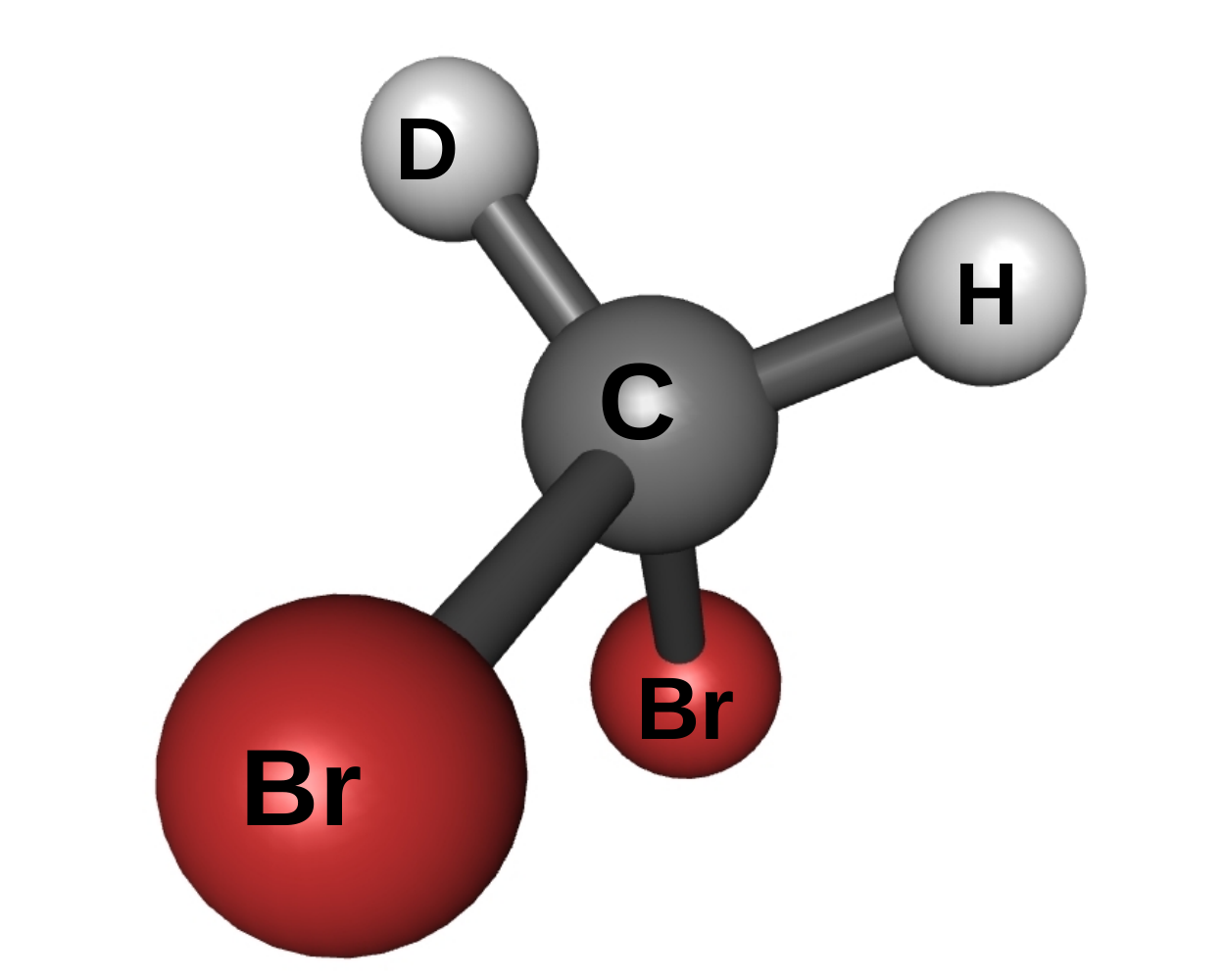}
\\
$^1$CHDBrI$^+$
&
$^1$CHDBr$_2^+$
\\
$TS = 1.30$~eV & $TS= 1.39$~eV
\\
\hline
\includegraphics[width=0.475\columnwidth, angle=0,scale=1,
draft=false,clip=true]{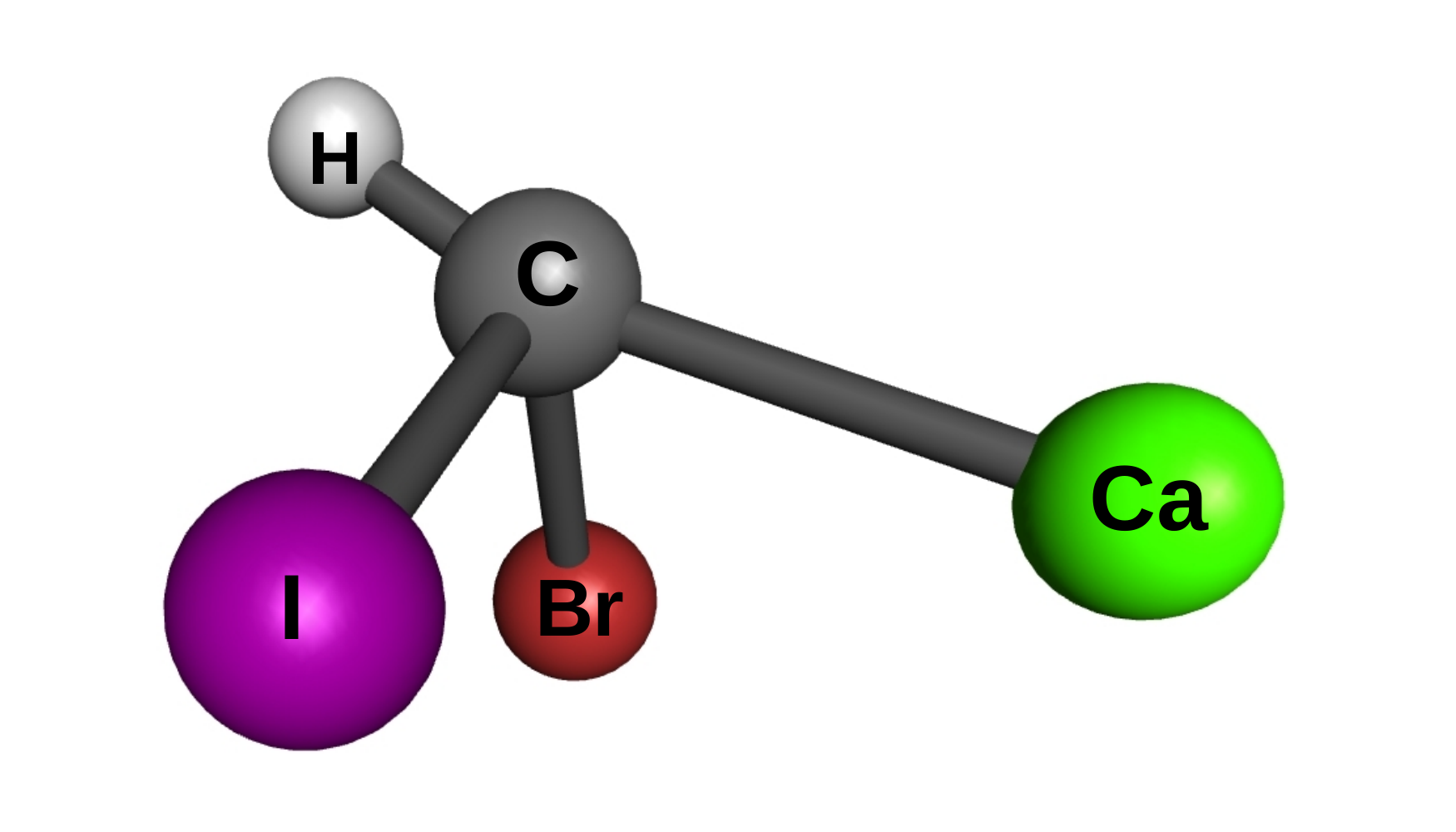} 
&
\includegraphics[width=0.475\columnwidth, angle=0,scale=1,
draft=false,clip=true]{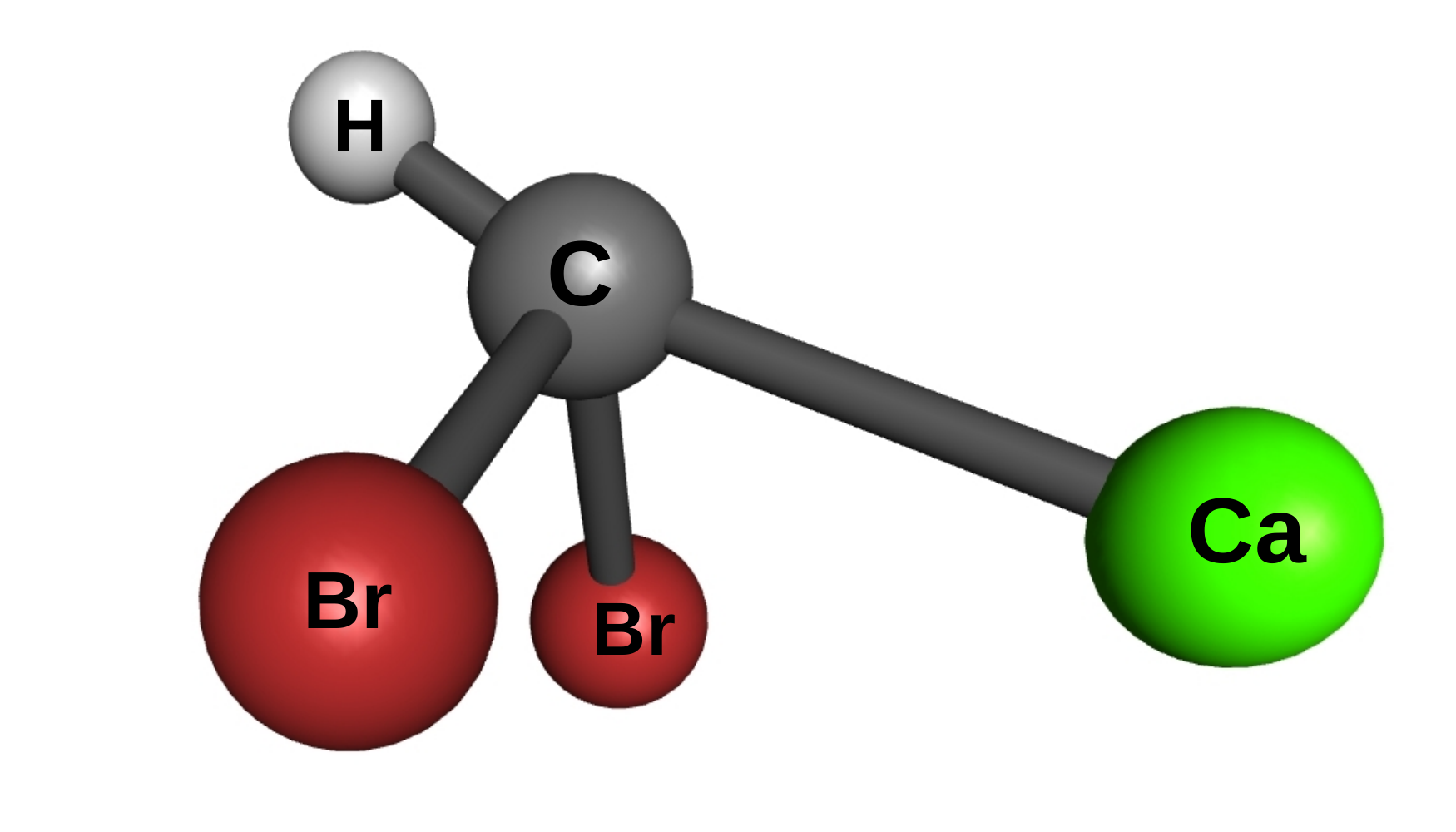}\
\\
$^2$CHCaBrI$^+$
&
$^2$CHCaBr$_2^+$
\\
$TS = 0.82$~eV & $TS = 0.78$~eV
\\
\hline

&
\includegraphics[width=0.475\columnwidth, angle=0,scale=1,
draft=false,clip=true]{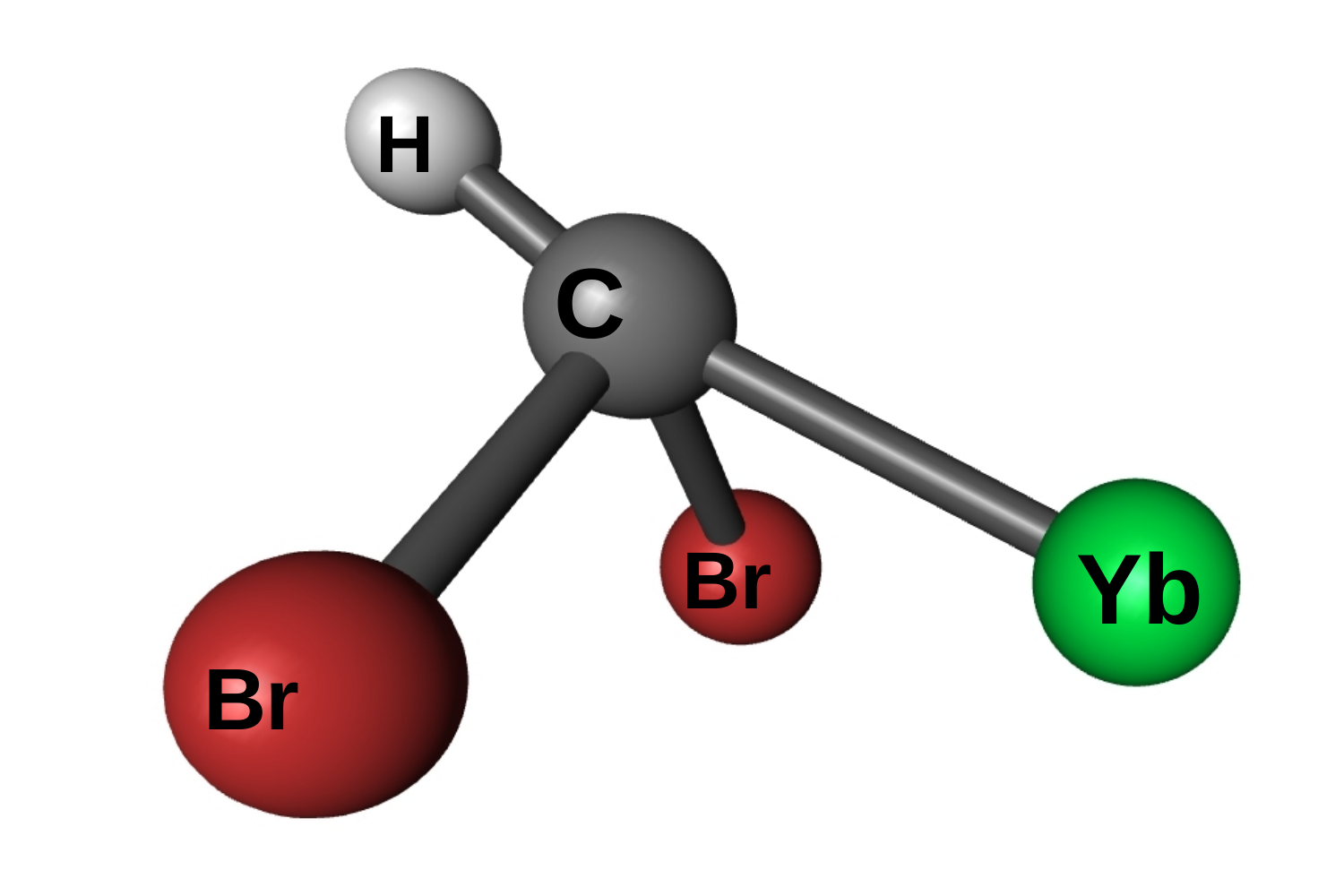}
\\
&
$^2$CHYbBr$_2^+$
\\
 & $TS = 0.70$~eV
\\
\hline

\includegraphics[width=0.475\columnwidth, angle=0,
draft=false,clip=true]{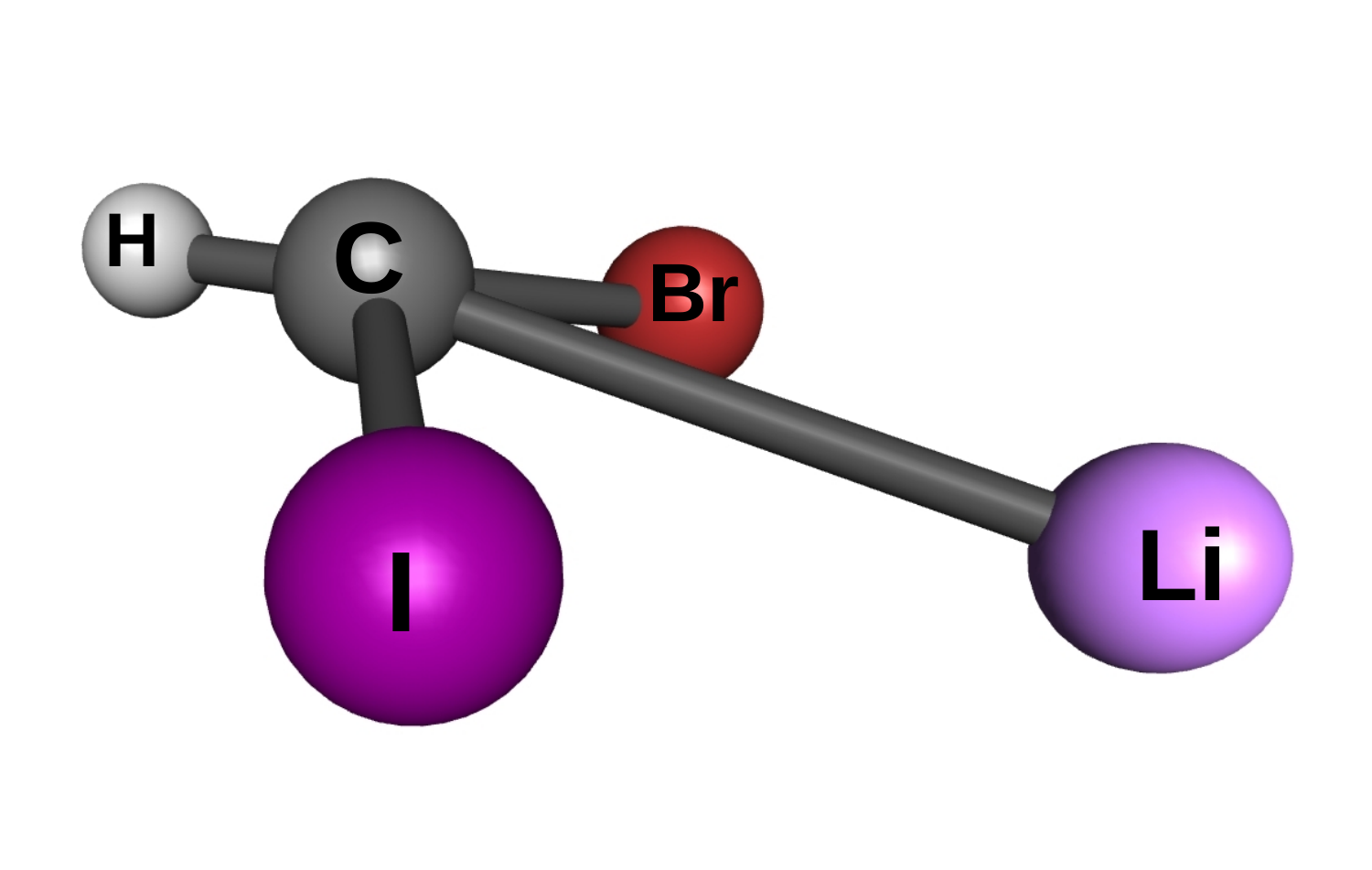}
&
\includegraphics[width=0.475\columnwidth, angle=0,
draft=false,clip=true]{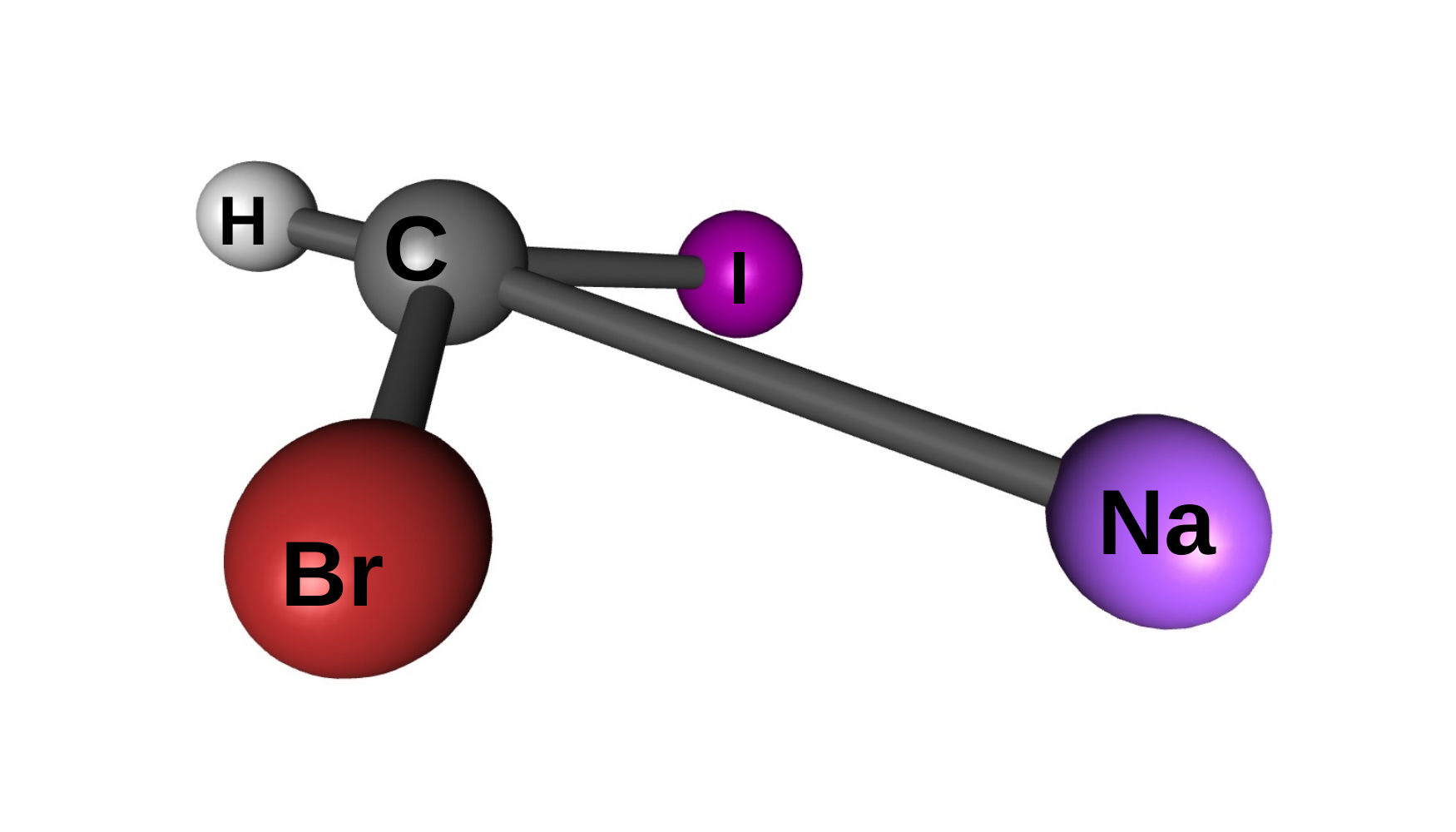} 

\\
$^1$CHLiBrI$^+$  & $^1$CHNaBrI$^+$
\\
$TS = 0.01$~eV   &  $TS=0.03$~eV
\\

\hline
   \end{tabular}  
\caption{Geometries, EOM-CC/QZ for doublet cations and MP2/QZ for singlets. The top five molecules have a significant transition state ($TS$) energy examined at the $\omega$B97M-V/TZ level. The Li and Na substituted molecules have near planar geometries with a very low $TS$ energy.}
    \label{tbl:geo_table}

\end{figure}

For our candidates, we search for  stable chiral geometry which can be determined if the transition state between the two enantiomers is sufficiently high.
First, we calculate the minimal-energy geometry of the molecular ion candidates to verify that it is indeed chiral. These geometries can be seen in Figure~\ref{tbl:geo_table}. While all of these molecules have a chiral structure, the geometries of both the lithium and sodium substituted molecules are very close to planar. 
This near planar geometry hints at a low barrier between the two enantiomers. Indeed a search for a planar transition state ($TS$) reveals that its energy is only 0.01~eV and 0.03 eV for CHLiBrI$^+$ and CHNaBrI$^+$ respectively. The transition states are planar and have a single imaginary frequency out of the 9 normal modes. Thus we can conclude through comparison with the zero point energies of the molecules that the chiral ground state of these molecules will not be time invariant.

On the other hand, the rest of the candidates whose chiral structure is more acute survive this test.
Their geometries and transition state energies are shown in Figure~\ref{tbl:geo_table}.
For CHDBrI$^+$ and CHDBr$_2^+$, the transition state energies at 1.30 and 1.39~eV are significantly higher than all the vibrational mode energies in the system including the first excited state of the C$-$H stretch mode.
For the other metal substituted candidates, the transition states are all above 0.7 eV, which is higher than the zero point energy for all the molecules.
In particular, all the bending modes that overlap with the enantiomer mutation coordinate have at least 5 states below the transition state barrier, and this barrier scale is approximately equal to the energy of the first excited state of the C$-$H stretch mode.

Another effect that might limit the lifetime of excited vibrational states is known as intra-molecular vibrational redistribution (IVR).\cite{Nesbitt1996} To a certain extent, the rate of IVR is determined by the vibrational state density at the excited energy. Taking all the different vibrational combinations, we find that the state density at the energy of the first excited state of the C$-$H stretch mode ($v_9=1$, where $v_9$ is the vibrational quantum number of mode \#9) is 1.3 and 0.9 states per cm$^{-1}$ for CHDBrI$^+$  and CHDBr$_2^+$ respectively. At the energy of the excited state of the C$-$D  stretch mode ($v_8=1$) the densities are 3 times lower. For CHCaBrI$^+$ and CHCaBr$_2^+$ the density at $v_9=1$ is significantly higher, ranging from 35 to 25 states per cm$^{-1}$, but for the C$-$H bend modes ($v_8=1$) the density drops below 1 state per cm$^{-1}$ for both calcium substituted molecules. These densities are much higher relative to the effective density that should be considered as redistribution is less likely to proceed to combinations of more than two modes. An accidental overlap is unlikely for these modes when comparing to the natural linewidth of the vibrational excited state even when considering the densities resulting from combinations of all modes.

These are good indications for long-lived excited vibrational states.
We do not explicitly estimate the state lifetimes in the current manuscript. However,  spectroscopy using lower energy vibrational modes would enhance the lifetimes with respect to dissociation.

Additionally, in order to support 3-wave mixing schemes such as the one presented in Ref.~\citenum{erez2022} we also need the rotational constants of these molecular ions. These are shown in Tab.~\ref{tbl:catrot}. The two deuterated molecules are approximately prolate symmetric tops while the Yb substituted molecule may be approximated as an oblate top.

 \begin{table}[]
        \caption{Rotational constants of the cations in MHz. Herein the isotopic effect is considered explicitly using the mass of deuterium and the two different Br isotope masses, i.e., Br$_2$=$^{79}$Br$^{81}$Br.}
        \label{tbl:catrot}
   \def\arraystretch{1.5}
 \begin{center}
   \begin{tabular}{c|c|c|c|c}
    \hline
    \hline
   CHDBrI$^+$ & CHDBr$_2^+$ & CHCaBrI$^+$ & CHCaBr$_2^+$ & CHYbBr$_2^+$ \\
     \hline
   14750   &  15610  & 2550  & 2705  &  1170  \\
   1170   &  1645  & 830  & 1175  &  1070  \\
   1095   &  1510  & 660  & 870  &  585  \\
  \hline
  \hline
   \end{tabular}  \\
   \end{center}
 \end{table}

Finally we discuss the sensitivity of our candidates to Zeeman shifts.
The chiral molecular ions have two spin multiplicities for the candidates.
Closed-shell molecules such as CHCaBrI$^+$ will have a small magnetic moment which makes them favorably immune to magnetic field drifts.

Although we have skipped the computations for CHYbBrI$^+$ in this work, we can compare the results between BrI-containing molecules and the Br$_2$-containing molecules. In particular, we observe the similarities  between CHCaBr$_2^+$ and CHCaBrI$^+$ as well as between CHCaBr$_2^+$ and CHYbBr$_2^+$ with respect to the ionization wavelengths, dissociation threshold and transition state energy. These similarities in  molecules that are explicitly examined, lead us to infer that CHYbBrI$^+$ is also a promising candidate from experimental perspective.

\section{Parity violating frequency shifts for selected candidate molecules}
The previous sections (\ref{sec:STPI}, \ref{sec:stability}) discussed pathways to create cold molecules, which affects the contrast, and the stability of the molecular ions, which relates to the accessible coherence time in Eq.~(\ref{eq:qpn}). However, to estimate the number of molecules that  need to be measured in a precision measurement assuming that the  quantum projection limit is achieved \cite{erez2022}, we must compare the magnitude of the expected shift due to PNC in the vibrational frequency to the expected precision $\delta f$.

Here we present the PNC calculations for the different vibrational modes of  CHCaBrI$^+$. Table~\ref{tbl:pv_comparison} shows the PNC shifts expected for the various modes, with 100~mHz and 29~mHz shifts for the most relevant modes for precision spectroscopy, the C$-$H bend and stretch modes (\# 8 and 9 in Tab.~\ref{tbl:catvib}), respectively.  We also computed the PNC shifts for the doubly isotopically chiral CHD$^{79}$Br$^{81}$Br$^+$, which are significantly lower, probably due to the relatively low mass of its constituents. However, its symmetric structure may simplify the molecule's spectrum, promoting other aspects of the experiment. In contrast, the PNC frequency shift in CHDBrI$^+$ is very large, on the order of 1~Hz for most of its higher energy modes, and is fully reported in Ref.~\citenum{eduardus2023}.

For the PNC calculations, the molecular geometry was optimized on the $\omega$B97M-V/Def2-TZVPP level of theory using Q-Chem 5.2.2.\cite{qchem}

\begin{table}[]
\caption{PNC frequency difference between for between enantiomers for each vibrational mode of CHCaBrI$^+$ and CHD$^{79}$Br$^{81}$Br$^+$ in Hz.}
\label{tbl:pv_comparison}
\def\arraystretch{1.5}
\begin{center}
\begin{tabular}{c|c|c}

\hline
\hline

Mode \# &  {  CHCaBrI$^+$   }       & {   CHD$^{79}$Br$^{81}$Br$^+$    }     \\ \hline
1                         & {-2.8E-2}     & 4.3E-5     \\ 
2                         & {-5.6E-2}     & -3.1E-6    \\ 
3                         & {1.5E-1}      & 3.3E-3     \\ 
4                         & {6.6E-2}      & -3.0E-3    \\ 
5                         & {-2.0E-1}     & 2.9E-3     \\ 
6                         & {-2.0E-3}     & -7.7E-3    \\ 
7                         & {1.9E-1}      & -4.8E-3    \\ 
8                         & {-1.0E-1}     & 1.0E-2     \\ 
9                         & {-2.9E-2}     & -1.3E-2    \\ \hline \hline
\end{tabular}
\end{center}
\end{table}

In order to obtain the PV contributions to the total energies, we carried out single point relativistic DFT calculations using the DIRAC23 program.\cite{DIRAC23} In order to conserve computational effort, we replaced the 4-component Dirac Hamiltonian by the exact 2-component (X2C) Hamiltonian, where the large and the small component are exactly decoupled and the positive energy spectrum of the 4-component Hamiltonian is reproduced within numerical precision. In this scheme, the spin-same-orbit interactions are introduced in a mean-field fashion by use of the AMFI procedure.\cite{AMFI} We used the CAM-B3LYP* functional, the parameters of which were adjusted by Thierfelder et al.\ to reproduce the PV energy shifts obtained using coupled cluster calculations.\cite{camb3lyp*} Dyall's v4z basis sets were used for all the elements.\cite{Dyall2006} Alongside the relativistic absolute energy at each geometry, these calculations also yield the PV energy contribution, $E_\mathrm{PV}$.

To calculate the vibrational parity violating frequency shifts, the parity violating shift of the vibrational ground and first excited states are needed; these we obtained as follows. We performed relativistic single-point calculations at 11 equally spaced points between -0.5 and 0.5 \AA { }along the selected normal mode. This yields the potential energy and the parity violating energy as a function of the normal coordinate $q$; we fitted polynomials to these points in order to create smooth potential and parity violating energy curves $V(q)$ and $V_\mathrm{PV}(q)$.

Next, we numerically solve the Schrödinger equation for $V(q)$ using the Numerov-Cooley procedure  as implemented by Bast and obtain the vibrational wavefunctions $|{n}\rangle$.\cite{numerov1,Cooley_1961,Bast_2017} The parity violating shift of the $n$th vibrational level in the first order of perturbation theory is then 
\begin{equation} \label{eq:E_pv_int}
  E^\mathrm{PV}_n=\langle{n}|V_\mathrm{PV}(q)|{n}\rangle.
\end{equation}
The difference between the enantiomers in the frequency of a transition from level $m$ to level $n$ is then given by 
\begin{equation}
    \Delta \nu^\mathrm{PV}_{m\rightarrow n}=\frac{2}{h}(E^\mathrm{PV}_n-E^\mathrm{PV}_m),
\end{equation}
with $h$ the Planck constant. The factor 2 arises since when in one enantiomer the energy shifts up by $E^\mathrm{PV}$, it shifts down by the same amount in the other enantiomer. 

\section{Molecule preparation strategies}
Any candidate discussed here requires a pathway for its creation if it is to be used in a precision measurement. For some of the candidates such as CHDBrI and CHDBr$_2$ the non-chiral counterpart CH$_2$BrI and CH$_2$Br$_2$ is commercially available. The natural abundance between the two bromine isotopes is 1:1, leaving 50\% of the molecules in the pro-chiral  mixed isotope form. The molecules with different bromine isotopes  can be chosen through mass selection. For the deuterated molecules selection is also an option but the very low natural abundance means that a synthesis method is preferable. 

The Ca substituted molecules would need to be generated in the vacuum chamber. For example, it may be possible to generate CHCaBrI by creation of a Ca plasma by laser ablation near a supersonic expansion seeded with CH$_2$BrI or CHBr$_2$I. A similar scheme\cite{brazier1987} has been used to generate CH$_3$Ca by ablation of Ca near CH$_3$Cl. Another similar scheme  has been proposed to create Yb substituted methanes,\cite{chamorro2022molecular} which may be a pathway to generate CHYbBrI.

\section{Summary and Outlook}
The search for PNC in molecules can benefit from long interrogation times accessible in trapped chiral molecular ions as well as the enhanced PNC shifts they are predicted to exhibit. 
However, for a successful precision spectroscopy experiment with chiral molecular ions a favorable candidate must also fulfill other criteria, including efficient state preparation, high quantum efficiency in detection, as well as resistance to predissociation when vibrationally excited.
In this work we investigate these properties for several five-atom, carbon-center, tetrahedral chiral cations via {\it ab initio} calculations and estimate the magnitude of the PNC shift for some of the candidates.

To this end, we calculate several electronic properties mainly using coupled-cluster based methods.
We validate the chirality of the optimized cation geometries and calculate vertical and adiabatic ionization energies, dissociation channel energies, isotopic vibrational modes, rotational constants, transition state energies, several excitation energies of the neutral systems and PNC frequency shifts where relevant.

Our in depth study in search of  candidates for the trapped chiral-molecular-ion precision spectroscopy has revealed that CHXBrI$^+$, where X $\in$ \{D, Ca\}, and isotopically chiral CHX$^{79}$Br$^{81}$Br$^+$, where X $\in$  \{D, Ca, Yb\}, are favorable candidates. These candidates have promising avenues toward their preparation with internally cold temperatures and are stable in the charged form. 
Moreover, the magnitude of vibrational frequency shifts due to PNC is shown to be significant for selected candidates.

\begin{acknowledgements}
    
We thank I. Gilary for useful discussions. E.E. thanks Peter Schwerdtfeger for useful discussions. This research was supported by the Israel Science Foundation Grant No.\ 1661/19 and the Israel Science Foundation Grant No.\ 1142/21. This research project was partially supported by the Helen Diller Quantum Center at the Technion. E.E. wishes to acknowledge Indonesia Endowment Fund for Education/\textit{Lembaga Pengelola Dana Pendidikan} \text{(LPDP)} for research funding. A.B., E.E., and L.F.P thank the Center for Information Technology of the University of Groningen for their support and for providing access to the Peregrine high-performance computing cluster.

\end{acknowledgements}

\section*{Data Availability Statement}
The data that support the findings of this study are available from the corresponding author upon reasonable request.

\section*{References}
\bibliography{citations}

\begin{thebibliography}{86}%
\makeatletter
\providecommand \@ifxundefined [1]{%
 \@ifx{#1\undefined}
}%
\providecommand \@ifnum [1]{%
 \ifnum #1\expandafter \@firstoftwo
 \else \expandafter \@secondoftwo
 \fi
}%
\providecommand \@ifx [1]{%
 \ifx #1\expandafter \@firstoftwo
 \else \expandafter \@secondoftwo
 \fi
}%
\providecommand \natexlab [1]{#1}%
\providecommand \enquote  [1]{``#1''}%
\providecommand \bibnamefont  [1]{#1}%
\providecommand \bibfnamefont [1]{#1}%
\providecommand \citenamefont [1]{#1}%
\providecommand \href@noop [0]{\@secondoftwo}%
\providecommand \href [0]{\begingroup \@sanitize@url \@href}%
\providecommand \@href[1]{\@@startlink{#1}\@@href}%
\providecommand \@@href[1]{\endgroup#1\@@endlink}%
\providecommand \@sanitize@url [0]{\catcode `\\12\catcode `\$12\catcode
  `\&12\catcode `\#12\catcode `\^12\catcode `\_12\catcode `\%12\relax}%
\providecommand \@@startlink[1]{}%
\providecommand \@@endlink[0]{}%
\providecommand \url  [0]{\begingroup\@sanitize@url \@url }%
\providecommand \@url [1]{\endgroup\@href {#1}{\urlprefix }}%
\providecommand \urlprefix  [0]{URL }%
\providecommand \Eprint [0]{\href }%
\providecommand \doibase [0]{https://doi.org/}%
\providecommand \selectlanguage [0]{\@gobble}%
\providecommand \bibinfo  [0]{\@secondoftwo}%
\providecommand \bibfield  [0]{\@secondoftwo}%
\providecommand \translation [1]{[#1]}%
\providecommand \BibitemOpen [0]{}%
\providecommand \bibitemStop [0]{}%
\providecommand \bibitemNoStop [0]{.\EOS\space}%
\providecommand \EOS [0]{\spacefactor3000\relax}%
\providecommand \BibitemShut  [1]{\csname bibitem#1\endcsname}%
\let\auto@bib@innerbib\@empty
\bibitem [{\citenamefont {Wu}\ \emph {et~al.}(1957)\citenamefont {Wu},
  \citenamefont {Ambler}, \citenamefont {Hayward}, \citenamefont {Hoppes},\
  and\ \citenamefont {Hudson}}]{wu1957}%
  \BibitemOpen
  \bibfield  {author} {\bibinfo {author} {\bibfnamefont {C.-S.}\ \bibnamefont
  {Wu}}, \bibinfo {author} {\bibfnamefont {E.}~\bibnamefont {Ambler}}, \bibinfo
  {author} {\bibfnamefont {R.~W.}\ \bibnamefont {Hayward}}, \bibinfo {author}
  {\bibfnamefont {D.~D.}\ \bibnamefont {Hoppes}},\ and\ \bibinfo {author}
  {\bibfnamefont {R.~P.}\ \bibnamefont {Hudson}},\ }\bibfield  {title}
  {\enquote {\bibinfo {title} {Experimental test of parity conservation in beta
  decay},}\ }\href@noop {} {\bibfield  {journal} {\bibinfo  {journal} {Physical
  review}\ }\textbf {\bibinfo {volume} {105}},\ \bibinfo {pages} {1413}
  (\bibinfo {year} {1957})}\BibitemShut {NoStop}%
\bibitem [{\citenamefont {Bouchiat}\ \emph {et~al.}(1982)\citenamefont
  {Bouchiat}, \citenamefont {Guena}, \citenamefont {Hunter},\ and\
  \citenamefont {Pottier}}]{BOUCHIAT1982358}%
  \BibitemOpen
  \bibfield  {author} {\bibinfo {author} {\bibfnamefont {M.}~\bibnamefont
  {Bouchiat}}, \bibinfo {author} {\bibfnamefont {J.}~\bibnamefont {Guena}},
  \bibinfo {author} {\bibfnamefont {L.}~\bibnamefont {Hunter}},\ and\ \bibinfo
  {author} {\bibfnamefont {L.}~\bibnamefont {Pottier}},\ }\bibfield  {title}
  {\enquote {\bibinfo {title} {Observation of a parity violation in cesium},}\
  }\href {https://doi.org/https://doi.org/10.1016/0370-2693(82)90736-5}
  {\bibfield  {journal} {\bibinfo  {journal} {Physics Letters B}\ }\textbf
  {\bibinfo {volume} {117}},\ \bibinfo {pages} {358--364} (\bibinfo {year}
  {1982})}\BibitemShut {NoStop}%
\bibitem [{\citenamefont {Wood}\ \emph {et~al.}(1997)\citenamefont {Wood},
  \citenamefont {Bennett}, \citenamefont {Cho}, \citenamefont {Masterson},
  \citenamefont {Roberts}, \citenamefont {Tanner},\ and\ \citenamefont
  {Wieman}}]{Wood1997}%
  \BibitemOpen
  \bibfield  {author} {\bibinfo {author} {\bibfnamefont {C.~S.}\ \bibnamefont
  {Wood}}, \bibinfo {author} {\bibfnamefont {S.~C.}\ \bibnamefont {Bennett}},
  \bibinfo {author} {\bibfnamefont {D.}~\bibnamefont {Cho}}, \bibinfo {author}
  {\bibfnamefont {B.~P.}\ \bibnamefont {Masterson}}, \bibinfo {author}
  {\bibfnamefont {J.~L.}\ \bibnamefont {Roberts}}, \bibinfo {author}
  {\bibfnamefont {C.~E.}\ \bibnamefont {Tanner}},\ and\ \bibinfo {author}
  {\bibfnamefont {C.~E.}\ \bibnamefont {Wieman}},\ }\bibfield  {title}
  {\enquote {\bibinfo {title} {Measurement of parity nonconservation and an
  anapole moment in cesium},}\ }\href
  {https://doi.org/10.1126/science.275.5307.1759} {\bibfield  {journal}
  {\bibinfo  {journal} {Science}\ }\textbf {\bibinfo {volume} {275}},\ \bibinfo
  {pages} {1759--1763} (\bibinfo {year} {1997})}\BibitemShut {NoStop}%
\bibitem [{\citenamefont {Yamagata}(1966)}]{YAMAGATA1966495}%
  \BibitemOpen
  \bibfield  {author} {\bibinfo {author} {\bibfnamefont {Y.}~\bibnamefont
  {Yamagata}},\ }\bibfield  {title} {\enquote {\bibinfo {title} {A hypothesis
  for the asymmetric appearance of biomolecules on earth},}\ }\href
  {https://doi.org/https://doi.org/10.1016/0022-5193(66)90110-X} {\bibfield
  {journal} {\bibinfo  {journal} {Journal of Theoretical Biology}\ }\textbf
  {\bibinfo {volume} {11}},\ \bibinfo {pages} {495--498} (\bibinfo {year}
  {1966})}\BibitemShut {NoStop}%
\bibitem [{\citenamefont {Quack}, \citenamefont {Stohner},\ and\ \citenamefont
  {Willeke}(2008)}]{Quack2008}%
  \BibitemOpen
  \bibfield  {author} {\bibinfo {author} {\bibfnamefont {M.}~\bibnamefont
  {Quack}}, \bibinfo {author} {\bibfnamefont {J.}~\bibnamefont {Stohner}},\
  and\ \bibinfo {author} {\bibfnamefont {M.}~\bibnamefont {Willeke}},\
  }\bibfield  {title} {\enquote {\bibinfo {title} {{High-Resolution
  Spectroscopic Studies and Theory of Parity Violation in Chiral Molecules}},}\
  }\href {https://doi.org/10.1146/annurev.physchem.58.032806.104511} {\bibfield
   {journal} {\bibinfo  {journal} {Annual Review of Physical Chemistry}\
  }\textbf {\bibinfo {volume} {59}},\ \bibinfo {pages} {741--769} (\bibinfo
  {year} {2008})}\BibitemShut {NoStop}%
\bibitem [{\citenamefont {Quack}, \citenamefont {Seyfang},\ and\ \citenamefont
  {Wichmann}(2022)}]{quack2022}%
  \BibitemOpen
  \bibfield  {author} {\bibinfo {author} {\bibfnamefont {M.}~\bibnamefont
  {Quack}}, \bibinfo {author} {\bibfnamefont {G.}~\bibnamefont {Seyfang}},\
  and\ \bibinfo {author} {\bibfnamefont {G.}~\bibnamefont {Wichmann}},\
  }\bibfield  {title} {\enquote {\bibinfo {title} {Perspectives on parity
  violation in chiral molecules: theory{,} spectroscopic experiment and
  biomolecular homochirality},}\ }\href {https://doi.org/10.1039/D2SC01323A}
  {\bibfield  {journal} {\bibinfo  {journal} {Chem. Sci.}\ }\textbf {\bibinfo
  {volume} {13}},\ \bibinfo {pages} {10598--10643} (\bibinfo {year}
  {2022})}\BibitemShut {NoStop}%
\bibitem [{\citenamefont {Hund}(1927)}]{Hund1927}%
  \BibitemOpen
  \bibfield  {author} {\bibinfo {author} {\bibfnamefont {F.}~\bibnamefont
  {Hund}},\ }\bibfield  {title} {\enquote {\bibinfo {title} {Zur deutung der
  molekelspektren. iii. - bemerkungen über das schwingungs- und
  rotationsspektrum bei molekeln mit mehr als zwei kernen},}\ }\href
  {https://doi.org/10.1007/BF01397249} {\bibfield  {journal} {\bibinfo
  {journal} {Zeitschrift für Physik}\ }\textbf {\bibinfo {volume} {43}},\
  \bibinfo {pages} {805--826} (\bibinfo {year} {1927})}\BibitemShut {NoStop}%
\bibitem [{\citenamefont {Letokhov}(1975)}]{LETOKHOV1975275}%
  \BibitemOpen
  \bibfield  {author} {\bibinfo {author} {\bibfnamefont {V.}~\bibnamefont
  {Letokhov}},\ }\bibfield  {title} {\enquote {\bibinfo {title} {On difference
  of energy levels of left and right molecules due to weak interactions},}\
  }\href {https://doi.org/https://doi.org/10.1016/0375-9601(75)90064-X}
  {\bibfield  {journal} {\bibinfo  {journal} {Physics Letters A}\ }\textbf
  {\bibinfo {volume} {53}},\ \bibinfo {pages} {275--276} (\bibinfo {year}
  {1975})}\BibitemShut {NoStop}%
\bibitem [{\citenamefont {Senami}\ and\ \citenamefont
  {Ito}(2019)}]{Senami2019}%
  \BibitemOpen
  \bibfield  {author} {\bibinfo {author} {\bibfnamefont {M.}~\bibnamefont
  {Senami}}\ and\ \bibinfo {author} {\bibfnamefont {K.}~\bibnamefont {Ito}},\
  }\bibfield  {title} {\enquote {\bibinfo {title} {Asymmetry of electron
  chirality between enantiomeric pair molecules and the origin of homochirality
  in nature},}\ }\href {https://doi.org/10.1103/PhysRevA.99.012509} {\bibfield
  {journal} {\bibinfo  {journal} {Physical Review A}\ }\textbf {\bibinfo
  {volume} {99}},\ \bibinfo {pages} {012509} (\bibinfo {year}
  {2019})}\BibitemShut {NoStop}%
\bibitem [{\citenamefont {Quack}(2002)}]{Quack2002}%
  \BibitemOpen
  \bibfield  {author} {\bibinfo {author} {\bibfnamefont {M.}~\bibnamefont
  {Quack}},\ }\bibfield  {title} {\enquote {\bibinfo {title} {How important is
  parity violation for molecular and biomolecular chirality?}}\ }\href
  {https://doi.org/10.1002/anie.200290005} {\bibfield  {journal} {\bibinfo
  {journal} {Angewandte Chemie International Edition}\ }\textbf {\bibinfo
  {volume} {41}},\ \bibinfo {pages} {4618--4630} (\bibinfo {year}
  {2002})}\BibitemShut {NoStop}%
\bibitem [{\citenamefont {Glavin}\ \emph {et~al.}(2020)\citenamefont {Glavin},
  \citenamefont {Burton}, \citenamefont {Elsila}, \citenamefont {Aponte},\ and\
  \citenamefont {Dworkin}}]{Glavin2019}%
  \BibitemOpen
  \bibfield  {author} {\bibinfo {author} {\bibfnamefont {D.~P.}\ \bibnamefont
  {Glavin}}, \bibinfo {author} {\bibfnamefont {A.~S.}\ \bibnamefont {Burton}},
  \bibinfo {author} {\bibfnamefont {J.~E.}\ \bibnamefont {Elsila}}, \bibinfo
  {author} {\bibfnamefont {J.~C.}\ \bibnamefont {Aponte}},\ and\ \bibinfo
  {author} {\bibfnamefont {J.~P.}\ \bibnamefont {Dworkin}},\ }\bibfield
  {title} {\enquote {\bibinfo {title} {The search for chiral asymmetry as a
  potential biosignature in our solar system},}\ }\href
  {https://doi.org/10.1021/acs.chemrev.9b00474} {\bibfield  {journal} {\bibinfo
   {journal} {Chemical Reviews}\ }\textbf {\bibinfo {volume} {120}},\ \bibinfo
  {pages} {4660--4689} (\bibinfo {year} {2020})}\BibitemShut {NoStop}%
\bibitem [{\citenamefont {Daussy}\ \emph {et~al.}(1999)\citenamefont {Daussy},
  \citenamefont {Marrel}, \citenamefont {Amy-Klein}, \citenamefont {Nguyen},
  \citenamefont {Bord{\'{e}}},\ and\ \citenamefont {Chardonnet}}]{Daussy1999}%
  \BibitemOpen
  \bibfield  {author} {\bibinfo {author} {\bibfnamefont {C.}~\bibnamefont
  {Daussy}}, \bibinfo {author} {\bibfnamefont {T.}~\bibnamefont {Marrel}},
  \bibinfo {author} {\bibfnamefont {A.}~\bibnamefont {Amy-Klein}}, \bibinfo
  {author} {\bibfnamefont {C.~T.}\ \bibnamefont {Nguyen}}, \bibinfo {author}
  {\bibfnamefont {C.~J.}\ \bibnamefont {Bord{\'{e}}}},\ and\ \bibinfo {author}
  {\bibfnamefont {C.}~\bibnamefont {Chardonnet}},\ }\bibfield  {title}
  {\enquote {\bibinfo {title} {{Limit on the Parity Nonconserving Energy
  Difference between the Enantiomers of a Chiral Molecule by Laser
  Spectroscopy}},}\ }\href {https://doi.org/10.1103/PhysRevLett.83.1554}
  {\bibfield  {journal} {\bibinfo  {journal} {Physical Review Letters}\
  }\textbf {\bibinfo {volume} {83}},\ \bibinfo {pages} {1554} (\bibinfo {year}
  {1999})}\BibitemShut {NoStop}%
\bibitem [{\citenamefont {Satterthwaite}\ \emph {et~al.}(2021)\citenamefont
  {Satterthwaite}, \citenamefont {Koumarianou}, \citenamefont {Sorensen},\ and\
  \citenamefont {Patterson}}]{Satterthwaite2021}%
  \BibitemOpen
  \bibfield  {author} {\bibinfo {author} {\bibfnamefont {L.}~\bibnamefont
  {Satterthwaite}}, \bibinfo {author} {\bibfnamefont {G.}~\bibnamefont
  {Koumarianou}}, \bibinfo {author} {\bibfnamefont {D.}~\bibnamefont
  {Sorensen}},\ and\ \bibinfo {author} {\bibfnamefont {D.}~\bibnamefont
  {Patterson}},\ }\bibfield  {title} {\enquote {\bibinfo {title} {{Sub-Hz
  Differential Rotational Spectroscopy of Enantiomers}},}\ }\href
  {https://doi.org/10.3390/SYM14010028} {\bibfield  {journal} {\bibinfo
  {journal} {Symmetry 2022, Vol. 14, Page 28}\ }\textbf {\bibinfo {volume}
  {14}},\ \bibinfo {pages} {28} (\bibinfo {year} {2021})}\BibitemShut {NoStop}%
\bibitem [{\citenamefont {Albert}\ \emph {et~al.}(2017)\citenamefont {Albert},
  \citenamefont {Bolotova}, \citenamefont {Chen}, \citenamefont {F{\'{a}}bri},
  \citenamefont {Quack}, \citenamefont {Seyfang},\ and\ \citenamefont
  {Zindel}}]{Albert2017}%
  \BibitemOpen
  \bibfield  {author} {\bibinfo {author} {\bibfnamefont {S.}~\bibnamefont
  {Albert}}, \bibinfo {author} {\bibfnamefont {I.}~\bibnamefont {Bolotova}},
  \bibinfo {author} {\bibfnamefont {Z.}~\bibnamefont {Chen}}, \bibinfo {author}
  {\bibfnamefont {C.}~\bibnamefont {F{\'{a}}bri}}, \bibinfo {author}
  {\bibfnamefont {M.}~\bibnamefont {Quack}}, \bibinfo {author} {\bibfnamefont
  {G.}~\bibnamefont {Seyfang}},\ and\ \bibinfo {author} {\bibfnamefont
  {D.}~\bibnamefont {Zindel}},\ }\bibfield  {title} {\enquote {\bibinfo {title}
  {{High-resolution FTIR spectroscopy of trisulfane HSSSH: a candidate for
  detecting parity violation in chiral molecules}},}\ }\href
  {https://doi.org/10.1039/C7CP01139C} {\bibfield  {journal} {\bibinfo
  {journal} {Physical Chemistry Chemical Physics}\ }\textbf {\bibinfo {volume}
  {19}},\ \bibinfo {pages} {11738--11743} (\bibinfo {year} {2017})}\BibitemShut
  {NoStop}%
\bibitem [{\citenamefont {Stoeffler}\ \emph {et~al.}(2010)\citenamefont
  {Stoeffler}, \citenamefont {Darqui{\'{e}}}, \citenamefont {Shelkovnikov},
  \citenamefont {Daussy}, \citenamefont {Amy-Klein}, \citenamefont
  {Chardonnet}, \citenamefont {Guy}, \citenamefont {Crassous}, \citenamefont
  {Huet}, \citenamefont {Soulard},\ and\ \citenamefont
  {Asselin}}]{Stoeffler2010}%
  \BibitemOpen
  \bibfield  {author} {\bibinfo {author} {\bibfnamefont {C.}~\bibnamefont
  {Stoeffler}}, \bibinfo {author} {\bibfnamefont {B.}~\bibnamefont
  {Darqui{\'{e}}}}, \bibinfo {author} {\bibfnamefont {A.}~\bibnamefont
  {Shelkovnikov}}, \bibinfo {author} {\bibfnamefont {C.}~\bibnamefont
  {Daussy}}, \bibinfo {author} {\bibfnamefont {A.}~\bibnamefont {Amy-Klein}},
  \bibinfo {author} {\bibfnamefont {C.}~\bibnamefont {Chardonnet}}, \bibinfo
  {author} {\bibfnamefont {L.}~\bibnamefont {Guy}}, \bibinfo {author}
  {\bibfnamefont {J.}~\bibnamefont {Crassous}}, \bibinfo {author}
  {\bibfnamefont {T.~R.}\ \bibnamefont {Huet}}, \bibinfo {author}
  {\bibfnamefont {P.}~\bibnamefont {Soulard}},\ and\ \bibinfo {author}
  {\bibfnamefont {P.}~\bibnamefont {Asselin}},\ }\bibfield  {title} {\enquote
  {\bibinfo {title} {{High resolution spectroscopy of methyltrioxorhenium:
  towards the observation of parity violation in chiral molecules}},}\ }\href
  {https://doi.org/10.1039/C0CP01806F} {\bibfield  {journal} {\bibinfo
  {journal} {Physical Chemistry Chemical Physics}\ }\textbf {\bibinfo {volume}
  {13}},\ \bibinfo {pages} {854--863} (\bibinfo {year} {2010})}\BibitemShut
  {NoStop}%
\bibitem [{\citenamefont {Cournol}\ \emph {et~al.}(2019)\citenamefont
  {Cournol}, \citenamefont {Manceau}, \citenamefont {Pierens}, \citenamefont
  {Lecordier}, \citenamefont {Tran}, \citenamefont {Santagata}, \citenamefont
  {Argence}, \citenamefont {Goncharov}, \citenamefont {Lopez}, \citenamefont
  {Abgrall}, \citenamefont {{Le Coq}}, \citenamefont {{Le Targat}},
  \citenamefont {{Alvarez Martinez}}, \citenamefont {Lee}, \citenamefont {Xu},
  \citenamefont {Pottie}, \citenamefont {Hendricks}, \citenamefont {Wall},
  \citenamefont {Bieniewska}, \citenamefont {Sauer}, \citenamefont {Tarbutt},
  \citenamefont {Amy-Klein}, \citenamefont {Tokunaga},\ and\ \citenamefont
  {Darqui{\'{e}}}}]{Cournol2019}%
  \BibitemOpen
  \bibfield  {author} {\bibinfo {author} {\bibfnamefont {A.}~\bibnamefont
  {Cournol}}, \bibinfo {author} {\bibfnamefont {M.}~\bibnamefont {Manceau}},
  \bibinfo {author} {\bibfnamefont {M.}~\bibnamefont {Pierens}}, \bibinfo
  {author} {\bibfnamefont {L.}~\bibnamefont {Lecordier}}, \bibinfo {author}
  {\bibfnamefont {D.~B.~A.}\ \bibnamefont {Tran}}, \bibinfo {author}
  {\bibfnamefont {R.}~\bibnamefont {Santagata}}, \bibinfo {author}
  {\bibfnamefont {B.}~\bibnamefont {Argence}}, \bibinfo {author} {\bibfnamefont
  {A.}~\bibnamefont {Goncharov}}, \bibinfo {author} {\bibfnamefont
  {O.}~\bibnamefont {Lopez}}, \bibinfo {author} {\bibfnamefont
  {M.}~\bibnamefont {Abgrall}}, \bibinfo {author} {\bibfnamefont
  {Y.}~\bibnamefont {{Le Coq}}}, \bibinfo {author} {\bibfnamefont
  {R.}~\bibnamefont {{Le Targat}}}, \bibinfo {author} {\bibfnamefont
  {H.}~\bibnamefont {{Alvarez Martinez}}}, \bibinfo {author} {\bibfnamefont
  {W.~K.}\ \bibnamefont {Lee}}, \bibinfo {author} {\bibfnamefont
  {D.}~\bibnamefont {Xu}}, \bibinfo {author} {\bibfnamefont {P.~E.}\
  \bibnamefont {Pottie}}, \bibinfo {author} {\bibfnamefont {R.~J.}\
  \bibnamefont {Hendricks}}, \bibinfo {author} {\bibfnamefont {T.~E.}\
  \bibnamefont {Wall}}, \bibinfo {author} {\bibfnamefont {J.~M.}\ \bibnamefont
  {Bieniewska}}, \bibinfo {author} {\bibfnamefont {B.~E.}\ \bibnamefont
  {Sauer}}, \bibinfo {author} {\bibfnamefont {M.~R.}\ \bibnamefont {Tarbutt}},
  \bibinfo {author} {\bibfnamefont {A.}~\bibnamefont {Amy-Klein}}, \bibinfo
  {author} {\bibfnamefont {S.~K.}\ \bibnamefont {Tokunaga}},\ and\ \bibinfo
  {author} {\bibfnamefont {B.}~\bibnamefont {Darqui{\'{e}}}},\ }\bibfield
  {title} {\enquote {\bibinfo {title} {{A new experiment to test parity
  symmetry in cold chiral molecules using vibrational spectroscopy}},}\ }\href
  {https://doi.org/10.1070/QEL16880} {\bibfield  {journal} {\bibinfo  {journal}
  {Quantum Electronics}\ }\textbf {\bibinfo {volume} {49}},\ \bibinfo {pages}
  {288--292} (\bibinfo {year} {2019})}\BibitemShut {NoStop}%
\bibitem [{\citenamefont {Fiechter}\ \emph {et~al.}(2022)\citenamefont
  {Fiechter}, \citenamefont {Haase}, \citenamefont {Saleh}, \citenamefont
  {Soulard}, \citenamefont {Tremblay}, \citenamefont {Havenith}, \citenamefont
  {Timmermans}, \citenamefont {Schwerdtfeger}, \citenamefont {Crassous},
  \citenamefont {Darquié}, \citenamefont {Pašteka},\ and\ \citenamefont
  {Borschevsky}}]{Fiechter2021}%
  \BibitemOpen
  \bibfield  {author} {\bibinfo {author} {\bibfnamefont {M.~R.}\ \bibnamefont
  {Fiechter}}, \bibinfo {author} {\bibfnamefont {P.~A.~B.}\ \bibnamefont
  {Haase}}, \bibinfo {author} {\bibfnamefont {N.}~\bibnamefont {Saleh}},
  \bibinfo {author} {\bibfnamefont {P.}~\bibnamefont {Soulard}}, \bibinfo
  {author} {\bibfnamefont {B.}~\bibnamefont {Tremblay}}, \bibinfo {author}
  {\bibfnamefont {R.~W.~A.}\ \bibnamefont {Havenith}}, \bibinfo {author}
  {\bibfnamefont {R.~G.~E.}\ \bibnamefont {Timmermans}}, \bibinfo {author}
  {\bibfnamefont {P.}~\bibnamefont {Schwerdtfeger}}, \bibinfo {author}
  {\bibfnamefont {J.}~\bibnamefont {Crassous}}, \bibinfo {author}
  {\bibfnamefont {B.}~\bibnamefont {Darquié}}, \bibinfo {author}
  {\bibfnamefont {L.~F.}\ \bibnamefont {Pašteka}},\ and\ \bibinfo {author}
  {\bibfnamefont {A.}~\bibnamefont {Borschevsky}},\ }\bibfield  {title}
  {\enquote {\bibinfo {title} {Toward detection of the molecular parity
  violation in chiral {Ru(acac)$_3$ and Os(acac)$_3$}},}\ }\href
  {https://doi.org/10.1021/acs.jpclett.2c02434} {\bibfield  {journal} {\bibinfo
   {journal} {The Journal of Physical Chemistry Letters}\ }\textbf {\bibinfo
  {volume} {13}},\ \bibinfo {pages} {10011--10017} (\bibinfo {year}
  {2022})}\BibitemShut {NoStop}%
\bibitem [{\citenamefont {Figgen}\ and\ \citenamefont
  {Schwerdtfeger}(2009)}]{Figgen2009}%
  \BibitemOpen
  \bibfield  {author} {\bibinfo {author} {\bibfnamefont {D.}~\bibnamefont
  {Figgen}}\ and\ \bibinfo {author} {\bibfnamefont {P.}~\bibnamefont
  {Schwerdtfeger}},\ }\bibfield  {title} {\enquote {\bibinfo {title}
  {{Structures, inversion barriers, and parity violation effects in chiral
  SeOXY molecules (X,Y=H, F, Cl, Br, or I)}},}\ }\href
  {https://doi.org/10.1063/1.3072370} {\bibfield  {journal} {\bibinfo
  {journal} {The Journal of Chemical Physics}\ }\textbf {\bibinfo {volume}
  {130}},\ \bibinfo {pages} {054306} (\bibinfo {year} {2009})}\BibitemShut
  {NoStop}%
\bibitem [{\citenamefont {Cairncross}\ \emph {et~al.}(2017)\citenamefont
  {Cairncross}, \citenamefont {Gresh}, \citenamefont {Grau}, \citenamefont
  {Cossel}, \citenamefont {Roussy}, \citenamefont {Ni}, \citenamefont {Zhou},
  \citenamefont {Ye},\ and\ \citenamefont {Cornell}}]{Cairncross2017}%
  \BibitemOpen
  \bibfield  {author} {\bibinfo {author} {\bibfnamefont {W.~B.}\ \bibnamefont
  {Cairncross}}, \bibinfo {author} {\bibfnamefont {D.~N.}\ \bibnamefont
  {Gresh}}, \bibinfo {author} {\bibfnamefont {M.}~\bibnamefont {Grau}},
  \bibinfo {author} {\bibfnamefont {K.~C.}\ \bibnamefont {Cossel}}, \bibinfo
  {author} {\bibfnamefont {T.~S.}\ \bibnamefont {Roussy}}, \bibinfo {author}
  {\bibfnamefont {Y.}~\bibnamefont {Ni}}, \bibinfo {author} {\bibfnamefont
  {Y.}~\bibnamefont {Zhou}}, \bibinfo {author} {\bibfnamefont {J.}~\bibnamefont
  {Ye}},\ and\ \bibinfo {author} {\bibfnamefont {E.~A.}\ \bibnamefont
  {Cornell}},\ }\bibfield  {title} {\enquote {\bibinfo {title} {{Precision
  Measurement of the Electron's Electric Dipole Moment Using Trapped Molecular
  Ions}},}\ }\href {https://doi.org/10.1103/PhysRevLett.119.153001} {\bibfield
  {journal} {\bibinfo  {journal} {Physical Review Letters}\ }\textbf {\bibinfo
  {volume} {119}},\ \bibinfo {pages} {153001} (\bibinfo {year}
  {2017})}\BibitemShut {NoStop}%
\bibitem [{\citenamefont {Zhou}\ \emph {et~al.}(2020)\citenamefont {Zhou},
  \citenamefont {Shagam}, \citenamefont {Cairncross}, \citenamefont {Ng},
  \citenamefont {Roussy}, \citenamefont {Grogan}, \citenamefont {Boyce},
  \citenamefont {Vigil}, \citenamefont {Pettine}, \citenamefont {Zelevinsky},
  \citenamefont {Ye},\ and\ \citenamefont {Cornell}}]{Zhou2020}%
  \BibitemOpen
  \bibfield  {author} {\bibinfo {author} {\bibfnamefont {Y.}~\bibnamefont
  {Zhou}}, \bibinfo {author} {\bibfnamefont {Y.}~\bibnamefont {Shagam}},
  \bibinfo {author} {\bibfnamefont {W.~B.}\ \bibnamefont {Cairncross}},
  \bibinfo {author} {\bibfnamefont {K.~B.}\ \bibnamefont {Ng}}, \bibinfo
  {author} {\bibfnamefont {T.~S.}\ \bibnamefont {Roussy}}, \bibinfo {author}
  {\bibfnamefont {T.}~\bibnamefont {Grogan}}, \bibinfo {author} {\bibfnamefont
  {K.}~\bibnamefont {Boyce}}, \bibinfo {author} {\bibfnamefont
  {A.}~\bibnamefont {Vigil}}, \bibinfo {author} {\bibfnamefont
  {M.}~\bibnamefont {Pettine}}, \bibinfo {author} {\bibfnamefont
  {T.}~\bibnamefont {Zelevinsky}}, \bibinfo {author} {\bibfnamefont
  {J.}~\bibnamefont {Ye}},\ and\ \bibinfo {author} {\bibfnamefont {E.~A.}\
  \bibnamefont {Cornell}},\ }\bibfield  {title} {\enquote {\bibinfo {title}
  {{Second-scale coherence measured at the quantum projection noise limit with
  hundreds of molecular ions}},}\ }\href
  {https://doi.org/10.1103/PhysRevLett.124.053201} {\bibfield  {journal}
  {\bibinfo  {journal} {Physical Review Letters}\ }\textbf {\bibinfo {volume}
  {124}},\ \bibinfo {pages} {053201} (\bibinfo {year} {2020})}\BibitemShut
  {NoStop}%
\bibitem [{\citenamefont {Roussy}\ \emph {et~al.}(2023)\citenamefont {Roussy},
  \citenamefont {Caldwell}, \citenamefont {Wright}, \citenamefont {Cairncross},
  \citenamefont {Shagam}, \citenamefont {Ng}, \citenamefont {Schlossberger},
  \citenamefont {Park}, \citenamefont {Wang}, \citenamefont {Ye},\ and\
  \citenamefont {Cornell}}]{roussy2022new}%
  \BibitemOpen
  \bibfield  {author} {\bibinfo {author} {\bibfnamefont {T.~S.}\ \bibnamefont
  {Roussy}}, \bibinfo {author} {\bibfnamefont {L.}~\bibnamefont {Caldwell}},
  \bibinfo {author} {\bibfnamefont {T.}~\bibnamefont {Wright}}, \bibinfo
  {author} {\bibfnamefont {W.~B.}\ \bibnamefont {Cairncross}}, \bibinfo
  {author} {\bibfnamefont {Y.}~\bibnamefont {Shagam}}, \bibinfo {author}
  {\bibfnamefont {K.~B.}\ \bibnamefont {Ng}}, \bibinfo {author} {\bibfnamefont
  {N.}~\bibnamefont {Schlossberger}}, \bibinfo {author} {\bibfnamefont {S.~Y.}\
  \bibnamefont {Park}}, \bibinfo {author} {\bibfnamefont {A.}~\bibnamefont
  {Wang}}, \bibinfo {author} {\bibfnamefont {J.}~\bibnamefont {Ye}},\ and\
  \bibinfo {author} {\bibfnamefont {E.~A.}\ \bibnamefont {Cornell}},\
  }\bibfield  {title} {\enquote {\bibinfo {title} {An improved bound on the
  electron’s electric dipole moment},}\ }\href
  {https://doi.org/10.1126/science.adg4084} {\bibfield  {journal} {\bibinfo
  {journal} {Science}\ }\textbf {\bibinfo {volume} {381}},\ \bibinfo {pages}
  {46--50} (\bibinfo {year} {2023})}\BibitemShut {NoStop}%
\bibitem [{\citenamefont {Hanneke}, \citenamefont {Carollo},\ and\
  \citenamefont {Lane}(2016)}]{Hanneke2016}%
  \BibitemOpen
  \bibfield  {author} {\bibinfo {author} {\bibfnamefont {D.}~\bibnamefont
  {Hanneke}}, \bibinfo {author} {\bibfnamefont {R.~A.}\ \bibnamefont
  {Carollo}},\ and\ \bibinfo {author} {\bibfnamefont {D.~A.}\ \bibnamefont
  {Lane}},\ }\bibfield  {title} {\enquote {\bibinfo {title} {{High sensitivity
  to variation in the proton-to-electron mass ratio in O{$_2^+$}}},}\ }\href
  {https://doi.org/10.1103/PHYSREVA.94.050101/FIGURES/3/MEDIUM} {\bibfield
  {journal} {\bibinfo  {journal} {Physical Review A}\ }\textbf {\bibinfo
  {volume} {94}},\ \bibinfo {pages} {050101} (\bibinfo {year}
  {2016})}\BibitemShut {NoStop}%
\bibitem [{\citenamefont {Fan}\ \emph {et~al.}(2021)\citenamefont {Fan},
  \citenamefont {Holliman}, \citenamefont {Shi}, \citenamefont {Zhang},
  \citenamefont {Straus}, \citenamefont {Li}, \citenamefont {Buechele},\ and\
  \citenamefont {Jayich}}]{Fan2021}%
  \BibitemOpen
  \bibfield  {author} {\bibinfo {author} {\bibfnamefont {M.}~\bibnamefont
  {Fan}}, \bibinfo {author} {\bibfnamefont {C.~A.}\ \bibnamefont {Holliman}},
  \bibinfo {author} {\bibfnamefont {X.}~\bibnamefont {Shi}}, \bibinfo {author}
  {\bibfnamefont {H.}~\bibnamefont {Zhang}}, \bibinfo {author} {\bibfnamefont
  {M.~W.}\ \bibnamefont {Straus}}, \bibinfo {author} {\bibfnamefont
  {X.}~\bibnamefont {Li}}, \bibinfo {author} {\bibfnamefont {S.~W.}\
  \bibnamefont {Buechele}},\ and\ \bibinfo {author} {\bibfnamefont {A.~M.}\
  \bibnamefont {Jayich}},\ }\bibfield  {title} {\enquote {\bibinfo {title}
  {Optical mass spectrometry of cold raoh+ and raoch3+},}\ }\href
  {https://doi.org/10.1103/PHYSREVLETT.126.023002/FIGURES/4/MEDIUM} {\bibfield
  {journal} {\bibinfo  {journal} {Physical Review Letters}\ }\textbf {\bibinfo
  {volume} {126}},\ \bibinfo {pages} {023002} (\bibinfo {year}
  {2021})}\BibitemShut {NoStop}%
\bibitem [{\citenamefont {Roussy}\ \emph {et~al.}(2021)\citenamefont {Roussy},
  \citenamefont {Palken}, \citenamefont {Cairncross}, \citenamefont {Brubaker},
  \citenamefont {Gresh}, \citenamefont {Grau}, \citenamefont {Cossel},
  \citenamefont {Ng}, \citenamefont {Shagam}, \citenamefont {Zhou},
  \citenamefont {Flambaum}, \citenamefont {Lehnert}, \citenamefont {Ye},\ and\
  \citenamefont {Cornell}}]{Roussy2021}%
  \BibitemOpen
  \bibfield  {author} {\bibinfo {author} {\bibfnamefont {T.~S.}\ \bibnamefont
  {Roussy}}, \bibinfo {author} {\bibfnamefont {D.~A.}\ \bibnamefont {Palken}},
  \bibinfo {author} {\bibfnamefont {W.~B.}\ \bibnamefont {Cairncross}},
  \bibinfo {author} {\bibfnamefont {B.~M.}\ \bibnamefont {Brubaker}}, \bibinfo
  {author} {\bibfnamefont {D.~N.}\ \bibnamefont {Gresh}}, \bibinfo {author}
  {\bibfnamefont {M.}~\bibnamefont {Grau}}, \bibinfo {author} {\bibfnamefont
  {K.~C.}\ \bibnamefont {Cossel}}, \bibinfo {author} {\bibfnamefont {K.~B.}\
  \bibnamefont {Ng}}, \bibinfo {author} {\bibfnamefont {Y.}~\bibnamefont
  {Shagam}}, \bibinfo {author} {\bibfnamefont {Y.}~\bibnamefont {Zhou}},
  \bibinfo {author} {\bibfnamefont {V.~V.}\ \bibnamefont {Flambaum}}, \bibinfo
  {author} {\bibfnamefont {K.~W.}\ \bibnamefont {Lehnert}}, \bibinfo {author}
  {\bibfnamefont {J.}~\bibnamefont {Ye}},\ and\ \bibinfo {author}
  {\bibfnamefont {E.~A.}\ \bibnamefont {Cornell}},\ }\bibfield  {title}
  {\enquote {\bibinfo {title} {Experimental constraint on axionlike particles
  over seven orders of magnitude in mass},}\ }\href
  {https://doi.org/10.1103/PHYSREVLETT.126.171301/FIGURES/3/MEDIUM} {\bibfield
  {journal} {\bibinfo  {journal} {Physical Review Letters}\ }\textbf {\bibinfo
  {volume} {126}},\ \bibinfo {pages} {171301} (\bibinfo {year}
  {2021})}\BibitemShut {NoStop}%
\bibitem [{\citenamefont {Chou}\ \emph {et~al.}(2020)\citenamefont {Chou},
  \citenamefont {Collopy}, \citenamefont {Kurz}, \citenamefont {Lin},
  \citenamefont {Harding}, \citenamefont {Plessow}, \citenamefont {Fortier},
  \citenamefont {Diddams}, \citenamefont {Leibfried},\ and\ \citenamefont
  {Leibrandt}}]{Chou2020}%
  \BibitemOpen
  \bibfield  {author} {\bibinfo {author} {\bibfnamefont {C.~W.}\ \bibnamefont
  {Chou}}, \bibinfo {author} {\bibfnamefont {A.~L.}\ \bibnamefont {Collopy}},
  \bibinfo {author} {\bibfnamefont {C.}~\bibnamefont {Kurz}}, \bibinfo {author}
  {\bibfnamefont {Y.}~\bibnamefont {Lin}}, \bibinfo {author} {\bibfnamefont
  {M.~E.}\ \bibnamefont {Harding}}, \bibinfo {author} {\bibfnamefont {P.~N.}\
  \bibnamefont {Plessow}}, \bibinfo {author} {\bibfnamefont {T.}~\bibnamefont
  {Fortier}}, \bibinfo {author} {\bibfnamefont {S.}~\bibnamefont {Diddams}},
  \bibinfo {author} {\bibfnamefont {D.}~\bibnamefont {Leibfried}},\ and\
  \bibinfo {author} {\bibfnamefont {D.~R.}\ \bibnamefont {Leibrandt}},\
  }\bibfield  {title} {\enquote {\bibinfo {title} {Frequency-comb spectroscopy
  on pure quantum states of a single molecular ion},}\ }\href
  {https://doi.org/10.1126/science.aba3628} {\bibfield  {journal} {\bibinfo
  {journal} {Science}\ }\textbf {\bibinfo {volume} {367}},\ \bibinfo {pages}
  {1458--1461} (\bibinfo {year} {2020})}\BibitemShut {NoStop}%
\bibitem [{\citenamefont {Tong}, \citenamefont {Winney},\ and\ \citenamefont
  {Willitsch}(2010)}]{Tong2010}%
  \BibitemOpen
  \bibfield  {author} {\bibinfo {author} {\bibfnamefont {X.}~\bibnamefont
  {Tong}}, \bibinfo {author} {\bibfnamefont {A.~H.}\ \bibnamefont {Winney}},\
  and\ \bibinfo {author} {\bibfnamefont {S.}~\bibnamefont {Willitsch}},\
  }\bibfield  {title} {\enquote {\bibinfo {title} {{Sympathetic cooling of
  molecular ions in selected rotational and vibrational states produced by
  threshold photoionization}},}\ }\href
  {https://doi.org/10.1103/PHYSREVLETT.105.143001/FIGURES/4/MEDIUM} {\bibfield
  {journal} {\bibinfo  {journal} {Physical Review Letters}\ }\textbf {\bibinfo
  {volume} {105}},\ \bibinfo {pages} {143001} (\bibinfo {year}
  {2010})}\BibitemShut {NoStop}%
\bibitem [{\citenamefont {Kuroda}\ \emph {et~al.}(2022)\citenamefont {Kuroda},
  \citenamefont {Oho}, \citenamefont {Senami},\ and\ \citenamefont
  {Sunaga}}]{Kuroda2022}%
  \BibitemOpen
  \bibfield  {author} {\bibinfo {author} {\bibfnamefont {N.}~\bibnamefont
  {Kuroda}}, \bibinfo {author} {\bibfnamefont {T.}~\bibnamefont {Oho}},
  \bibinfo {author} {\bibfnamefont {M.}~\bibnamefont {Senami}},\ and\ \bibinfo
  {author} {\bibfnamefont {A.}~\bibnamefont {Sunaga}},\ }\bibfield  {title}
  {\enquote {\bibinfo {title} {{Enhancement of the parity-violating energy
  difference of H2X2 molecules by electronic excitation}},}\ }\href
  {https://doi.org/10.1103/PHYSREVA.105.012820/FIGURES/5/MEDIUM} {\bibfield
  {journal} {\bibinfo  {journal} {Physical Review A}\ }\textbf {\bibinfo
  {volume} {105}},\ \bibinfo {pages} {012820} (\bibinfo {year}
  {2022})}\BibitemShut {NoStop}%
\bibitem [{\citenamefont {Stohner}(2004)}]{Stohner2004}%
  \BibitemOpen
  \bibfield  {author} {\bibinfo {author} {\bibfnamefont {J.}~\bibnamefont
  {Stohner}},\ }\bibfield  {title} {\enquote {\bibinfo {title} {Parity
  violating effects in the molecular anion cbrclf{$^-$}},}\ }\href@noop {}
  {\bibfield  {journal} {\bibinfo  {journal} {International Journal of Mass
  Spectrometry}\ }\textbf {\bibinfo {volume} {233}},\ \bibinfo {pages}
  {385--394} (\bibinfo {year} {2004})}\BibitemShut {NoStop}%
\bibitem [{\citenamefont {Gottselig}\ \emph {et~al.}(2004)\citenamefont
  {Gottselig}, \citenamefont {Quack}, \citenamefont {Stohner},\ and\
  \citenamefont {Willeke}}]{Gottselig2004}%
  \BibitemOpen
  \bibfield  {author} {\bibinfo {author} {\bibfnamefont {M.}~\bibnamefont
  {Gottselig}}, \bibinfo {author} {\bibfnamefont {M.}~\bibnamefont {Quack}},
  \bibinfo {author} {\bibfnamefont {J.}~\bibnamefont {Stohner}},\ and\ \bibinfo
  {author} {\bibfnamefont {M.}~\bibnamefont {Willeke}},\ }\bibfield  {title}
  {\enquote {\bibinfo {title} {Mode-selective stereomutation tunneling and
  parity violation in {HOClH$^+$} and {H$_2$Te$_2$} isotopomers},}\ }\href@noop
  {} {\bibfield  {journal} {\bibinfo  {journal} {International Journal of Mass
  Spectrometry}\ }\textbf {\bibinfo {volume} {233}} (\bibinfo {year}
  {2004})}\BibitemShut {NoStop}%
\bibitem [{\citenamefont {Segal}\ \emph {et~al.}(2018)\citenamefont {Segal},
  \citenamefont {Lorent}, \citenamefont {Dubessy},\ and\ \citenamefont
  {Darqui{\'{e}}}}]{Segal2017}%
  \BibitemOpen
  \bibfield  {author} {\bibinfo {author} {\bibfnamefont {D.~M.}\ \bibnamefont
  {Segal}}, \bibinfo {author} {\bibfnamefont {V.}~\bibnamefont {Lorent}},
  \bibinfo {author} {\bibfnamefont {R.}~\bibnamefont {Dubessy}},\ and\ \bibinfo
  {author} {\bibfnamefont {B.}~\bibnamefont {Darqui{\'{e}}}},\ }\bibfield
  {title} {\enquote {\bibinfo {title} {{Studying fundamental physics using
  quantum enabled technologies with trapped molecular ions}},}\ }\href
  {https://doi.org/10.1080/09500340.2017.1402962} {\bibfield  {journal}
  {\bibinfo  {journal} {Journal of Modern Optics}\ }\textbf {\bibinfo {volume}
  {65}},\ \bibinfo {pages} {490--500} (\bibinfo {year} {2018})}\BibitemShut
  {NoStop}%
\bibitem [{\citenamefont {Loh}\ \emph {et~al.}(2012)\citenamefont {Loh},
  \citenamefont {Stutz}, \citenamefont {Yahn}, \citenamefont {Looser},
  \citenamefont {Field}, \citenamefont {Cornell},\ and\ \citenamefont
  {Cornell}}]{Loh2012}%
  \BibitemOpen
  \bibfield  {author} {\bibinfo {author} {\bibfnamefont {H.}~\bibnamefont
  {Loh}}, \bibinfo {author} {\bibfnamefont {R.~P.}\ \bibnamefont {Stutz}},
  \bibinfo {author} {\bibfnamefont {T.~S.}\ \bibnamefont {Yahn}}, \bibinfo
  {author} {\bibfnamefont {H.}~\bibnamefont {Looser}}, \bibinfo {author}
  {\bibfnamefont {R.~W.}\ \bibnamefont {Field}}, \bibinfo {author}
  {\bibfnamefont {E.~A.}\ \bibnamefont {Cornell}},\ and\ \bibinfo {author}
  {\bibfnamefont {E.~A.}\ \bibnamefont {Cornell}},\ }\bibfield  {title}
  {\enquote {\bibinfo {title} {{REMPI spectroscopy of HfF}},}\ }\href
  {https://doi.org/10.1016/j.jms.2012.06.014} {\bibfield  {journal} {\bibinfo
  {journal} {Journal of Molecular Spectroscopy}\ }\textbf {\bibinfo {volume}
  {276-277}},\ \bibinfo {pages} {49--56} (\bibinfo {year} {2012})}\BibitemShut
  {NoStop}%
\bibitem [{\citenamefont {Schmidt}\ \emph {et~al.}(2020)\citenamefont
  {Schmidt}, \citenamefont {Louvradoux}, \citenamefont {Heinrich},
  \citenamefont {Sillitoe}, \citenamefont {Simpson}, \citenamefont {Karr},\
  and\ \citenamefont {Hilico}}]{Schmidt2020}%
  \BibitemOpen
  \bibfield  {author} {\bibinfo {author} {\bibfnamefont {J.}~\bibnamefont
  {Schmidt}}, \bibinfo {author} {\bibfnamefont {T.}~\bibnamefont {Louvradoux}},
  \bibinfo {author} {\bibfnamefont {J.}~\bibnamefont {Heinrich}}, \bibinfo
  {author} {\bibfnamefont {N.}~\bibnamefont {Sillitoe}}, \bibinfo {author}
  {\bibfnamefont {M.}~\bibnamefont {Simpson}}, \bibinfo {author} {\bibfnamefont
  {J.~P.}\ \bibnamefont {Karr}},\ and\ \bibinfo {author} {\bibfnamefont
  {L.}~\bibnamefont {Hilico}},\ }\bibfield  {title} {\enquote {\bibinfo {title}
  {{Trapping, cooling, and photodissociation analysis of state-selected
  H{$_2^+$} ions produced by (3+1) multiphoton ionization}},}\ }\href
  {https://doi.org/10.1103/PhysRevApplied.14.024053} {\bibfield  {journal}
  {\bibinfo  {journal} {Physical Review Applied}\ }\textbf {\bibinfo {volume}
  {14}},\ \bibinfo {pages} {024053} (\bibinfo {year} {2020})}\BibitemShut
  {NoStop}%
\bibitem [{\citenamefont {Isaev}\ and\ \citenamefont
  {Berger}(2016)}]{BergerChiralLaserCooling2016}%
  \BibitemOpen
  \bibfield  {author} {\bibinfo {author} {\bibfnamefont {T.~A.}\ \bibnamefont
  {Isaev}}\ and\ \bibinfo {author} {\bibfnamefont {R.}~\bibnamefont {Berger}},\
  }\bibfield  {title} {\enquote {\bibinfo {title} {Polyatomic candidates for
  cooling of molecules with lasers from simple theoretical concepts},}\ }\href
  {https://doi.org/10.1103/PhysRevLett.116.063006} {\bibfield  {journal}
  {\bibinfo  {journal} {Phys. Rev. Lett.}\ }\textbf {\bibinfo {volume} {116}},\
  \bibinfo {pages} {063006} (\bibinfo {year} {2016})}\BibitemShut {NoStop}%
\bibitem [{\citenamefont {Kozyryev}\ \emph {et~al.}(2016)\citenamefont
  {Kozyryev}, \citenamefont {Baum}, \citenamefont {Matsuda},\ and\
  \citenamefont {Doyle}}]{Kozyryev2016}%
  \BibitemOpen
  \bibfield  {author} {\bibinfo {author} {\bibfnamefont {I.}~\bibnamefont
  {Kozyryev}}, \bibinfo {author} {\bibfnamefont {L.}~\bibnamefont {Baum}},
  \bibinfo {author} {\bibfnamefont {K.}~\bibnamefont {Matsuda}},\ and\ \bibinfo
  {author} {\bibfnamefont {J.~M.}\ \bibnamefont {Doyle}},\ }\bibfield  {title}
  {\enquote {\bibinfo {title} {Proposal for laser cooling of complex polyatomic
  molecules},}\ }\href {https://doi.org/https://doi.org/10.1002/cphc.201601051}
  {\bibfield  {journal} {\bibinfo  {journal} {ChemPhysChem}\ }\textbf {\bibinfo
  {volume} {17}},\ \bibinfo {pages} {3641--3648} (\bibinfo {year}
  {2016})}\BibitemShut {NoStop}%
\bibitem [{\citenamefont {Kozyryev}\ \emph {et~al.}(2017)\citenamefont
  {Kozyryev}, \citenamefont {Baum}, \citenamefont {Matsuda}, \citenamefont
  {Augenbraun}, \citenamefont {Anderegg}, \citenamefont {Sedlack},\ and\
  \citenamefont {Doyle}}]{Kozyryev2017}%
  \BibitemOpen
  \bibfield  {author} {\bibinfo {author} {\bibfnamefont {I.}~\bibnamefont
  {Kozyryev}}, \bibinfo {author} {\bibfnamefont {L.}~\bibnamefont {Baum}},
  \bibinfo {author} {\bibfnamefont {K.}~\bibnamefont {Matsuda}}, \bibinfo
  {author} {\bibfnamefont {B.~L.}\ \bibnamefont {Augenbraun}}, \bibinfo
  {author} {\bibfnamefont {L.}~\bibnamefont {Anderegg}}, \bibinfo {author}
  {\bibfnamefont {A.~P.}\ \bibnamefont {Sedlack}},\ and\ \bibinfo {author}
  {\bibfnamefont {J.~M.}\ \bibnamefont {Doyle}},\ }\bibfield  {title} {\enquote
  {\bibinfo {title} {Sisyphus laser cooling of a polyatomic molecule},}\ }\href
  {https://doi.org/10.1103/PhysRevLett.118.173201} {\bibfield  {journal}
  {\bibinfo  {journal} {Phys. Rev. Lett.}\ }\textbf {\bibinfo {volume} {118}},\
  \bibinfo {pages} {173201} (\bibinfo {year} {2017})}\BibitemShut {NoStop}%
\bibitem [{\citenamefont {Mitra}\ \emph {et~al.}(2022)\citenamefont {Mitra},
  \citenamefont {Lasner}, \citenamefont {Zhu}, \citenamefont {Dickerson},
  \citenamefont {Augenbraun}, \citenamefont {Bailey}, \citenamefont
  {Alexandrova}, \citenamefont {Campbell}, \citenamefont {Caram}, \citenamefont
  {Hudson},\ and\ \citenamefont {Doyle}}]{Mitra2022}%
  \BibitemOpen
  \bibfield  {author} {\bibinfo {author} {\bibfnamefont {D.}~\bibnamefont
  {Mitra}}, \bibinfo {author} {\bibfnamefont {Z.~D.}\ \bibnamefont {Lasner}},
  \bibinfo {author} {\bibfnamefont {G.~Z.}\ \bibnamefont {Zhu}}, \bibinfo
  {author} {\bibfnamefont {C.~E.}\ \bibnamefont {Dickerson}}, \bibinfo {author}
  {\bibfnamefont {B.~L.}\ \bibnamefont {Augenbraun}}, \bibinfo {author}
  {\bibfnamefont {A.~D.}\ \bibnamefont {Bailey}}, \bibinfo {author}
  {\bibfnamefont {A.~N.}\ \bibnamefont {Alexandrova}}, \bibinfo {author}
  {\bibfnamefont {W.~C.}\ \bibnamefont {Campbell}}, \bibinfo {author}
  {\bibfnamefont {J.~R.}\ \bibnamefont {Caram}}, \bibinfo {author}
  {\bibfnamefont {E.~R.}\ \bibnamefont {Hudson}},\ and\ \bibinfo {author}
  {\bibfnamefont {J.~M.}\ \bibnamefont {Doyle}},\ }\bibfield  {title} {\enquote
  {\bibinfo {title} {Pathway toward optical cycling and laser cooling of
  functionalized arenes},}\ }\href
  {https://doi.org/10.1021/ACS.JPCLETT.2C01430/ASSET/IMAGES/MEDIUM/JZ2C01430_0004.GIF}
  {\bibfield  {journal} {\bibinfo  {journal} {Journal of Physical Chemistry
  Letters}\ }\textbf {\bibinfo {volume} {13}},\ \bibinfo {pages} {7029--7035}
  (\bibinfo {year} {2022})}\BibitemShut {NoStop}%
\bibitem [{\citenamefont {Erez}, \citenamefont {Wallach},\ and\ \citenamefont
  {Shagam}(2022)}]{erez2022}%
  \BibitemOpen
  \bibfield  {author} {\bibinfo {author} {\bibfnamefont {I.}~\bibnamefont
  {Erez}}, \bibinfo {author} {\bibfnamefont {E.~R.}\ \bibnamefont {Wallach}},\
  and\ \bibinfo {author} {\bibfnamefont {Y.}~\bibnamefont {Shagam}},\
  }\bibfield  {title} {\enquote {\bibinfo {title} {Simultaneous
  enantiomer-resolved ramsey spectroscopy for chiral molecules},}\ }\href@noop
  {} {\bibfield  {journal} {\bibinfo  {journal} {arXiv preprint
  arXiv:2206.03699}\ } (\bibinfo {year} {2022})}\BibitemShut {NoStop}%
\bibitem [{\citenamefont {Changala}\ \emph {et~al.}(2017)\citenamefont
  {Changala}, \citenamefont {Nguyen}, \citenamefont {Baraban}, \citenamefont
  {Ellison}, \citenamefont {Stanton}, \citenamefont {Bross},\ and\
  \citenamefont {Ruscic}}]{Changala2017}%
  \BibitemOpen
  \bibfield  {author} {\bibinfo {author} {\bibfnamefont {P.~B.}\ \bibnamefont
  {Changala}}, \bibinfo {author} {\bibfnamefont {T.~L.}\ \bibnamefont
  {Nguyen}}, \bibinfo {author} {\bibfnamefont {J.~H.}\ \bibnamefont {Baraban}},
  \bibinfo {author} {\bibfnamefont {G.~B.}\ \bibnamefont {Ellison}}, \bibinfo
  {author} {\bibfnamefont {J.~F.}\ \bibnamefont {Stanton}}, \bibinfo {author}
  {\bibfnamefont {D.~H.}\ \bibnamefont {Bross}},\ and\ \bibinfo {author}
  {\bibfnamefont {B.}~\bibnamefont {Ruscic}},\ }\bibfield  {title} {\enquote
  {\bibinfo {title} {Active thermochemical tables: The adiabatic ionization
  energy of hydrogen peroxide},}\ }\href
  {https://doi.org/10.1021/acs.jpca.7b06221} {\bibfield  {journal} {\bibinfo
  {journal} {The Journal of Physical Chemistry A}\ }\textbf {\bibinfo {volume}
  {121}},\ \bibinfo {pages} {8799--8806} (\bibinfo {year} {2017})}\BibitemShut
  {NoStop}%
\bibitem [{\citenamefont {Eduardus}\ \emph {et~al.}(2023)\citenamefont
  {Eduardus}, \citenamefont {Shagam}, \citenamefont {Landau}, \citenamefont
  {Faraji}, \citenamefont {Schwerdtfeger}, \citenamefont {Borschevsky},\ and\
  \citenamefont {Pa\v{s}teka}}]{eduardus2023}%
  \BibitemOpen
  \bibfield  {author} {\bibinfo {author} {\bibnamefont {Eduardus}}, \bibinfo
  {author} {\bibfnamefont {Y.}~\bibnamefont {Shagam}}, \bibinfo {author}
  {\bibfnamefont {A.}~\bibnamefont {Landau}}, \bibinfo {author} {\bibfnamefont
  {S.}~\bibnamefont {Faraji}}, \bibinfo {author} {\bibfnamefont
  {P.}~\bibnamefont {Schwerdtfeger}}, \bibinfo {author} {\bibfnamefont
  {A.}~\bibnamefont {Borschevsky}},\ and\ \bibinfo {author} {\bibfnamefont
  {L.~F.}\ \bibnamefont {Pa\v{s}teka}},\ }\bibfield  {title} {\enquote
  {\bibinfo {title} {Vibrationally induced large parity violation effects in
  {CHDBrI$^+$} - a promising candidate for future experiments},}\ }\href@noop
  {} {\bibfield  {journal} {\bibinfo  {journal} {arXiv:2306.09763}\ } (\bibinfo
  {year} {2023})}\BibitemShut {NoStop}%
\bibitem [{\citenamefont {Morse}(1996)}]{morse1996supersonic}%
  \BibitemOpen
  \bibfield  {author} {\bibinfo {author} {\bibfnamefont {M.~D.}\ \bibnamefont
  {Morse}},\ }\bibfield  {title} {\enquote {\bibinfo {title} {Supersonic beam
  sources},}\ }\href@noop {} {\bibfield  {journal} {\bibinfo  {journal}
  {Experimental methods in the physical sciences}\ }\textbf {\bibinfo {volume}
  {29}},\ \bibinfo {pages} {21--47} (\bibinfo {year} {1996})}\BibitemShut
  {NoStop}%
\bibitem [{\citenamefont {Hutzler}, \citenamefont {Lu},\ and\ \citenamefont
  {Doyle}(2012)}]{Hutzler2012}%
  \BibitemOpen
  \bibfield  {author} {\bibinfo {author} {\bibfnamefont {N.~R.}\ \bibnamefont
  {Hutzler}}, \bibinfo {author} {\bibfnamefont {H.-I.}\ \bibnamefont {Lu}},\
  and\ \bibinfo {author} {\bibfnamefont {J.~M.}\ \bibnamefont {Doyle}},\
  }\bibfield  {title} {\enquote {\bibinfo {title} {The buffer gas beam: an
  intense, cold, and slow source for atoms and molecules.}}\ }\href
  {https://doi.org/10.1021/cr200362u} {\bibfield  {journal} {\bibinfo
  {journal} {Chemical reviews}\ }\textbf {\bibinfo {volume} {112}},\ \bibinfo
  {pages} {4803--27} (\bibinfo {year} {2012})}\BibitemShut {NoStop}%
\bibitem [{\citenamefont {Vilas}\ \emph {et~al.}(2022)\citenamefont {Vilas},
  \citenamefont {Hallas}, \citenamefont {Anderegg}, \citenamefont {Robichaud},
  \citenamefont {Winnicki}, \citenamefont {Mitra},\ and\ \citenamefont
  {Doyle}}]{vilas2022magneto}%
  \BibitemOpen
  \bibfield  {author} {\bibinfo {author} {\bibfnamefont {N.~B.}\ \bibnamefont
  {Vilas}}, \bibinfo {author} {\bibfnamefont {C.}~\bibnamefont {Hallas}},
  \bibinfo {author} {\bibfnamefont {L.}~\bibnamefont {Anderegg}}, \bibinfo
  {author} {\bibfnamefont {P.}~\bibnamefont {Robichaud}}, \bibinfo {author}
  {\bibfnamefont {A.}~\bibnamefont {Winnicki}}, \bibinfo {author}
  {\bibfnamefont {D.}~\bibnamefont {Mitra}},\ and\ \bibinfo {author}
  {\bibfnamefont {J.~M.}\ \bibnamefont {Doyle}},\ }\bibfield  {title} {\enquote
  {\bibinfo {title} {Magneto-optical trapping and sub-doppler cooling of a
  polyatomic molecule},}\ }\href@noop {} {\bibfield  {journal} {\bibinfo
  {journal} {Nature}\ }\textbf {\bibinfo {volume} {606}},\ \bibinfo {pages}
  {70--74} (\bibinfo {year} {2022})}\BibitemShut {NoStop}%
\bibitem [{\citenamefont {Zhu}\ \emph {et~al.}(2022)\citenamefont {Zhu},
  \citenamefont {Mitra}, \citenamefont {Augenbraun}, \citenamefont {Dickerson},
  \citenamefont {Frim}, \citenamefont {Lao}, \citenamefont {Lasner},
  \citenamefont {Alexandrova}, \citenamefont {Campbell}, \citenamefont {Caram}
  \emph {et~al.}}]{zhu2022functionalizing}%
  \BibitemOpen
  \bibfield  {author} {\bibinfo {author} {\bibfnamefont {G.-Z.}\ \bibnamefont
  {Zhu}}, \bibinfo {author} {\bibfnamefont {D.}~\bibnamefont {Mitra}}, \bibinfo
  {author} {\bibfnamefont {B.~L.}\ \bibnamefont {Augenbraun}}, \bibinfo
  {author} {\bibfnamefont {C.~E.}\ \bibnamefont {Dickerson}}, \bibinfo {author}
  {\bibfnamefont {M.~J.}\ \bibnamefont {Frim}}, \bibinfo {author}
  {\bibfnamefont {G.}~\bibnamefont {Lao}}, \bibinfo {author} {\bibfnamefont
  {Z.~D.}\ \bibnamefont {Lasner}}, \bibinfo {author} {\bibfnamefont {A.~N.}\
  \bibnamefont {Alexandrova}}, \bibinfo {author} {\bibfnamefont {W.~C.}\
  \bibnamefont {Campbell}}, \bibinfo {author} {\bibfnamefont {J.~R.}\
  \bibnamefont {Caram}}, \emph {et~al.},\ }\bibfield  {title} {\enquote
  {\bibinfo {title} {Functionalizing aromatic compounds with optical cycling
  centres},}\ }\href@noop {} {\bibfield  {journal} {\bibinfo  {journal} {Nature
  chemistry}\ }\textbf {\bibinfo {volume} {14}},\ \bibinfo {pages} {995--999}
  (\bibinfo {year} {2022})}\BibitemShut {NoStop}%
\bibitem [{\citenamefont {Lee}\ \emph {et~al.}(2022)\citenamefont {Lee},
  \citenamefont {Bischoff}, \citenamefont {Hernandez-Castillo}, \citenamefont
  {Sartakov}, \citenamefont {Meijer},\ and\ \citenamefont
  {Eibenberger-Arias}}]{Lee2022}%
  \BibitemOpen
  \bibfield  {author} {\bibinfo {author} {\bibfnamefont {J.}~\bibnamefont
  {Lee}}, \bibinfo {author} {\bibfnamefont {J.}~\bibnamefont {Bischoff}},
  \bibinfo {author} {\bibfnamefont {A.~O.}\ \bibnamefont {Hernandez-Castillo}},
  \bibinfo {author} {\bibfnamefont {B.}~\bibnamefont {Sartakov}}, \bibinfo
  {author} {\bibfnamefont {G.}~\bibnamefont {Meijer}},\ and\ \bibinfo {author}
  {\bibfnamefont {S.}~\bibnamefont {Eibenberger-Arias}},\ }\bibfield  {title}
  {\enquote {\bibinfo {title} {Quantitative study of enantiomer-specific state
  transfer},}\ }\href
  {https://doi.org/10.1103/PHYSREVLETT.128.173001/FIGURES/4/MEDIUM} {\bibfield
  {journal} {\bibinfo  {journal} {Physical Review Letters}\ }\textbf {\bibinfo
  {volume} {128}},\ \bibinfo {pages} {173001} (\bibinfo {year}
  {2022})}\BibitemShut {NoStop}%
\bibitem [{\citenamefont {Leibscher}\ \emph {et~al.}(2022)\citenamefont
  {Leibscher}, \citenamefont {Pozzoli}, \citenamefont {Pérez}, \citenamefont
  {Schnell}, \citenamefont {Sigalotti}, \citenamefont {Boscain},\ and\
  \citenamefont {Koch}}]{Leibscher2022}%
  \BibitemOpen
  \bibfield  {author} {\bibinfo {author} {\bibfnamefont {M.}~\bibnamefont
  {Leibscher}}, \bibinfo {author} {\bibfnamefont {E.}~\bibnamefont {Pozzoli}},
  \bibinfo {author} {\bibfnamefont {C.}~\bibnamefont {Pérez}}, \bibinfo
  {author} {\bibfnamefont {M.}~\bibnamefont {Schnell}}, \bibinfo {author}
  {\bibfnamefont {M.}~\bibnamefont {Sigalotti}}, \bibinfo {author}
  {\bibfnamefont {U.}~\bibnamefont {Boscain}},\ and\ \bibinfo {author}
  {\bibfnamefont {C.~P.}\ \bibnamefont {Koch}},\ }\bibfield  {title} {\enquote
  {\bibinfo {title} {Full quantum control of enantiomer-selective state
  transfer in chiral molecules despite degeneracy},}\ }\href
  {https://doi.org/10.1038/s42005-022-00883-6} {\bibfield  {journal} {\bibinfo
  {journal} {Communications Physics}\ }\textbf {\bibinfo {volume} {5}},\
  \bibinfo {pages} {110} (\bibinfo {year} {2022})}\BibitemShut {NoStop}%
\bibitem [{\citenamefont {Ni}\ \emph {et~al.}(2014)\citenamefont {Ni},
  \citenamefont {Loh}, \citenamefont {Grau}, \citenamefont {Cossel},
  \citenamefont {Ye},\ and\ \citenamefont {Cornell}}]{Ni2014}%
  \BibitemOpen
  \bibfield  {author} {\bibinfo {author} {\bibfnamefont {K.~K.}\ \bibnamefont
  {Ni}}, \bibinfo {author} {\bibfnamefont {H.}~\bibnamefont {Loh}}, \bibinfo
  {author} {\bibfnamefont {M.}~\bibnamefont {Grau}}, \bibinfo {author}
  {\bibfnamefont {K.~C.}\ \bibnamefont {Cossel}}, \bibinfo {author}
  {\bibfnamefont {J.}~\bibnamefont {Ye}},\ and\ \bibinfo {author}
  {\bibfnamefont {E.~A.}\ \bibnamefont {Cornell}},\ }\bibfield  {title}
  {\enquote {\bibinfo {title} {{State-specific detection of trapped HfF{$^+$}
  by photodissociation}},}\ }\href {https://doi.org/10.1016/j.jms.2014.02.001}
  {\bibfield  {journal} {\bibinfo  {journal} {Journal of Molecular
  Spectroscopy}\ }\textbf {\bibinfo {volume} {300}},\ \bibinfo {pages} {12--15}
  (\bibinfo {year} {2014})}\BibitemShut {NoStop}%
\bibitem [{\citenamefont {Karr}, \citenamefont {Douillet},\ and\ \citenamefont
  {Hilico}(2012)}]{Karr2012}%
  \BibitemOpen
  \bibfield  {author} {\bibinfo {author} {\bibfnamefont {J.~P.}\ \bibnamefont
  {Karr}}, \bibinfo {author} {\bibfnamefont {A.}~\bibnamefont {Douillet}},\
  and\ \bibinfo {author} {\bibfnamefont {L.}~\bibnamefont {Hilico}},\
  }\bibfield  {title} {\enquote {\bibinfo {title} {{Photodissociation of
  trapped H{$_2^+$} ions for REMPD spectroscopy}},}\ }\href
  {https://doi.org/10.1007/s00340-011-4757-z} {\bibfield  {journal} {\bibinfo
  {journal} {Applied Physics B: Lasers and Optics}\ }\textbf {\bibinfo {volume}
  {107}},\ \bibinfo {pages} {1043--1052} (\bibinfo {year} {2012})}\BibitemShut
  {NoStop}%
\bibitem [{\citenamefont {Carollo}, \citenamefont {Frenett},\ and\
  \citenamefont {Hanneke}(2018)}]{carollo2018two}%
  \BibitemOpen
  \bibfield  {author} {\bibinfo {author} {\bibfnamefont {R.}~\bibnamefont
  {Carollo}}, \bibinfo {author} {\bibfnamefont {A.}~\bibnamefont {Frenett}},\
  and\ \bibinfo {author} {\bibfnamefont {D.}~\bibnamefont {Hanneke}},\
  }\bibfield  {title} {\enquote {\bibinfo {title} {Two-photon vibrational
  transitions in {{$^{16}$}O{$_2^+$}} as probes of variation of the
  proton-to-electron mass ratio},}\ }\href@noop {} {\bibfield  {journal}
  {\bibinfo  {journal} {Atoms}\ }\textbf {\bibinfo {volume} {7}},\ \bibinfo
  {pages} {1} (\bibinfo {year} {2018})}\BibitemShut {NoStop}%
\bibitem [{\citenamefont {Z{\'a}dor}\ \emph {et~al.}(2009)\citenamefont
  {Z{\'a}dor}, \citenamefont {Fernandes}, \citenamefont {Georgievskii},
  \citenamefont {Meloni}, \citenamefont {Taatjes},\ and\ \citenamefont
  {Miller}}]{zador2009reaction}%
  \BibitemOpen
  \bibfield  {author} {\bibinfo {author} {\bibfnamefont {J.}~\bibnamefont
  {Z{\'a}dor}}, \bibinfo {author} {\bibfnamefont {R.~X.}\ \bibnamefont
  {Fernandes}}, \bibinfo {author} {\bibfnamefont {Y.}~\bibnamefont
  {Georgievskii}}, \bibinfo {author} {\bibfnamefont {G.}~\bibnamefont
  {Meloni}}, \bibinfo {author} {\bibfnamefont {C.~A.}\ \bibnamefont
  {Taatjes}},\ and\ \bibinfo {author} {\bibfnamefont {J.~A.}\ \bibnamefont
  {Miller}},\ }\bibfield  {title} {\enquote {\bibinfo {title} {The reaction of
  hydroxyethyl radicals with o2: A theoretical analysis and experimental
  product study},}\ }\href@noop {} {\bibfield  {journal} {\bibinfo  {journal}
  {Proceedings of the Combustion Institute}\ }\textbf {\bibinfo {volume}
  {32}},\ \bibinfo {pages} {271--277} (\bibinfo {year} {2009})}\BibitemShut
  {NoStop}%
\bibitem [{\citenamefont {Qi}\ \emph {et~al.}(2005)\citenamefont {Qi},
  \citenamefont {Liu}, \citenamefont {Fu},\ and\ \citenamefont
  {Guo}}]{qi2005hydrogen}%
  \BibitemOpen
  \bibfield  {author} {\bibinfo {author} {\bibfnamefont {X.-J.}\ \bibnamefont
  {Qi}}, \bibinfo {author} {\bibfnamefont {L.}~\bibnamefont {Liu}}, \bibinfo
  {author} {\bibfnamefont {Y.}~\bibnamefont {Fu}},\ and\ \bibinfo {author}
  {\bibfnamefont {Q.-X.}\ \bibnamefont {Guo}},\ }\bibfield  {title} {\enquote
  {\bibinfo {title} {Hydrogen bonding interactions of radicals},}\ }\href@noop
  {} {\bibfield  {journal} {\bibinfo  {journal} {Structural Chemistry}\
  }\textbf {\bibinfo {volume} {16}},\ \bibinfo {pages} {347--353} (\bibinfo
  {year} {2005})}\BibitemShut {NoStop}%
\bibitem [{\citenamefont {Luo}\ \emph {et~al.}(2018)\citenamefont {Luo},
  \citenamefont {Gao}, \citenamefont {Wei}, \citenamefont {Spinney},
  \citenamefont {Dionysiou}, \citenamefont {Hu}, \citenamefont {Chai},\ and\
  \citenamefont {Xiao}}]{luo2018kinetic}%
  \BibitemOpen
  \bibfield  {author} {\bibinfo {author} {\bibfnamefont {S.}~\bibnamefont
  {Luo}}, \bibinfo {author} {\bibfnamefont {L.}~\bibnamefont {Gao}}, \bibinfo
  {author} {\bibfnamefont {Z.}~\bibnamefont {Wei}}, \bibinfo {author}
  {\bibfnamefont {R.}~\bibnamefont {Spinney}}, \bibinfo {author} {\bibfnamefont
  {D.~D.}\ \bibnamefont {Dionysiou}}, \bibinfo {author} {\bibfnamefont {W.-P.}\
  \bibnamefont {Hu}}, \bibinfo {author} {\bibfnamefont {L.}~\bibnamefont
  {Chai}},\ and\ \bibinfo {author} {\bibfnamefont {R.}~\bibnamefont {Xiao}},\
  }\bibfield  {title} {\enquote {\bibinfo {title} {Kinetic and mechanistic
  aspects of hydroxyl radical--mediated degradation of naproxen and reaction
  intermediates},}\ }\href@noop {} {\bibfield  {journal} {\bibinfo  {journal}
  {Water research}\ }\textbf {\bibinfo {volume} {137}},\ \bibinfo {pages}
  {233--241} (\bibinfo {year} {2018})}\BibitemShut {NoStop}%
\bibitem [{\citenamefont {Xu}\ \emph {et~al.}(2019)\citenamefont {Xu},
  \citenamefont {Xi}, \citenamefont {Wang},\ and\ \citenamefont
  {Li}}]{xu2019theoretical}%
  \BibitemOpen
  \bibfield  {author} {\bibinfo {author} {\bibfnamefont {Y.}~\bibnamefont
  {Xu}}, \bibinfo {author} {\bibfnamefont {S.}~\bibnamefont {Xi}}, \bibinfo
  {author} {\bibfnamefont {F.}~\bibnamefont {Wang}},\ and\ \bibinfo {author}
  {\bibfnamefont {X.}~\bibnamefont {Li}},\ }\bibfield  {title} {\enquote
  {\bibinfo {title} {Theoretical study on reactions of alkylperoxy radicals},}\
  }\href@noop {} {\bibfield  {journal} {\bibinfo  {journal} {The Journal of
  Physical Chemistry A}\ }\textbf {\bibinfo {volume} {123}},\ \bibinfo {pages}
  {3949--3958} (\bibinfo {year} {2019})}\BibitemShut {NoStop}%
\bibitem [{\citenamefont {Xiao}\ \emph {et~al.}(2020)\citenamefont {Xiao},
  \citenamefont {He}, \citenamefont {Luo}, \citenamefont {Spinney},
  \citenamefont {Wei}, \citenamefont {Dionysiou},\ and\ \citenamefont
  {Zhao}}]{xiao2020experimental}%
  \BibitemOpen
  \bibfield  {author} {\bibinfo {author} {\bibfnamefont {R.}~\bibnamefont
  {Xiao}}, \bibinfo {author} {\bibfnamefont {L.}~\bibnamefont {He}}, \bibinfo
  {author} {\bibfnamefont {Z.}~\bibnamefont {Luo}}, \bibinfo {author}
  {\bibfnamefont {R.}~\bibnamefont {Spinney}}, \bibinfo {author} {\bibfnamefont
  {Z.}~\bibnamefont {Wei}}, \bibinfo {author} {\bibfnamefont {D.~D.}\
  \bibnamefont {Dionysiou}},\ and\ \bibinfo {author} {\bibfnamefont
  {F.}~\bibnamefont {Zhao}},\ }\bibfield  {title} {\enquote {\bibinfo {title}
  {An experimental and theoretical study on the degradation of clonidine by
  hydroxyl and sulfate radicals},}\ }\href@noop {} {\bibfield  {journal}
  {\bibinfo  {journal} {Science of The Total Environment}\ }\textbf {\bibinfo
  {volume} {710}},\ \bibinfo {pages} {136333} (\bibinfo {year}
  {2020})}\BibitemShut {NoStop}%
\bibitem [{\citenamefont {Lisovskaya}, \citenamefont {Carmichael},\ and\
  \citenamefont {Harriman}(2021)}]{lisovskaya2021pulse}%
  \BibitemOpen
  \bibfield  {author} {\bibinfo {author} {\bibfnamefont {A.}~\bibnamefont
  {Lisovskaya}}, \bibinfo {author} {\bibfnamefont {I.}~\bibnamefont
  {Carmichael}},\ and\ \bibinfo {author} {\bibfnamefont {A.}~\bibnamefont
  {Harriman}},\ }\bibfield  {title} {\enquote {\bibinfo {title} {Pulse
  radiolysis investigation of radicals derived from water-soluble cyanine dyes:
  Implications for super-resolution microscopy},}\ }\href@noop {} {\bibfield
  {journal} {\bibinfo  {journal} {The Journal of Physical Chemistry A}\
  }\textbf {\bibinfo {volume} {125}},\ \bibinfo {pages} {5779--5793} (\bibinfo
  {year} {2021})}\BibitemShut {NoStop}%
\bibitem [{\citenamefont {Goddard}(1989)}]{goddard1989computational}%
  \BibitemOpen
  \bibfield  {author} {\bibinfo {author} {\bibfnamefont {J.~D.}\ \bibnamefont
  {Goddard}},\ }\bibfield  {title} {\enquote {\bibinfo {title} {A computational
  study of the hcco and hccs radicals},}\ }\href@noop {} {\bibfield  {journal}
  {\bibinfo  {journal} {Chemical physics letters}\ }\textbf {\bibinfo {volume}
  {154}},\ \bibinfo {pages} {387--390} (\bibinfo {year} {1989})}\BibitemShut
  {NoStop}%
\bibitem [{\citenamefont {Rauk}, \citenamefont {Yu},\ and\ \citenamefont
  {Armstrong}(1994)}]{rauk1994carboxyl}%
  \BibitemOpen
  \bibfield  {author} {\bibinfo {author} {\bibfnamefont {A.}~\bibnamefont
  {Rauk}}, \bibinfo {author} {\bibfnamefont {D.}~\bibnamefont {Yu}},\ and\
  \bibinfo {author} {\bibfnamefont {D.~A.}\ \bibnamefont {Armstrong}},\
  }\bibfield  {title} {\enquote {\bibinfo {title} {Carboxyl free radicals:
  formyloxyl (hcoo. bul.) and acetyloxyl (ch3coo. bul.) revisited},}\
  }\href@noop {} {\bibfield  {journal} {\bibinfo  {journal} {Journal of the
  American Chemical Society}\ }\textbf {\bibinfo {volume} {116}},\ \bibinfo
  {pages} {8222--8228} (\bibinfo {year} {1994})}\BibitemShut {NoStop}%
\bibitem [{\citenamefont {Krylov}(2008)}]{Krylov:EOM}%
  \BibitemOpen
  \bibfield  {author} {\bibinfo {author} {\bibfnamefont {A.~I.}\ \bibnamefont
  {Krylov}},\ }\bibfield  {title} {\enquote {\bibinfo {title}
  {Equation-of-motion coupled-cluster methods for open-shell and electronically
  excited species: The hitchhiker's guide to {F}ock space},}\ }\href@noop {}
  {\bibfield  {journal} {\bibinfo  {journal} {Physical Chemistry}\ }\textbf
  {\bibinfo {volume} {59}},\ \bibinfo {pages} {433--462} (\bibinfo {year}
  {2008})}\BibitemShut {NoStop}%
\bibitem [{\citenamefont {Pieniazek}, \citenamefont {Bradforth},\ and\
  \citenamefont {Krylov}(2008)}]{pieniazek2008charge}%
  \BibitemOpen
  \bibfield  {author} {\bibinfo {author} {\bibfnamefont {P.~A.}\ \bibnamefont
  {Pieniazek}}, \bibinfo {author} {\bibfnamefont {S.~E.}\ \bibnamefont
  {Bradforth}},\ and\ \bibinfo {author} {\bibfnamefont {A.~I.}\ \bibnamefont
  {Krylov}},\ }\bibfield  {title} {\enquote {\bibinfo {title} {Charge
  localization and jahn--teller distortions in the benzene dimer cation},}\
  }\href@noop {} {\bibfield  {journal} {\bibinfo  {journal} {The Journal of
  chemical physics}\ }\textbf {\bibinfo {volume} {129}},\ \bibinfo {pages}
  {074104} (\bibinfo {year} {2008})}\BibitemShut {NoStop}%
\bibitem [{\citenamefont {Helgaker}, \citenamefont {Jorgensen},\ and\
  \citenamefont {Olsen}(2014)}]{helgaker2014molecular}%
  \BibitemOpen
  \bibfield  {author} {\bibinfo {author} {\bibfnamefont {T.}~\bibnamefont
  {Helgaker}}, \bibinfo {author} {\bibfnamefont {P.}~\bibnamefont
  {Jorgensen}},\ and\ \bibinfo {author} {\bibfnamefont {J.}~\bibnamefont
  {Olsen}},\ }\href@noop {} {\emph {\bibinfo {title} {Molecular
  electronic-structure theory}}}\ (\bibinfo  {publisher} {John Wiley \& Sons},\
  \bibinfo {year} {2014})\BibitemShut {NoStop}%
\bibitem [{\citenamefont {Feng}\ \emph {et~al.}(2019)\citenamefont {Feng},
  \citenamefont {Epifanovsky}, \citenamefont {Gauss},\ and\ \citenamefont
  {Krylov}}]{feng2019implementation}%
  \BibitemOpen
  \bibfield  {author} {\bibinfo {author} {\bibfnamefont {X.}~\bibnamefont
  {Feng}}, \bibinfo {author} {\bibfnamefont {E.}~\bibnamefont {Epifanovsky}},
  \bibinfo {author} {\bibfnamefont {J.}~\bibnamefont {Gauss}},\ and\ \bibinfo
  {author} {\bibfnamefont {A.~I.}\ \bibnamefont {Krylov}},\ }\bibfield  {title}
  {\enquote {\bibinfo {title} {Implementation of analytic gradients for ccsd
  and eom-ccsd using cholesky decomposition of the electron-repulsion integrals
  and their derivatives: Theory and benchmarks},}\ }\href@noop {} {\bibfield
  {journal} {\bibinfo  {journal} {The Journal of chemical physics}\ }\textbf
  {\bibinfo {volume} {151}},\ \bibinfo {pages} {014110} (\bibinfo {year}
  {2019})}\BibitemShut {NoStop}%
\bibitem [{\citenamefont {Dunning~Jr}(1989)}]{dunning1989gaussian}%
  \BibitemOpen
  \bibfield  {author} {\bibinfo {author} {\bibfnamefont {T.~H.}\ \bibnamefont
  {Dunning~Jr}},\ }\bibfield  {title} {\enquote {\bibinfo {title} {Gaussian
  basis sets for use in correlated molecular calculations. i. the atoms boron
  through neon and hydrogen},}\ }\href@noop {} {\bibfield  {journal} {\bibinfo
  {journal} {The Journal of chemical physics}\ }\textbf {\bibinfo {volume}
  {90}},\ \bibinfo {pages} {1007--1023} (\bibinfo {year} {1989})}\BibitemShut
  {NoStop}%
\bibitem [{\citenamefont {Woon}\ and\ \citenamefont
  {Dunning~Jr}(1993)}]{woon1993gaussian}%
  \BibitemOpen
  \bibfield  {author} {\bibinfo {author} {\bibfnamefont {D.~E.}\ \bibnamefont
  {Woon}}\ and\ \bibinfo {author} {\bibfnamefont {T.~H.}\ \bibnamefont
  {Dunning~Jr}},\ }\bibfield  {title} {\enquote {\bibinfo {title} {Gaussian
  basis sets for use in correlated molecular calculations. iii. the atoms
  aluminum through argon},}\ }\href@noop {} {\bibfield  {journal} {\bibinfo
  {journal} {The Journal of chemical physics}\ }\textbf {\bibinfo {volume}
  {98}},\ \bibinfo {pages} {1358--1371} (\bibinfo {year} {1993})}\BibitemShut
  {NoStop}%
\bibitem [{\citenamefont {Kendall}, \citenamefont {Dunning~Jr},\ and\
  \citenamefont {Harrison}(1992)}]{dunning1992augccpvxz}%
  \BibitemOpen
  \bibfield  {author} {\bibinfo {author} {\bibfnamefont {R.~A.}\ \bibnamefont
  {Kendall}}, \bibinfo {author} {\bibfnamefont {T.~H.}\ \bibnamefont
  {Dunning~Jr}},\ and\ \bibinfo {author} {\bibfnamefont {R.~J.}\ \bibnamefont
  {Harrison}},\ }\bibfield  {title} {\enquote {\bibinfo {title} {Electron
  affinities of the first-row atoms revisited. systematic basis sets and wave
  functions},}\ }\href@noop {} {\bibfield  {journal} {\bibinfo  {journal} {The
  Journal of chemical physics}\ }\textbf {\bibinfo {volume} {96}},\ \bibinfo
  {pages} {6796--6806} (\bibinfo {year} {1992})}\BibitemShut {NoStop}%
\bibitem [{\citenamefont {Peterson}(2003)}]{peterson2003pp1}%
  \BibitemOpen
  \bibfield  {author} {\bibinfo {author} {\bibfnamefont {K.~A.}\ \bibnamefont
  {Peterson}},\ }\bibfield  {title} {\enquote {\bibinfo {title} {Systematically
  convergent basis sets with relativistic pseudopotentials. i. correlation
  consistent basis sets for the post-d group 13--15 elements},}\ }\href@noop {}
  {\bibfield  {journal} {\bibinfo  {journal} {The Journal of chemical physics}\
  }\textbf {\bibinfo {volume} {119}},\ \bibinfo {pages} {11099--11112}
  (\bibinfo {year} {2003})}\BibitemShut {NoStop}%
\bibitem [{\citenamefont {Peterson}\ \emph {et~al.}(2003)\citenamefont
  {Peterson}, \citenamefont {Figgen}, \citenamefont {Goll}, \citenamefont
  {Stoll},\ and\ \citenamefont {Dolg}}]{peterson2003pp2}%
  \BibitemOpen
  \bibfield  {author} {\bibinfo {author} {\bibfnamefont {K.~A.}\ \bibnamefont
  {Peterson}}, \bibinfo {author} {\bibfnamefont {D.}~\bibnamefont {Figgen}},
  \bibinfo {author} {\bibfnamefont {E.}~\bibnamefont {Goll}}, \bibinfo {author}
  {\bibfnamefont {H.}~\bibnamefont {Stoll}},\ and\ \bibinfo {author}
  {\bibfnamefont {M.}~\bibnamefont {Dolg}},\ }\bibfield  {title} {\enquote
  {\bibinfo {title} {Systematically convergent basis sets with relativistic
  pseudopotentials. ii. small-core pseudopotentials and correlation consistent
  basis sets for the post-d group 16--18 elements},}\ }\href@noop {} {\bibfield
   {journal} {\bibinfo  {journal} {The Journal of chemical physics}\ }\textbf
  {\bibinfo {volume} {119}},\ \bibinfo {pages} {11113--11123} (\bibinfo {year}
  {2003})}\BibitemShut {NoStop}%
\bibitem [{\citenamefont {Gulde}, \citenamefont {Pollak},\ and\ \citenamefont
  {Weigend}(2012)}]{gulde2012error}%
  \BibitemOpen
  \bibfield  {author} {\bibinfo {author} {\bibfnamefont {R.}~\bibnamefont
  {Gulde}}, \bibinfo {author} {\bibfnamefont {P.}~\bibnamefont {Pollak}},\ and\
  \bibinfo {author} {\bibfnamefont {F.}~\bibnamefont {Weigend}},\ }\bibfield
  {title} {\enquote {\bibinfo {title} {Error-balanced segmented contracted
  basis sets of double-$\zeta$ to quadruple-$\zeta$ valence quality for the
  lanthanides},}\ }\href@noop {} {\bibfield  {journal} {\bibinfo  {journal}
  {Journal of chemical theory and computation}\ }\textbf {\bibinfo {volume}
  {8}},\ \bibinfo {pages} {4062--4068} (\bibinfo {year} {2012})}\BibitemShut
  {NoStop}%
\bibitem [{\citenamefont {Dyall}(1994)}]{Dyall1994spinFree}%
  \BibitemOpen
  \bibfield  {author} {\bibinfo {author} {\bibfnamefont {K.~G.}\ \bibnamefont
  {Dyall}},\ }\bibfield  {title} {\enquote {\bibinfo {title} {An exact
  separation of the spin-free and spin-dependent terms of the
  dirac--coulomb--breit hamiltonian},}\ }\href@noop {} {\bibfield  {journal}
  {\bibinfo  {journal} {The Journal of chemical physics}\ }\textbf {\bibinfo
  {volume} {100}},\ \bibinfo {pages} {2118--2127} (\bibinfo {year}
  {1994})}\BibitemShut {NoStop}%
\bibitem [{\citenamefont {Saue}\ \emph {et~al.}(2020)\citenamefont {Saue},
  \citenamefont {Bast}, \citenamefont {Gomes}, \citenamefont {Jensen},
  \citenamefont {Visscher}, \citenamefont {Aucar}, \citenamefont {Di~Remigio},
  \citenamefont {Dyall}, \citenamefont {Eliav}, \citenamefont {Fasshauer} \emph
  {et~al.}}]{code2020dirac}%
  \BibitemOpen
  \bibfield  {author} {\bibinfo {author} {\bibfnamefont {T.}~\bibnamefont
  {Saue}}, \bibinfo {author} {\bibfnamefont {R.}~\bibnamefont {Bast}}, \bibinfo
  {author} {\bibfnamefont {A.~S.~P.}\ \bibnamefont {Gomes}}, \bibinfo {author}
  {\bibfnamefont {H.~J.~A.}\ \bibnamefont {Jensen}}, \bibinfo {author}
  {\bibfnamefont {L.}~\bibnamefont {Visscher}}, \bibinfo {author}
  {\bibfnamefont {I.~A.}\ \bibnamefont {Aucar}}, \bibinfo {author}
  {\bibfnamefont {R.}~\bibnamefont {Di~Remigio}}, \bibinfo {author}
  {\bibfnamefont {K.~G.}\ \bibnamefont {Dyall}}, \bibinfo {author}
  {\bibfnamefont {E.}~\bibnamefont {Eliav}}, \bibinfo {author} {\bibfnamefont
  {E.}~\bibnamefont {Fasshauer}}, \emph {et~al.},\ }\bibfield  {title}
  {\enquote {\bibinfo {title} {The dirac code for relativistic molecular
  calculations},}\ }\href@noop {} {\bibfield  {journal} {\bibinfo  {journal}
  {The Journal of chemical physics}\ }\textbf {\bibinfo {volume} {152}},\
  \bibinfo {pages} {204104} (\bibinfo {year} {2020})},\ \bibinfo {note}
  {http://diracprogram.org}\BibitemShut {NoStop}%
\bibitem [{\citenamefont {Roberts}\ and\ \citenamefont
  {Lehman}(2022)}]{Roberts2022}%
  \BibitemOpen
  \bibfield  {author} {\bibinfo {author} {\bibfnamefont {F.~C.}\ \bibnamefont
  {Roberts}}\ and\ \bibinfo {author} {\bibfnamefont {J.~H.}\ \bibnamefont
  {Lehman}},\ }\bibfield  {title} {\enquote {\bibinfo {title} {Infrared
  frequency comb spectroscopy of {CH$_2$I$_2$}: Influence of hot bands and
  pressure broadening on the {$\nu$}1 and {$\nu$}6 fundamental transitions},}\
  }\href {https://doi.org/10.1063/5.0081836} {\bibfield  {journal} {\bibinfo
  {journal} {The Journal of Chemical Physics}\ }\textbf {\bibinfo {volume}
  {156}},\ \bibinfo {pages} {114301} (\bibinfo {year} {2022})}\BibitemShut
  {NoStop}%
\bibitem [{\citenamefont {Irikura}, \citenamefont {Johnson},\ and\
  \citenamefont {Kacker}(2005)}]{Irikura2005}%
  \BibitemOpen
  \bibfield  {author} {\bibinfo {author} {\bibfnamefont {K.~K.}\ \bibnamefont
  {Irikura}}, \bibinfo {author} {\bibfnamefont {R.~D.}\ \bibnamefont
  {Johnson}},\ and\ \bibinfo {author} {\bibfnamefont {R.~N.}\ \bibnamefont
  {Kacker}},\ }\bibfield  {title} {\enquote {\bibinfo {title} {Uncertainties in
  scaling factors for ab initio vibrational frequencies},}\ }\href
  {https://doi.org/10.1021/jp052793n} {\bibfield  {journal} {\bibinfo
  {journal} {The Journal of Physical Chemistry A}\ }\textbf {\bibinfo {volume}
  {109}},\ \bibinfo {pages} {8430--8437} (\bibinfo {year} {2005})},\ \bibinfo
  {note} {pMID: 16834237},\ \Eprint
  {https://arxiv.org/abs/https://doi.org/10.1021/jp052793n}
  {https://doi.org/10.1021/jp052793n} \BibitemShut {NoStop}%
\bibitem [{\citenamefont {NIST}(2022)}]{NIST2022}%
  \BibitemOpen
  \bibfield  {author} {\bibinfo {author} {\bibnamefont {NIST}},\ }\href@noop {}
  {\enquote {\bibinfo {title} {Computational chemistry comparison and benchmark
  database},}\ } (\bibinfo {year} {2022}),\ \Eprint
  {https://arxiv.org/abs/https://cccbdb.nist.gov/vibscalejustx.asp}
  {https://cccbdb.nist.gov/vibscalejustx.asp} \BibitemShut {NoStop}%
\bibitem [{\citenamefont {Schowalter}\ \emph {et~al.}(2012)\citenamefont
  {Schowalter}, \citenamefont {Chen}, \citenamefont {Rellergert}, \citenamefont
  {Sullivan},\ and\ \citenamefont {Hudson}}]{Schowalter2012}%
  \BibitemOpen
  \bibfield  {author} {\bibinfo {author} {\bibfnamefont {S.~J.}\ \bibnamefont
  {Schowalter}}, \bibinfo {author} {\bibfnamefont {K.}~\bibnamefont {Chen}},
  \bibinfo {author} {\bibfnamefont {W.~G.}\ \bibnamefont {Rellergert}},
  \bibinfo {author} {\bibfnamefont {S.~T.}\ \bibnamefont {Sullivan}},\ and\
  \bibinfo {author} {\bibfnamefont {E.~R.}\ \bibnamefont {Hudson}},\ }\bibfield
   {title} {\enquote {\bibinfo {title} {An integrated ion trap and
  time-of-flight mass spectrometer for chemical and photo- reaction dynamics
  studies},}\ }\href {https://doi.org/10.1063/1.3700216} {\bibfield  {journal}
  {\bibinfo  {journal} {Review of Scientific Instruments}\ }\textbf {\bibinfo
  {volume} {83}},\ \bibinfo {pages} {043103} (\bibinfo {year}
  {2012})}\BibitemShut {NoStop}%
\bibitem [{\citenamefont {Schmid}\ \emph {et~al.}(2017)\citenamefont {Schmid},
  \citenamefont {Greenberg}, \citenamefont {Miller}, \citenamefont {Loeffler},\
  and\ \citenamefont {Lewandowski}}]{Schmid2017}%
  \BibitemOpen
  \bibfield  {author} {\bibinfo {author} {\bibfnamefont {P.~C.}\ \bibnamefont
  {Schmid}}, \bibinfo {author} {\bibfnamefont {J.}~\bibnamefont {Greenberg}},
  \bibinfo {author} {\bibfnamefont {M.~I.}\ \bibnamefont {Miller}}, \bibinfo
  {author} {\bibfnamefont {K.}~\bibnamefont {Loeffler}},\ and\ \bibinfo
  {author} {\bibfnamefont {H.~J.}\ \bibnamefont {Lewandowski}},\ }\bibfield
  {title} {\enquote {\bibinfo {title} {High resolution ion trap time-of-flight
  mass spectrometer for cold trapped ion experiments},}\ }\href
  {https://doi.org/10.1063/1.4996911} {\bibfield  {journal} {\bibinfo
  {journal} {Review of Scientific Instruments}\ }\textbf {\bibinfo {volume}
  {88}},\ \bibinfo {pages} {123107} (\bibinfo {year} {2017})}\BibitemShut
  {NoStop}%
\bibitem [{\citenamefont {Nesbitt}\ and\ \citenamefont
  {Field}(1996)}]{Nesbitt1996}%
  \BibitemOpen
  \bibfield  {author} {\bibinfo {author} {\bibfnamefont {D.~J.}\ \bibnamefont
  {Nesbitt}}\ and\ \bibinfo {author} {\bibfnamefont {R.~W.}\ \bibnamefont
  {Field}},\ }\bibfield  {title} {\enquote {\bibinfo {title} {Vibrational
  energy flow in highly excited molecules: Role of intramolecular vibrational
  redistribution},}\ }\href {https://doi.org/10.1021/jp960698w} {\bibfield
  {journal} {\bibinfo  {journal} {The Journal of Physical Chemistry}\ }\textbf
  {\bibinfo {volume} {100}},\ \bibinfo {pages} {12735--12756} (\bibinfo {year}
  {1996})}\BibitemShut {NoStop}%
\bibitem [{\citenamefont {Epifanovsky}\ \emph {et~al.}(2021)\citenamefont
  {Epifanovsky}, \citenamefont {Gilbert}, \citenamefont {Feng}, \citenamefont
  {Lee}, \citenamefont {Mao}, \citenamefont {Mardirossian}, \citenamefont
  {Pokhilko}, \citenamefont {White}, \citenamefont {Coons}, \citenamefont
  {Dempwolff}, \citenamefont {Gan}, \citenamefont {Hait}, \citenamefont {Horn},
  \citenamefont {Jacobson}, \citenamefont {Kaliman}, \citenamefont {Kussmann},
  \citenamefont {Lange}, \citenamefont {Lao}, \citenamefont {Levine},
  \citenamefont {Liu}, \citenamefont {McKenzie}, \citenamefont {Morrison},
  \citenamefont {Nanda}, \citenamefont {Plasser}, \citenamefont {Rehn},
  \citenamefont {Vidal}, \citenamefont {You}, \citenamefont {Zhu},
  \citenamefont {Alam}, \citenamefont {Albrecht}, \citenamefont {Aldossary},
  \citenamefont {Alguire}, \citenamefont {Andersen}, \citenamefont {Athavale},
  \citenamefont {Barton}, \citenamefont {Begam}, \citenamefont {Behn},
  \citenamefont {Bellonzi}, \citenamefont {Bernard}, \citenamefont {Berquist},
  \citenamefont {Burton}, \citenamefont {Carreras}, \citenamefont
  {Carter-Fenk}, \citenamefont {Chakraborty}, \citenamefont {Chien},
  \citenamefont {Closser}, \citenamefont {Cofer-Shabica}, \citenamefont
  {Dasgupta}, \citenamefont {Wergifosse}, \citenamefont {Deng}, \citenamefont
  {Diedenhofen}, \citenamefont {Do}, \citenamefont {Ehlert}, \citenamefont
  {Fang}, \citenamefont {Fatehi}, \citenamefont {Feng}, \citenamefont
  {Friedhoff}, \citenamefont {Gayvert}, \citenamefont {Ge}, \citenamefont
  {Gidofalvi}, \citenamefont {Goldey}, \citenamefont {Gomes}, \citenamefont
  {González-Espinoza}, \citenamefont {Gulania}, \citenamefont {Gunina},
  \citenamefont {Hanson-Heine}, \citenamefont {Harbach}, \citenamefont
  {Hauser}, \citenamefont {Herbst}, \citenamefont {Vera}, \citenamefont
  {Hodecker}, \citenamefont {Holden}, \citenamefont {Houck}, \citenamefont
  {Huang}, \citenamefont {Hui}, \citenamefont {Huynh}, \citenamefont {Ivanov},
  \citenamefont {Ádám Jász}, \citenamefont {Ji}, \citenamefont {Jiang},
  \citenamefont {Kaduk}, \citenamefont {Kähler}, \citenamefont {Khistyaev},
  \citenamefont {Kim}, \citenamefont {Kis}, \citenamefont {Klunzinger},
  \citenamefont {Koczor-Benda}, \citenamefont {Koh}, \citenamefont {Kosenkov},
  \citenamefont {Koulias}, \citenamefont {Kowalczyk}, \citenamefont {Krauter},
  \citenamefont {Kue}, \citenamefont {Kunitsa}, \citenamefont {Kus},
  \citenamefont {Ladjánszki}, \citenamefont {Landau}, \citenamefont {Lawler},
  \citenamefont {Lefrancois}, \citenamefont {Lehtola}, \citenamefont {Li},
  \citenamefont {Li}, \citenamefont {Liang}, \citenamefont {Liebenthal},
  \citenamefont {Lin}, \citenamefont {Lin}, \citenamefont {Liu}, \citenamefont
  {Liu}, \citenamefont {Loipersberger}, \citenamefont {Luenser}, \citenamefont
  {Manjanath}, \citenamefont {Manohar}, \citenamefont {Mansoor}, \citenamefont
  {Manzer}, \citenamefont {Mao}, \citenamefont {Marenich}, \citenamefont
  {Markovich}, \citenamefont {Mason}, \citenamefont {Maurer}, \citenamefont
  {McLaughlin}, \citenamefont {Menger}, \citenamefont {Mewes}, \citenamefont
  {Mewes}, \citenamefont {Morgante}, \citenamefont {Mullinax}, \citenamefont
  {Oosterbaan}, \citenamefont {Paran}, \citenamefont {Paul}, \citenamefont
  {Paul}, \citenamefont {Pavošević}, \citenamefont {Pei}, \citenamefont
  {Prager}, \citenamefont {Proynov}, \citenamefont {Ádám Rák}, \citenamefont
  {Ramos-Cordoba}, \citenamefont {Rana}, \citenamefont {Rask}, \citenamefont
  {Rettig}, \citenamefont {Richard}, \citenamefont {Rob}, \citenamefont
  {Rossomme}, \citenamefont {Scheele}, \citenamefont {Scheurer}, \citenamefont
  {Schneider}, \citenamefont {Sergueev}, \citenamefont {Sharada}, \citenamefont
  {Skomorowski}, \citenamefont {Small}, \citenamefont {Stein}, \citenamefont
  {Su}, \citenamefont {Sundstrom}, \citenamefont {Tao}, \citenamefont
  {Thirman}, \citenamefont {Tornai}, \citenamefont {Tsuchimochi}, \citenamefont
  {Tubman}, \citenamefont {Veccham}, \citenamefont {Vydrov}, \citenamefont
  {Wenzel}, \citenamefont {Witte}, \citenamefont {Yamada}, \citenamefont {Yao},
  \citenamefont {Yeganeh}, \citenamefont {Yost}, \citenamefont {Zech},
  \citenamefont {Zhang}, \citenamefont {Zhang}, \citenamefont {Zhang},
  \citenamefont {Zuev}, \citenamefont {Aspuru-Guzik}, \citenamefont {Bell},
  \citenamefont {Besley}, \citenamefont {Bravaya}, \citenamefont {Brooks},
  \citenamefont {Casanova}, \citenamefont {Chai}, \citenamefont {Coriani},
  \citenamefont {Cramer}, \citenamefont {Cserey}, \citenamefont {Deprince},
  \citenamefont {Distasio}, \citenamefont {Dreuw}, \citenamefont {Dunietz},
  \citenamefont {Furlani}, \citenamefont {Goddard}, \citenamefont
  {Hammes-Schiffer}, \citenamefont {Head-Gordon}, \citenamefont {Hehre},
  \citenamefont {Hsu}, \citenamefont {Jagau}, \citenamefont {Jung},
  \citenamefont {Klamt}, \citenamefont {Kong}, \citenamefont {Lambrecht},
  \citenamefont {Liang}, \citenamefont {Mayhall}, \citenamefont {McCurdy},
  \citenamefont {Neaton}, \citenamefont {Ochsenfeld}, \citenamefont {Parkhill},
  \citenamefont {Peverati}, \citenamefont {Rassolov}, \citenamefont {Shao},
  \citenamefont {Slipchenko}, \citenamefont {Stauch}, \citenamefont {Steele},
  \citenamefont {Subotnik}, \citenamefont {Thom}, \citenamefont {Tkatchenko},
  \citenamefont {Truhlar}, \citenamefont {Voorhis}, \citenamefont {Wesolowski},
  \citenamefont {Whaley}, \citenamefont {Woodcock}, \citenamefont {Zimmerman},
  \citenamefont {Faraji}, \citenamefont {Gill}, \citenamefont {Head-Gordon},
  \citenamefont {Herbert},\ and\ \citenamefont {Krylov}}]{qchem}%
  \BibitemOpen
  \bibfield  {author} {\bibinfo {author} {\bibfnamefont {E.}~\bibnamefont
  {Epifanovsky}}, \bibinfo {author} {\bibfnamefont {A.~T.}\ \bibnamefont
  {Gilbert}}, \bibinfo {author} {\bibfnamefont {X.}~\bibnamefont {Feng}},
  \bibinfo {author} {\bibfnamefont {J.}~\bibnamefont {Lee}}, \bibinfo {author}
  {\bibfnamefont {Y.}~\bibnamefont {Mao}}, \bibinfo {author} {\bibfnamefont
  {N.}~\bibnamefont {Mardirossian}}, \bibinfo {author} {\bibfnamefont
  {P.}~\bibnamefont {Pokhilko}}, \bibinfo {author} {\bibfnamefont {A.~F.}\
  \bibnamefont {White}}, \bibinfo {author} {\bibfnamefont {M.~P.}\ \bibnamefont
  {Coons}}, \bibinfo {author} {\bibfnamefont {A.~L.}\ \bibnamefont
  {Dempwolff}}, \bibinfo {author} {\bibfnamefont {Z.}~\bibnamefont {Gan}},
  \bibinfo {author} {\bibfnamefont {D.}~\bibnamefont {Hait}}, \bibinfo {author}
  {\bibfnamefont {P.~R.}\ \bibnamefont {Horn}}, \bibinfo {author}
  {\bibfnamefont {L.~D.}\ \bibnamefont {Jacobson}}, \bibinfo {author}
  {\bibfnamefont {I.}~\bibnamefont {Kaliman}}, \bibinfo {author} {\bibfnamefont
  {J.}~\bibnamefont {Kussmann}}, \bibinfo {author} {\bibfnamefont {A.~W.}\
  \bibnamefont {Lange}}, \bibinfo {author} {\bibfnamefont {K.~U.}\ \bibnamefont
  {Lao}}, \bibinfo {author} {\bibfnamefont {D.~S.}\ \bibnamefont {Levine}},
  \bibinfo {author} {\bibfnamefont {J.}~\bibnamefont {Liu}}, \bibinfo {author}
  {\bibfnamefont {S.~C.}\ \bibnamefont {McKenzie}}, \bibinfo {author}
  {\bibfnamefont {A.~F.}\ \bibnamefont {Morrison}}, \bibinfo {author}
  {\bibfnamefont {K.~D.}\ \bibnamefont {Nanda}}, \bibinfo {author}
  {\bibfnamefont {F.}~\bibnamefont {Plasser}}, \bibinfo {author} {\bibfnamefont
  {D.~R.}\ \bibnamefont {Rehn}}, \bibinfo {author} {\bibfnamefont {M.~L.}\
  \bibnamefont {Vidal}}, \bibinfo {author} {\bibfnamefont {Z.~Q.}\ \bibnamefont
  {You}}, \bibinfo {author} {\bibfnamefont {Y.}~\bibnamefont {Zhu}}, \bibinfo
  {author} {\bibfnamefont {B.}~\bibnamefont {Alam}}, \bibinfo {author}
  {\bibfnamefont {B.~J.}\ \bibnamefont {Albrecht}}, \bibinfo {author}
  {\bibfnamefont {A.}~\bibnamefont {Aldossary}}, \bibinfo {author}
  {\bibfnamefont {E.}~\bibnamefont {Alguire}}, \bibinfo {author} {\bibfnamefont
  {J.~H.}\ \bibnamefont {Andersen}}, \bibinfo {author} {\bibfnamefont
  {V.}~\bibnamefont {Athavale}}, \bibinfo {author} {\bibfnamefont
  {D.}~\bibnamefont {Barton}}, \bibinfo {author} {\bibfnamefont
  {K.}~\bibnamefont {Begam}}, \bibinfo {author} {\bibfnamefont
  {A.}~\bibnamefont {Behn}}, \bibinfo {author} {\bibfnamefont {N.}~\bibnamefont
  {Bellonzi}}, \bibinfo {author} {\bibfnamefont {Y.~A.}\ \bibnamefont
  {Bernard}}, \bibinfo {author} {\bibfnamefont {E.~J.}\ \bibnamefont
  {Berquist}}, \bibinfo {author} {\bibfnamefont {H.~G.}\ \bibnamefont
  {Burton}}, \bibinfo {author} {\bibfnamefont {A.}~\bibnamefont {Carreras}},
  \bibinfo {author} {\bibfnamefont {K.}~\bibnamefont {Carter-Fenk}}, \bibinfo
  {author} {\bibfnamefont {R.}~\bibnamefont {Chakraborty}}, \bibinfo {author}
  {\bibfnamefont {A.~D.}\ \bibnamefont {Chien}}, \bibinfo {author}
  {\bibfnamefont {K.~D.}\ \bibnamefont {Closser}}, \bibinfo {author}
  {\bibfnamefont {V.}~\bibnamefont {Cofer-Shabica}}, \bibinfo {author}
  {\bibfnamefont {S.}~\bibnamefont {Dasgupta}}, \bibinfo {author}
  {\bibfnamefont {M.~D.}\ \bibnamefont {Wergifosse}}, \bibinfo {author}
  {\bibfnamefont {J.}~\bibnamefont {Deng}}, \bibinfo {author} {\bibfnamefont
  {M.}~\bibnamefont {Diedenhofen}}, \bibinfo {author} {\bibfnamefont
  {H.}~\bibnamefont {Do}}, \bibinfo {author} {\bibfnamefont {S.}~\bibnamefont
  {Ehlert}}, \bibinfo {author} {\bibfnamefont {P.~T.}\ \bibnamefont {Fang}},
  \bibinfo {author} {\bibfnamefont {S.}~\bibnamefont {Fatehi}}, \bibinfo
  {author} {\bibfnamefont {Q.}~\bibnamefont {Feng}}, \bibinfo {author}
  {\bibfnamefont {T.}~\bibnamefont {Friedhoff}}, \bibinfo {author}
  {\bibfnamefont {J.}~\bibnamefont {Gayvert}}, \bibinfo {author} {\bibfnamefont
  {Q.}~\bibnamefont {Ge}}, \bibinfo {author} {\bibfnamefont {G.}~\bibnamefont
  {Gidofalvi}}, \bibinfo {author} {\bibfnamefont {M.}~\bibnamefont {Goldey}},
  \bibinfo {author} {\bibfnamefont {J.}~\bibnamefont {Gomes}}, \bibinfo
  {author} {\bibfnamefont {C.~E.}\ \bibnamefont {González-Espinoza}}, \bibinfo
  {author} {\bibfnamefont {S.}~\bibnamefont {Gulania}}, \bibinfo {author}
  {\bibfnamefont {A.~O.}\ \bibnamefont {Gunina}}, \bibinfo {author}
  {\bibfnamefont {M.~W.}\ \bibnamefont {Hanson-Heine}}, \bibinfo {author}
  {\bibfnamefont {P.~H.}\ \bibnamefont {Harbach}}, \bibinfo {author}
  {\bibfnamefont {A.}~\bibnamefont {Hauser}}, \bibinfo {author} {\bibfnamefont
  {M.~F.}\ \bibnamefont {Herbst}}, \bibinfo {author} {\bibfnamefont {M.~H.}\
  \bibnamefont {Vera}}, \bibinfo {author} {\bibfnamefont {M.}~\bibnamefont
  {Hodecker}}, \bibinfo {author} {\bibfnamefont {Z.~C.}\ \bibnamefont
  {Holden}}, \bibinfo {author} {\bibfnamefont {S.}~\bibnamefont {Houck}},
  \bibinfo {author} {\bibfnamefont {X.}~\bibnamefont {Huang}}, \bibinfo
  {author} {\bibfnamefont {K.}~\bibnamefont {Hui}}, \bibinfo {author}
  {\bibfnamefont {B.~C.}\ \bibnamefont {Huynh}}, \bibinfo {author}
  {\bibfnamefont {M.}~\bibnamefont {Ivanov}}, \bibinfo {author} {\bibnamefont
  {Ádám Jász}}, \bibinfo {author} {\bibfnamefont {H.}~\bibnamefont {Ji}},
  \bibinfo {author} {\bibfnamefont {H.}~\bibnamefont {Jiang}}, \bibinfo
  {author} {\bibfnamefont {B.}~\bibnamefont {Kaduk}}, \bibinfo {author}
  {\bibfnamefont {S.}~\bibnamefont {Kähler}}, \bibinfo {author} {\bibfnamefont
  {K.}~\bibnamefont {Khistyaev}}, \bibinfo {author} {\bibfnamefont
  {J.}~\bibnamefont {Kim}}, \bibinfo {author} {\bibfnamefont {G.}~\bibnamefont
  {Kis}}, \bibinfo {author} {\bibfnamefont {P.}~\bibnamefont {Klunzinger}},
  \bibinfo {author} {\bibfnamefont {Z.}~\bibnamefont {Koczor-Benda}}, \bibinfo
  {author} {\bibfnamefont {J.~H.}\ \bibnamefont {Koh}}, \bibinfo {author}
  {\bibfnamefont {D.}~\bibnamefont {Kosenkov}}, \bibinfo {author}
  {\bibfnamefont {L.}~\bibnamefont {Koulias}}, \bibinfo {author} {\bibfnamefont
  {T.}~\bibnamefont {Kowalczyk}}, \bibinfo {author} {\bibfnamefont {C.~M.}\
  \bibnamefont {Krauter}}, \bibinfo {author} {\bibfnamefont {K.}~\bibnamefont
  {Kue}}, \bibinfo {author} {\bibfnamefont {A.}~\bibnamefont {Kunitsa}},
  \bibinfo {author} {\bibfnamefont {T.}~\bibnamefont {Kus}}, \bibinfo {author}
  {\bibfnamefont {I.}~\bibnamefont {Ladjánszki}}, \bibinfo {author}
  {\bibfnamefont {A.}~\bibnamefont {Landau}}, \bibinfo {author} {\bibfnamefont
  {K.~V.}\ \bibnamefont {Lawler}}, \bibinfo {author} {\bibfnamefont
  {D.}~\bibnamefont {Lefrancois}}, \bibinfo {author} {\bibfnamefont
  {S.}~\bibnamefont {Lehtola}}, \bibinfo {author} {\bibfnamefont {R.~R.}\
  \bibnamefont {Li}}, \bibinfo {author} {\bibfnamefont {Y.~P.}\ \bibnamefont
  {Li}}, \bibinfo {author} {\bibfnamefont {J.}~\bibnamefont {Liang}}, \bibinfo
  {author} {\bibfnamefont {M.}~\bibnamefont {Liebenthal}}, \bibinfo {author}
  {\bibfnamefont {H.~H.}\ \bibnamefont {Lin}}, \bibinfo {author} {\bibfnamefont
  {Y.~S.}\ \bibnamefont {Lin}}, \bibinfo {author} {\bibfnamefont
  {F.}~\bibnamefont {Liu}}, \bibinfo {author} {\bibfnamefont {K.~Y.}\
  \bibnamefont {Liu}}, \bibinfo {author} {\bibfnamefont {M.}~\bibnamefont
  {Loipersberger}}, \bibinfo {author} {\bibfnamefont {A.}~\bibnamefont
  {Luenser}}, \bibinfo {author} {\bibfnamefont {A.}~\bibnamefont {Manjanath}},
  \bibinfo {author} {\bibfnamefont {P.}~\bibnamefont {Manohar}}, \bibinfo
  {author} {\bibfnamefont {E.}~\bibnamefont {Mansoor}}, \bibinfo {author}
  {\bibfnamefont {S.~F.}\ \bibnamefont {Manzer}}, \bibinfo {author}
  {\bibfnamefont {S.~P.}\ \bibnamefont {Mao}}, \bibinfo {author} {\bibfnamefont
  {A.~V.}\ \bibnamefont {Marenich}}, \bibinfo {author} {\bibfnamefont
  {T.}~\bibnamefont {Markovich}}, \bibinfo {author} {\bibfnamefont
  {S.}~\bibnamefont {Mason}}, \bibinfo {author} {\bibfnamefont {S.~A.}\
  \bibnamefont {Maurer}}, \bibinfo {author} {\bibfnamefont {P.~F.}\
  \bibnamefont {McLaughlin}}, \bibinfo {author} {\bibfnamefont {M.~F.}\
  \bibnamefont {Menger}}, \bibinfo {author} {\bibfnamefont {J.~M.}\
  \bibnamefont {Mewes}}, \bibinfo {author} {\bibfnamefont {S.~A.}\ \bibnamefont
  {Mewes}}, \bibinfo {author} {\bibfnamefont {P.}~\bibnamefont {Morgante}},
  \bibinfo {author} {\bibfnamefont {J.~W.}\ \bibnamefont {Mullinax}}, \bibinfo
  {author} {\bibfnamefont {K.~J.}\ \bibnamefont {Oosterbaan}}, \bibinfo
  {author} {\bibfnamefont {G.}~\bibnamefont {Paran}}, \bibinfo {author}
  {\bibfnamefont {A.~C.}\ \bibnamefont {Paul}}, \bibinfo {author}
  {\bibfnamefont {S.~K.}\ \bibnamefont {Paul}}, \bibinfo {author}
  {\bibfnamefont {F.}~\bibnamefont {Pavošević}}, \bibinfo {author}
  {\bibfnamefont {Z.}~\bibnamefont {Pei}}, \bibinfo {author} {\bibfnamefont
  {S.}~\bibnamefont {Prager}}, \bibinfo {author} {\bibfnamefont {E.~I.}\
  \bibnamefont {Proynov}}, \bibinfo {author} {\bibnamefont {Ádám Rák}},
  \bibinfo {author} {\bibfnamefont {E.}~\bibnamefont {Ramos-Cordoba}}, \bibinfo
  {author} {\bibfnamefont {B.}~\bibnamefont {Rana}}, \bibinfo {author}
  {\bibfnamefont {A.~E.}\ \bibnamefont {Rask}}, \bibinfo {author}
  {\bibfnamefont {A.}~\bibnamefont {Rettig}}, \bibinfo {author} {\bibfnamefont
  {R.~M.}\ \bibnamefont {Richard}}, \bibinfo {author} {\bibfnamefont
  {F.}~\bibnamefont {Rob}}, \bibinfo {author} {\bibfnamefont {E.}~\bibnamefont
  {Rossomme}}, \bibinfo {author} {\bibfnamefont {T.}~\bibnamefont {Scheele}},
  \bibinfo {author} {\bibfnamefont {M.}~\bibnamefont {Scheurer}}, \bibinfo
  {author} {\bibfnamefont {M.}~\bibnamefont {Schneider}}, \bibinfo {author}
  {\bibfnamefont {N.}~\bibnamefont {Sergueev}}, \bibinfo {author}
  {\bibfnamefont {S.~M.}\ \bibnamefont {Sharada}}, \bibinfo {author}
  {\bibfnamefont {W.}~\bibnamefont {Skomorowski}}, \bibinfo {author}
  {\bibfnamefont {D.~W.}\ \bibnamefont {Small}}, \bibinfo {author}
  {\bibfnamefont {C.~J.}\ \bibnamefont {Stein}}, \bibinfo {author}
  {\bibfnamefont {Y.~C.}\ \bibnamefont {Su}}, \bibinfo {author} {\bibfnamefont
  {E.~J.}\ \bibnamefont {Sundstrom}}, \bibinfo {author} {\bibfnamefont
  {Z.}~\bibnamefont {Tao}}, \bibinfo {author} {\bibfnamefont {J.}~\bibnamefont
  {Thirman}}, \bibinfo {author} {\bibfnamefont {G.~J.}\ \bibnamefont {Tornai}},
  \bibinfo {author} {\bibfnamefont {T.}~\bibnamefont {Tsuchimochi}}, \bibinfo
  {author} {\bibfnamefont {N.~M.}\ \bibnamefont {Tubman}}, \bibinfo {author}
  {\bibfnamefont {S.~P.}\ \bibnamefont {Veccham}}, \bibinfo {author}
  {\bibfnamefont {O.}~\bibnamefont {Vydrov}}, \bibinfo {author} {\bibfnamefont
  {J.}~\bibnamefont {Wenzel}}, \bibinfo {author} {\bibfnamefont
  {J.}~\bibnamefont {Witte}}, \bibinfo {author} {\bibfnamefont
  {A.}~\bibnamefont {Yamada}}, \bibinfo {author} {\bibfnamefont
  {K.}~\bibnamefont {Yao}}, \bibinfo {author} {\bibfnamefont {S.}~\bibnamefont
  {Yeganeh}}, \bibinfo {author} {\bibfnamefont {S.~R.}\ \bibnamefont {Yost}},
  \bibinfo {author} {\bibfnamefont {A.}~\bibnamefont {Zech}}, \bibinfo {author}
  {\bibfnamefont {I.~Y.}\ \bibnamefont {Zhang}}, \bibinfo {author}
  {\bibfnamefont {X.}~\bibnamefont {Zhang}}, \bibinfo {author} {\bibfnamefont
  {Y.}~\bibnamefont {Zhang}}, \bibinfo {author} {\bibfnamefont
  {D.}~\bibnamefont {Zuev}}, \bibinfo {author} {\bibfnamefont {A.}~\bibnamefont
  {Aspuru-Guzik}}, \bibinfo {author} {\bibfnamefont {A.~T.}\ \bibnamefont
  {Bell}}, \bibinfo {author} {\bibfnamefont {N.~A.}\ \bibnamefont {Besley}},
  \bibinfo {author} {\bibfnamefont {K.~B.}\ \bibnamefont {Bravaya}}, \bibinfo
  {author} {\bibfnamefont {B.~R.}\ \bibnamefont {Brooks}}, \bibinfo {author}
  {\bibfnamefont {D.}~\bibnamefont {Casanova}}, \bibinfo {author}
  {\bibfnamefont {J.~D.}\ \bibnamefont {Chai}}, \bibinfo {author}
  {\bibfnamefont {S.}~\bibnamefont {Coriani}}, \bibinfo {author} {\bibfnamefont
  {C.~J.}\ \bibnamefont {Cramer}}, \bibinfo {author} {\bibfnamefont
  {G.}~\bibnamefont {Cserey}}, \bibinfo {author} {\bibfnamefont {A.~E.}\
  \bibnamefont {Deprince}}, \bibinfo {author} {\bibfnamefont {R.~A.}\
  \bibnamefont {Distasio}}, \bibinfo {author} {\bibfnamefont {A.}~\bibnamefont
  {Dreuw}}, \bibinfo {author} {\bibfnamefont {B.~D.}\ \bibnamefont {Dunietz}},
  \bibinfo {author} {\bibfnamefont {T.~R.}\ \bibnamefont {Furlani}}, \bibinfo
  {author} {\bibfnamefont {W.~A.}\ \bibnamefont {Goddard}}, \bibinfo {author}
  {\bibfnamefont {S.}~\bibnamefont {Hammes-Schiffer}}, \bibinfo {author}
  {\bibfnamefont {T.}~\bibnamefont {Head-Gordon}}, \bibinfo {author}
  {\bibfnamefont {W.~J.}\ \bibnamefont {Hehre}}, \bibinfo {author}
  {\bibfnamefont {C.~P.}\ \bibnamefont {Hsu}}, \bibinfo {author} {\bibfnamefont
  {T.~C.}\ \bibnamefont {Jagau}}, \bibinfo {author} {\bibfnamefont
  {Y.}~\bibnamefont {Jung}}, \bibinfo {author} {\bibfnamefont {A.}~\bibnamefont
  {Klamt}}, \bibinfo {author} {\bibfnamefont {J.}~\bibnamefont {Kong}},
  \bibinfo {author} {\bibfnamefont {D.~S.}\ \bibnamefont {Lambrecht}}, \bibinfo
  {author} {\bibfnamefont {W.}~\bibnamefont {Liang}}, \bibinfo {author}
  {\bibfnamefont {N.~J.}\ \bibnamefont {Mayhall}}, \bibinfo {author}
  {\bibfnamefont {C.~W.}\ \bibnamefont {McCurdy}}, \bibinfo {author}
  {\bibfnamefont {J.~B.}\ \bibnamefont {Neaton}}, \bibinfo {author}
  {\bibfnamefont {C.}~\bibnamefont {Ochsenfeld}}, \bibinfo {author}
  {\bibfnamefont {J.~A.}\ \bibnamefont {Parkhill}}, \bibinfo {author}
  {\bibfnamefont {R.}~\bibnamefont {Peverati}}, \bibinfo {author}
  {\bibfnamefont {V.~A.}\ \bibnamefont {Rassolov}}, \bibinfo {author}
  {\bibfnamefont {Y.}~\bibnamefont {Shao}}, \bibinfo {author} {\bibfnamefont
  {L.~V.}\ \bibnamefont {Slipchenko}}, \bibinfo {author} {\bibfnamefont
  {T.}~\bibnamefont {Stauch}}, \bibinfo {author} {\bibfnamefont {R.~P.}\
  \bibnamefont {Steele}}, \bibinfo {author} {\bibfnamefont {J.~E.}\
  \bibnamefont {Subotnik}}, \bibinfo {author} {\bibfnamefont {A.~J.}\
  \bibnamefont {Thom}}, \bibinfo {author} {\bibfnamefont {A.}~\bibnamefont
  {Tkatchenko}}, \bibinfo {author} {\bibfnamefont {D.~G.}\ \bibnamefont
  {Truhlar}}, \bibinfo {author} {\bibfnamefont {T.~V.}\ \bibnamefont
  {Voorhis}}, \bibinfo {author} {\bibfnamefont {T.~A.}\ \bibnamefont
  {Wesolowski}}, \bibinfo {author} {\bibfnamefont {K.~B.}\ \bibnamefont
  {Whaley}}, \bibinfo {author} {\bibfnamefont {H.~L.}\ \bibnamefont
  {Woodcock}}, \bibinfo {author} {\bibfnamefont {P.~M.}\ \bibnamefont
  {Zimmerman}}, \bibinfo {author} {\bibfnamefont {S.}~\bibnamefont {Faraji}},
  \bibinfo {author} {\bibfnamefont {P.~M.}\ \bibnamefont {Gill}}, \bibinfo
  {author} {\bibfnamefont {M.}~\bibnamefont {Head-Gordon}}, \bibinfo {author}
  {\bibfnamefont {J.~M.}\ \bibnamefont {Herbert}},\ and\ \bibinfo {author}
  {\bibfnamefont {A.~I.}\ \bibnamefont {Krylov}},\ }\bibfield  {title}
  {\enquote {\bibinfo {title} {Software for the frontiers of quantum chemistry:
  An overview of developments in the q-chem 5 package},}\ }\href
  {https://doi.org/10.1063/5.0055522} {\bibfield  {journal} {\bibinfo
  {journal} {Journal of Chemical Physics}\ }\textbf {\bibinfo {volume} {155}},\
  \bibinfo {pages} {084801} (\bibinfo {year} {2021})}\BibitemShut {NoStop}%
\bibitem [{DIR()}]{DIRAC23}%
  \BibitemOpen
  \href@noop {} {}\bibinfo {note} {{DIRAC}, a relativistic ab initio electronic
  structure program, Release {DIRAC23} (2023), written by R.~numerov,
  A.~S.~P.~Gomes, T.~Saue and L.~Visscher and H.~J.~{\relax Aa}.~Jensen, with
  contributions from I.~A.~Aucar, V.~Bakken, C.~Chibueze, J.~Creutzberg,
  K.~G.~Dyall, S.~Dubillard, U.~Ekstr{\"o}m, E.~Eliav, T.~Enevoldsen,
  E.~Fa{\ss}hauer, T.~Fleig, O.~Fossgaard, L.~Halbert, E.~D.~Hedeg{\aa}rd,
  T.~Helgaker, B.~Helmich--Paris, J.~Henriksson, M.~van~Horn, M.~Ilia{\v{s}},
  Ch.~R.~Jacob, S.~Knecht, S.~Komorovsk{\'y}, O.~Kullie, J.~K.~L{\ae}rdahl,
  C.~V.~Larsen, Y.~S.~Lee, N.~H.~List, H.~S.~Nataraj, M.~K.~Nayak, P.~Norman,
  A.~Nyvang, G.~Olejniczak, J.~Olsen, J.~M.~H.~Olsen, A.~Papadopoulos,
  Y.~C.~Park, J.~K.~Pedersen, M.~Pernpointner, J.~V.~Pototschnig,
  R.~di~Remigio, M.~Repisky, K.~Ruud, P.~Sa{\l}ek, B.~Schimmelpfennig,
  B.~Senjean, A.~Shee, J.~Sikkema, A.~Sunaga, A.~J.~Thorvaldsen, J.~Thyssen,
  J.~van~Stralen, M.~L.~Vidal, S.~Villaume, O.~Visser, T.~Winther, S.~Yamamoto
  and X.~Yuan (available at \url{https://doi.org/10.5281/zenodo.7670749}, see
  also \url{https://www.diracprogram.org})}\BibitemShut {NoStop}%
\bibitem [{\citenamefont {Schimmelpfennig}(1999)}]{AMFI}%
  \BibitemOpen
  \bibfield  {author} {\bibinfo {author} {\bibfnamefont {B.}~\bibnamefont
  {Schimmelpfennig}},\ }\href@noop {} {\enquote {\bibinfo {title} {{AMFI}, an
  atomic mean-field spin-orbit integral program},}\ } (\bibinfo {year} {1996
  and 1999}),\ \bibinfo {note} {{B}ernd Schimmelpfennig, University of
  Stockholm}\BibitemShut {NoStop}%
\bibitem [{\citenamefont {Thierfelder}, \citenamefont {Rauhut},\ and\
  \citenamefont {Schwerdtfeger}(2010)}]{camb3lyp*}%
  \BibitemOpen
  \bibfield  {author} {\bibinfo {author} {\bibfnamefont {C.}~\bibnamefont
  {Thierfelder}}, \bibinfo {author} {\bibfnamefont {G.}~\bibnamefont
  {Rauhut}},\ and\ \bibinfo {author} {\bibfnamefont {P.}~\bibnamefont
  {Schwerdtfeger}},\ }\bibfield  {title} {\enquote {\bibinfo {title}
  {Relativistic coupled-cluster study of the parity-violation energy shift of
  chfclbr},}\ }\href {https://doi.org/10.1103/PhysRevA.81.032513} {\bibfield
  {journal} {\bibinfo  {journal} {Phys. Rev. A}\ }\textbf {\bibinfo {volume}
  {81}},\ \bibinfo {pages} {032513} (\bibinfo {year} {2010})}\BibitemShut
  {NoStop}%
\bibitem [{\citenamefont {Dyall}(2006)}]{Dyall2006}%
  \BibitemOpen
  \bibfield  {author} {\bibinfo {author} {\bibfnamefont {K.~G.}\ \bibnamefont
  {Dyall}},\ }\bibfield  {title} {\enquote {\bibinfo {title} {Relativistic
  quadruple-zeta and revised triple-zeta and double-zeta basis sets for the 4p,
  5p, and 6p elements},}\ }\href {https://doi.org/10.1007/s00214-006-0126-0}
  {\bibfield  {journal} {\bibinfo  {journal} {Theoretical Chemistry Accounts}\
  }\textbf {\bibinfo {volume} {115}},\ \bibinfo {pages} {441--447} (\bibinfo
  {year} {2006})}\BibitemShut {NoStop}%
\bibitem [{\citenamefont {Noumerov}(1924)}]{numerov1}%
  \BibitemOpen
  \bibfield  {author} {\bibinfo {author} {\bibfnamefont {B.~V.}\ \bibnamefont
  {Noumerov}},\ }\bibfield  {title} {\enquote {\bibinfo {title} {{A Method of
  Extrapolation of Perturbations}},}\ }\href
  {https://doi.org/10.1093/mnras/84.8.592} {\bibfield  {journal} {\bibinfo
  {journal} {Monthly Notices of the Royal Astronomical Society}\ }\textbf
  {\bibinfo {volume} {84}},\ \bibinfo {pages} {592--602} (\bibinfo {year}
  {1924})},\ \Eprint
  {https://arxiv.org/abs/https://academic.oup.com/mnras/article-pdf/84/8/592/3661174/mnras84-0592.pdf}
  {https://academic.oup.com/mnras/article-pdf/84/8/592/3661174/mnras84-0592.pdf}
  \BibitemShut {NoStop}%
\bibitem [{\citenamefont {Cooley}(1961)}]{Cooley_1961}%
  \BibitemOpen
  \bibfield  {author} {\bibinfo {author} {\bibfnamefont {J.~W.}\ \bibnamefont
  {Cooley}},\ }\bibfield  {title} {\enquote {\bibinfo {title} {{An Improved
  Eigenvalue Corrector Formula for Solving the Schrodinger Equation for Central
  Fields}},}\ }\href {https://doi.org/10.2307/2003025} {\bibfield  {journal}
  {\bibinfo  {journal} {Mathematics of Computation}\ } (\bibinfo {year}
  {1961}),\ 10.2307/2003025}\BibitemShut {NoStop}%
\bibitem [{\citenamefont {Bast}(2017)}]{Bast_2017}%
  \BibitemOpen
  \bibfield  {author} {\bibinfo {author} {\bibfnamefont {R.}~\bibnamefont
  {Bast}},\ }\bibfield  {title} {\enquote {\bibinfo {title} {bast/numerov
  v0.5.0},}\ }\href {https://doi.org/10.5281/zenodo.1000406} {\  (\bibinfo
  {year} {2017}),\ 10.5281/zenodo.1000406}\BibitemShut {NoStop}%
\bibitem [{\citenamefont {Brazier}\ and\ \citenamefont
  {Bernath}(1987)}]{brazier1987}%
  \BibitemOpen
  \bibfield  {author} {\bibinfo {author} {\bibfnamefont {C.~R.}\ \bibnamefont
  {Brazier}}\ and\ \bibinfo {author} {\bibfnamefont {P.~F.}\ \bibnamefont
  {Bernath}},\ }\bibfield  {title} {\enquote {\bibinfo {title} {{Observation of
  gas phase organometallic free radicals: Monomethyl derivatives of calcium and
  strontium}},}\ }\href {https://doi.org/10.1063/1.452476} {\bibfield
  {journal} {\bibinfo  {journal} {The Journal of Chemical Physics}\ }\textbf
  {\bibinfo {volume} {86}},\ \bibinfo {pages} {5918--5922} (\bibinfo {year}
  {1987})}\BibitemShut {NoStop}%
\bibitem [{\citenamefont {Chamorro}\ \emph {et~al.}(2022)\citenamefont
  {Chamorro}, \citenamefont {Borschevsky}, \citenamefont {Eliav}, \citenamefont
  {Hutzler}, \citenamefont {Hoekstra},\ and\ \citenamefont
  {Pa{\v{s}}teka}}]{chamorro2022molecular}%
  \BibitemOpen
  \bibfield  {author} {\bibinfo {author} {\bibfnamefont {Y.}~\bibnamefont
  {Chamorro}}, \bibinfo {author} {\bibfnamefont {A.}~\bibnamefont
  {Borschevsky}}, \bibinfo {author} {\bibfnamefont {E.}~\bibnamefont {Eliav}},
  \bibinfo {author} {\bibfnamefont {N.~R.}\ \bibnamefont {Hutzler}}, \bibinfo
  {author} {\bibfnamefont {S.}~\bibnamefont {Hoekstra}},\ and\ \bibinfo
  {author} {\bibfnamefont {L.~F.}\ \bibnamefont {Pa{\v{s}}teka}},\ }\bibfield
  {title} {\enquote {\bibinfo {title} {Molecular enhancement factors for the p,
  t-violating electric dipole moment of the electron in bach 3 and ybch 3
  symmetric top molecules},}\ }\href@noop {} {\bibfield  {journal} {\bibinfo
  {journal} {Physical Review A}\ }\textbf {\bibinfo {volume} {106}},\ \bibinfo
  {pages} {052811} (\bibinfo {year} {2022})}\BibitemShut {NoStop}%
\bibitem [{\citenamefont {Pieniazek}\ \emph {et~al.}(2007)\citenamefont
  {Pieniazek}, \citenamefont {Arnstein}, \citenamefont {Bradforth},
  \citenamefont {Krylov},\ and\ \citenamefont
  {Sherrill}}]{eomip_benchmark_krylov2007}%
  \BibitemOpen
  \bibfield  {author} {\bibinfo {author} {\bibfnamefont {P.}~\bibnamefont
  {Pieniazek}}, \bibinfo {author} {\bibfnamefont {S.}~\bibnamefont {Arnstein}},
  \bibinfo {author} {\bibfnamefont {S.}~\bibnamefont {Bradforth}}, \bibinfo
  {author} {\bibfnamefont {A.}~\bibnamefont {Krylov}},\ and\ \bibinfo {author}
  {\bibfnamefont {C.}~\bibnamefont {Sherrill}},\ }\bibfield  {title} {\enquote
  {\bibinfo {title} {Benchmark full configuration interaction and eom-ip-ccsd
  results for prototypical charge transfer systems: noncovalent ionized
  dimers},}\ }\href@noop {} {\bibfield  {journal} {\bibinfo  {journal} {J.
  Chem. Phys}\ }\textbf {\bibinfo {volume} {127}},\ \bibinfo {pages} {164110}
  (\bibinfo {year} {2007})}\BibitemShut {NoStop}%
\bibitem [{\citenamefont {Wu}(1942)}]{CH2BrIvibFreq1942}%
  \BibitemOpen
  \bibfield  {author} {\bibinfo {author} {\bibfnamefont {T.-Y.}\ \bibnamefont
  {Wu}},\ }\bibfield  {title} {\enquote {\bibinfo {title} {Systematics in the
  vibrational spectra of the halogen derivatives of methane},}\ }\href@noop {}
  {\bibfield  {journal} {\bibinfo  {journal} {The Journal of Chemical Physics}\
  }\textbf {\bibinfo {volume} {10}},\ \bibinfo {pages} {116--124} (\bibinfo
  {year} {1942})}\BibitemShut {NoStop}%
\end{thebibliography}%

\clearpage

\section{Supplemental information}

\subsection{Benchmark calculations}
\label{banchmark}

As detailed in Section~\ref{comp}, we tested two methodologies, first, geometries are obtained via MP2 and energies via CCSD(T), second, geometries and energies via EOM-CCSD.
In addition we use two basis sets, TZ (triple$-\zeta$) and a larger  QZ (quadruple$-\zeta$).
Tables~\ref{tbl:ips1}-~\ref{tbl:ips2} present these calculations for the VIPs and AIPs, respectively. 

We analyse the different contributions using the following parameters: $\Delta^{XZ}=E^{XZ}_{CCSD(T)}-E^{XZ}_{EOMCCSD}$ (X=T, Q); $\Delta^{Trpl}=E^{TZ}_{CCSD}-E^{TZ}_{CCSD(T)}$; $\Delta^{TZ}_{QZ}=E^{TZ}_{EOMCCSD}-E^{QZ}_{EOMCCSD}$.
In addition, Tables~\ref{tbl:ips1}-~\ref{tbl:ips2} present the absolute mean value,  $|\overline{\Delta}|$, and the absolute maximal energy difference value, $|{\Delta^{Max}}|$.
Notice that the VIP and AIP values are in eV, whereas the $\Delta$'s are in meV. 
Table~\ref{tbl:ips1} suggests that the perturbative triples contribution to the CCSD VIPs is very small.
In addition, $\Delta^{XZ}$ is small, i.e., the two methods provide similar results.
The largest parameter for the VIPs is $\Delta^{TZ}_{QZ}$, which is also quite small, with 0.08 eV mean value and 0.2 eV maximal difference.
Table~\ref{tbl:ips2} also suggests that the triples contribution to the CCSD AIPs and that $\Delta^{TZ}$ are very small, i.e., the two methods provide similar results when the smaller TZ basis set is used.
Contrary, for the QZ basis set  $\Delta^{QZ}$ appear larger with maximal difference value larger than 0.5 eV.

The reason for that is false MP2 geometries when using the larger, QZ, basis set.
In two cases the MP2/QZ optimization yields very different structures then EOM-IP-CCSD/QZ, for $^2$CH$_2$Br$_2^+$ and $^2$CH$_2$BrI$^+$, and as a result the  QZ  AIPs with EOM-IP-CCSD and CCSD(T) are quite different. 
For the other molecular systems the two methods yield very similar structures.
{Notice that the IP/EA EOM-CCSD variants are specifically intended for studying radical electronic structures since they describes the ground (reference) and doublet radicals (target states) on equal footing.~\cite{eomip_benchmark_krylov2007}}

Moreover, this inconsistency occurs only for the larger, QZ, basis set, i.e., using the TZ basis the two method are in agreement. 
Table~\ref{tbl:geom_cmpr} presents the Z-matrix of  $^2$CH$_2$BrI$^+$ on top and of $^2$CH$_2$Br$_2^+$ on the bottom panel.
The most pronounced differences are marked in bold in Table~\ref{tbl:geom_cmpr}.
For $^2$CH$_2$BrI$^+$ are the angles:
$\angle$HCI, 111\textdegree~(EOM-CCSD) vs 102\textdegree~(MP2), $\angle$BrCI, 93\textdegree vs 114\textdegree; and the dihedral angle  $\angle$HCIH, 131\textdegree vs 115\textdegree.
For $^2$CH$_2$Br$_2$$^+$:
the $\angle$BrCBr angle, 91.5\textdegree vs 120\textdegree and the dihedral angle  $\angle$HCBrH, 133\textdegree vs 113\textdegree, respectively.
In addition, Table~\ref{tbl:geom_cmpr} presents the Z-matrix of the nine cations at the EOM-CCSD/QZ for doublet state cations and MP2/QZ for singlets.

Tables~\ref{tbl:dis1}$-$\ref{tbl:dis3} present similar analysis for the calculated dissociation energies.
Summarizing all the dissociation channel energy differences from these three tables yields the following variations in the results:
\begin{itemize}
    \item $|\overline{\Delta}|^{TZ}=59$ meV and $|{\Delta^{Max}}|^{TZ}=103$ meV (26 values). 
    \item $|\overline{\Delta}|^{QZ}=160$ meV and $|{\Delta^{Max}}|^{QZ}=580$ meV  (23 values). 
    \item $|\overline{\Delta}|^{TZ}_{QZ}=124$ meV and $|{\Delta^{Max}}|^{TZ}_{QZ}=299$ meV (26 values).
\end{itemize}
Clearly the issues with the MP2/QZ optimization of radical electronic structures emerges here too.
In addition, it is desirable to use the larger basis set.
Therefore, above we report the EOM-CCSD/QZ calculated values.

An exception to this are the results in Table~\ref{tbl:dis.nogood.2}, which presents the CCSD(T) (at the MP2 geometries) dissociation energies for the candidates that are unlikely to be suitable.
The problematic MP2/QZ geometries (Table~\ref{tbl:geom_cmpr}) manifest in a relatively large difference (in the order of 0.5 eV, Table~\ref{tbl:dis1}) between CCSD(T)/TZ and CCSD(T)/QZ energies, however, in the case of Table~\ref{tbl:dis.nogood.2} the consistency between the two basis sets indicates the validity of the MP2 optimization.
Moreover, since these values are well below our 0.8 eV cutoff, we did not perform the EOM-CCSD calculations as done for the candidates in Table~\ref{tbl:dis.soc}.

Table~\ref{ee_basis} demonstrates the importance of adding diffuse basis functions when calculating excitation energies.
Inclusion of diffuse function via aug-XZ (X=T,Q) reduces the EEs with up to 1 eV for X=T and 0.7 eV for X=Q for the higher states of $^1$CHD$^{79}$Br$^{81}$Br.
These values are calculated via EOM-EE-CCSD.
For the $^1$CHD$^{79}$Br$^{81}$Br excitations, which are calculated via  EOM-EA-CCSD, the additional diffuse function play a much smaller role, where the largest effect (on the highest level) is 0.26 eV for X=T and 0.15 eV for X=Q.
Nevertheless, we use the aug-QZ calculated EEs values for all the candidates reported herein.

Table~\ref{freq_exprt} presents the vibrational frequency modes of the neutral achiral CH$_2$BrI and  CH$_2$Br$_2$ molecules calculated at the CCSD/TZ and  $\omega$B97M-V/TZ in comparison with experimental values.~\cite{CH2BrIvibFreq1942}
In order to better match the experimental vibrational frequencies for  CH$_2$BrI (CH$_2$Br$_2$) vibrational scaling factors of 0.942 (0.946) for CCSD and 0.946 (0.949) for  $\omega$B97M-V multiply each mode.
These scaling factors yield  absolute mean errors (with respect to the experimental values) of 8.8 (10.6) and 13.1 (14.4) cm$^{-1}$  for  CH$_2$BrI (CH$_2$Br$_2$) for the CCSD and  $\omega$B97M-V, respectively. 
A similar scaling should be applied to the cation results in this work.

 \begin{table*}
 	\caption{Vertical ionization potentials (VIPs). CCSD and CCSD(T) are calculated at the MP2 geometry with the same basis set. The EOM-CCSD energies are calculated at the EOM-CCSD geometry with the same basis set. $\Delta^{XZ}=E^{XZ}_{CCSD(T)}-E^{XZ}_{EOMCCSD}$ (X=T, Q); $\Delta^{Trpl}=E^{TZ}_{CCSD}-E^{TZ}_{CCSD(T)}$; $\Delta^{TZ}_{QZ}=E^{TZ}_{EOMCCSD}-E^{QZ}_{EOMCCSD}$. $|\overline{\Delta}|$ is the absolute mean value and $|{\Delta^{Max}}|$ is the absolute maximal energy difference value. VIPs are in eV, whereas the $\Delta$'s are in meV. 
 	}
   \label{tbl:ips1}
   \def\arraystretch{1.2}
 \begin{center}
   \begin{tabular}{l|c|c|c|c|c|c|c|c|c}
    \cline{1-9}
    \cline{1-9}
          VIP        &   \multicolumn{5}{c|}{TZ, eV ($\Delta$'s in meV)}                     &       \multicolumn{3}{c|}{QZ, eV ($\Delta$'s in meV)} \\
                     \hline
 Molecule     & {CCSD}     & {CCSD(T)}  &$\Delta^{tripls} $& {EOMCCSD} & $\Delta^{TZ}$  & {CCSD(T)}   & {EOMCCSD}   &  $\Delta^{QZ}$ &  $\Delta^{TZ}_{QZ}$ \\
   \hline                                                                                    
  {$^1$CH$_2$Br$_2$}  &  10.576   &   10.524  &   52 &  10.549   &   -25   &   10.816      &  10.680   &  136  & -131\\ 
 {$^1$CH$_2$BrI }     &   9.799   &    9.818  &  -19 &   9.837   &   -20   &   10.039      &   9.983   &   56  & -146\\
 {$^1$CHLiBr$_2$}  &   8.103   &    8.111  &   -8 &   8.250   &  -139   &    8.214      &   8.269   &  -55  &  -18\\
 {$^1$CHLiBrI}     &   7.941   &    7.954  &  -13 &   8.068   &  -114   &    8.076      &   8.123   &  -47  &  -55\\
 {$^1$CHNaBr$_2$}  &   7.487   &    7.493  &   -6 &   7.667   &  -174   &    7.785      &   7.851   &  -66  & -184\\
 {$^1$CHNaBrI}     &   7.361   &    7.376  &  -15 &   7.530   &  -154   &    7.684      &   7.743   &  -59  & -214\\
 {$^2$CHCaBr$_2$}  &   5.520   &    5.520  &    1 &   5.579   &   -60   &    5.502      &   5.557   &  -55  &   22\\
 {$^2$CHCaBrI}     &   5.497   &    5.497  &   -0 &   5.556   &   -58   &    5.475      &   5.534   &  -58  &   22\\ 
 {$^2$CHYbBr$_2$}  &   5.486   &    5.481  &    4 &   5.588   &  -106   &    5.508      &   5.603   &  -94  &  -15\\ 
  \hline
   \multicolumn{3}{c|}{$|\overline{\Delta}|; \quad |{\Delta^{Max}}|; \quad$ meV}              &   18; 66        &      &  92; 174    &      \multicolumn{2}{c|}{}    &    70; 136  &  82; 214    \\ 
  \hline
  \hline
   \end{tabular}  \\
   \end{center}
 \end{table*}

 \begin{table*}
 	\caption{Adiabatic ionization potentials (AIPs). CCSD and CCSD(T) are calculated at the MP2 geometry with the same basis set. The EOM-CCSD energies are calculated at the EOM-CCSD geometry with the same basis set. $\Delta^{XZ}=E^{XZ}_{CCSD(T)}-E^{XZ}_{EOMCCSD}$ (X=T, Q); $\Delta^{Trpl}=E^{TZ}_{CCSD}-E^{TZ}_{CCSD(T)}$; $\Delta^{TZ}_{QZ}=E^{TZ}_{EOMCCSD}-E^{QZ}_{EOMCCSD}$. $|\overline{\Delta}|$ is the absolute mean value and $|{\Delta^{Max}}|$ is the absolute maximal energy difference value. AIPs are in eV, whereas the $\Delta$'s are in meV. 
 	}
   \label{tbl:ips2}
   \def\arraystretch{1.2}
 \begin{center}
   \begin{tabular}{l|c|c|c|c|c|c|c|c|c}
    \cline{1-9}
    \cline{1-9}
           AIP        &   \multicolumn{5}{c|}{TZ, eV ($\Delta$'s in meV)}                     &       \multicolumn{3}{c|}{QZ, eV ($\Delta$'s in meV)} \\
                     \hline
 Molecule        & {CCSD}      & {CCSD(T)}  &$\Delta^{tripls}$& {EOMCCSD}  & $\Delta^{TZ}$  & {CCSD(T)}   & {EOMCCSD}   &  $\Delta^{QZ}$ &  $\Delta^{TZ}_{QZ}$ \\
   \hline                                                                                    
 {$^1$CH$_2$Br$_2$}   &  10.138    &   10.102  &   37          &  10.095   &    7         & 10.764      &  10.219   &  546   & -124      \\ 
 {$^1$CH$_2$BrI}      &   9.573    &    9.546  &   27          &   9.519   &   27         &  9.986      &   9.645   &  341   & -126      \\ 
 {$^1$CHLiBr$_2$}  &   6.990    &    7.091  & -101          &   7.074   &   17         &  7.183      &   7.167   &   17   &  -93      \\ 
 {$^1$CHNaBr$_2$}  &   6.350    &    6.471  & -120          &   6.424   &   47         &  6.964      &   6.912   &   52   & -489      \\ 
 {$^1$CHNaBrI}     &   6.358    &    6.465  & -107          &   6.412   &   53         &  6.959      &   6.920   &   38   & -508      \\ 
 {$^2$CHCaBr$_2$}  &   5.352    &    5.345  &    7          &   5.395   &  -51         &  5.327      &   5.378   &  -51   &   17      \\ 
 {$^2$CHCaBrI}     &   5.325    &    5.318  &    7          &   5.369   &  -51         &  5.302      &   5.354   &  -51   &   15      \\ 
 {$^2$CHYbBr$_2$}  &   5.308    &    5.372  &  -65          &   5.390   &  -17         &  5.401      &   5.431   &  -30   &  -42      \\ 
  \hline
   \multicolumn{3}{c|}{$|\overline{\Delta}|; \quad |{\Delta^{Max}}|; \quad$ meV}              &   57; 120       &      &  36; 53    &      \multicolumn{2}{c|}{}    &   140; 546  & 161; 508    \\ 
  \hline
  \hline
   \end{tabular}  \\
   \end{center}
 \end{table*}

  \begin{table}[!]
        \caption{Geometry optimization, EOM-IP-CCSD/QZ vs. MP2/QZ for $^2$CH$_2$BrI$^+$ (top) and  $^2$CH$_2$Br$_2^+$ (bottom). The geometries are presented as a Z-matrix in \AA.}
   \label{tbl:geom_cmpr}
 \begin{center}
   \begin{tabular}{ccccccc||ccccccc}
    \hline
    \hline
    \multicolumn{7}{c||}{EOM-CCSD/QZ}  &            \multicolumn{7}{c}{MP2/QZ}                \\
    \hline
I  &   &       &   &          &   &           &   I  &   &       &   &         &   &           \\
C  & 1 & 2.13 &   &           &   &           &   C  & 1 & 2.15 &   &         &   &           \\
H  & 2 & 1.08 & 1 & 111.1     &   &           &   H  & 2 & 1.10 & 1 & 102.4 &   &           \\
H  & 2 & 1.08 & 1 & 111.1 & 3 & {\bf -131.0} &   H  & 2 & 1.10 & 1 & 102.4 & 3 &  {\bf -115.4}  \\
Br & 2 & 1.91 & 1 &  {\bf 93.3} & 3 &  114.5 &   Br & 2 & 1.88 & 1 & {\bf 114.4} & 3 & 122.3  \\
   \hline
Br &   &       &   &         &   &           &  Br &   &       &   &         &   &            \\
C  & 1 & 1.93 &   &         &   &           &  C  & 1 & 1.86 &   &         &   &            \\
H  & 2 & 1.08 & 1 & 111.4 &   &           &  H  & 2 & 1.10 & 1 & 107.6 &   &            \\
H  & 2 & 1.08 & 1 & 111.4 & 3 & {\bf -132.6}  &  H  & 2 & 1.10 & 1 & 107.6 & 3 &  {\bf -113.4}   \\
Br & 2 & 1.93 & 1 &  {\bf 91.5} & 3 &  113.7  &  Br & 2 & 1.86 & 1 & {\bf 119.8} & 3 &   123.3   \\
 \hline
  \hline
   \end{tabular}  \\
   \end{center}
 \end{table}

\begin{table}[ht!]
        \caption{Geometry optimization, singlet cations via EOM-IP-CCSD/QZ and doublet cations via MP2/QZ. The geometries are presented as a Z-matrix in \AA ~and degrees\textdegree.}
   \label{tbl:geom}
 \begin{center}
   \begin{tabular}{ccccccc||ccccccc||ccccccc}
    \hline
    \hline
I  &   &      &   &       &   &         &  I  &    &       &   &        &   &         &  I  &   &      &   &       &   &        \\
C  & 1 & 2.13 &   &       &   &         &  C  & 1  & 2.04  &   &        &   &         &  C  & 1 & 2.04 &   &       &   &        \\
H  & 2 & 1.08 & 1 & 111.1 &   &         &  H  & 2  & 1.08  & 1 & 120.33 &   &         &  H  & 2 & 1.08 & 1 & 118.7 &   &        \\
H  & 2 & 1.08 & 1 & 111.1 & 3 & -131.0  &  Br & 2  & 1.86  & 1 & 118.13 & 3 & -154.7  &  Br & 2 & 1.86 & 1 & 121.5 & 3 &  156.8 \\
Br & 2 & 1.91 & 1 &  93.3 & 3 &  114.5  &  Li & 4  & 2.52  & 2 &  85.86 & 1 &   -2.6  &  Na & 4 & 2.87 & 2 &  86.2 & 1 &   18.9 \\
   \hline               
Br &   &      &   &       &   &         &  Br &    &       &   &        &   &         &  Br &   &      &   &       &   &        \\
C  & 1 & 1.93 &   &       &   &         &  C  & 1  &  1.87 &   &        &   &         &  C  & 1 & 1.85 &   &       &   &        \\
H  & 2 & 1.08 & 1 & 111.4 &   &         &  H  & 2  &  1.07 & 1 & 117.94 &   &         &  H  & 2 & 1.08 & 1 & 116.8 &   &        \\
H  & 2 & 1.08 & 1 & 111.4 & 3 & -132.6  &  Br & 2  &  1.87 & 1 & 117.83 & 3 & -151.5  &  Br & 2 & 1.85 & 1 & 119.6 & 3 &  150.0 \\
Br & 2 & 1.92 & 1 &  91.5 & 3 &  113.7  &  Li & 1  &  2.52 & 2 &  81.3  & 3 &  162.6  &  Na & 1 & 2.88 & 2 &  82.5 & 3 & -177.5 \\
    \hline                             
    \hline
Br &   &      &   &       &   &         &  Yb &    &       &   &        &   &        & I  &   &      &   &       &   &          \\
C  & 1 & 1.99 &   &       &   &         &  C  &  1 & 2.25  &   &        &   &        & C  & 1 & 2.18 &   &       &   &          \\
H  & 2 & 1.08 & 1 & 106.3 &   &         &  H  &  2 & 1.09  & 1 & 167.66 &   &        & H  & 2 & 1.09 & 1 & 106.2 &   &          \\
Br & 2 & 1.99 & 1 & 106.7 & 3 &-113.11  &  Br &  2 & 2.00  & 1 &  81.01 & 3 &  125.8 & Br & 2 & 2.00 & 1 & 108.0 & 3 &-113.29   \\
Ca & 2 & 2.24 & 1 &  81.7 & 3 & 168.14  &  Br &  2 & 2.00  & 1 &  81.01 & 3 &  125.8 & Ca & 2 & 2.25 & 1 &  84.0 & 3 & 167.69   \\    \hline        
  \hline
   \end{tabular}  \\
   \end{center}
 \end{table}

\clearpage

 \begin{longtable*}{l|c|c|c|c|c|c|c}
\caption{Dissociation energies. CCSD and CCSD(T) are calculated at the MP2 geometry with the same basis set. The EOM-IP-CCSD energies are calculated at the EOM-CCSD geometry with the same basis set. $\Delta^{XZ}=E^{XZ}_{CCSD(T)}-E^{XZ}_{EOMCCSD}$ (X=T, Q); $\Delta^{Trpl}=E^{TZ}_{CCSD}-E^{TZ}_{CCSD(T)}$; $\Delta^{TZ}_{QZ}=E^{TZ}_{EOMCCSD}-E^{QZ}_{EOMCCSD}$. $|\overline{\Delta}|$ is the absolute meaalue and $|{\Delta^{Max}}|$ is the absolute maximal energy difference value. Energies are in eV, whereas the $\Delta$'s are in meV.}\\
   \cline{2-7}
    \cline{2-7}
   \label{tbl:dis1}
   \def\arraystretch{1.2}
      &   \multicolumn{3}{c|}{TZ, eV ($\Delta$, meV)}   &       \multicolumn{3}{c|}{QZ, eV ($\Delta$, meV)} \\
   \hline                                                                                    
	   \text{Channel}            &       CCSD(T)       &IP-CCSD      & $\Delta^{TZ}$ &  CCSD(T)     &IP-CCSD      &  $\Delta^{QZ}$ &  $\Delta^{TZ}_{QZ}$ \\
   \hline                                                                                    
\multicolumn{1}{c}{{$^2$CH$_2$Br$_2^+   \rightarrow$}} & \multicolumn{5}{c|}{} \\                                                                          
	      \text{$^1$CH$_2$Br$^+$ + $^2$Br }          &      1.304    &  1.312    &       8     &  0.843    &  1.433    &       590   &    121   \\
              \text{$^2$CHBr$^+$ + $^1$HBr }             &      2.403    &  2.474    &      72     &  1.896    &  2.476    &       580   &      1   \\
  \hline         
\multicolumn{2}{c}{$^2${CH$_2$BrI$^+ \quad \rightarrow$}} & \multicolumn{5}{c}{} \\                                                                    
              \text{$^1$CH$_2$Br$^+$ + $^2$I  }          &      1.380    &  1.382    &       3     &  1.137    &  1.501    &       363   &    118   \\
              \text{$^1$CH$_2$I$^+$ + $^2$Br  }          &      1.588    &  1.643    &      55     &  1.347    &  1.763    &       415   &    119   \\
              \text{$^2$CHI$^+$ + $^1$HBr  }             &      2.686    &  2.750    &      64     &  2.399    &  2.740    &       341   &    -10   \\
              \text{$^2$CHBr$^+$ + $^1$HI  }             &      3.058    &  3.142    &      84     &  2.754    &  3.120    &       365   &    -22   \\
\hline           
\multicolumn{2}{c}{$^2${CHLiBr$_2^+ \quad  \rightarrow$}} & \multicolumn{5}{c}{} \\                                                                          
	      \text{$^2$CLiBr$^+$ + $^1$HBr}             &       2.5758   &  2.514     &     -61       &   2.692     &  2.667     &       -25   &   153    \\
              \text{$^1$CHLiBr$^+$ + $^2$Br}             &       2.5761   &  2.622     &      46       &   2.733     &  2.715     &       -18   &    93    \\
              \text{$^2$CHBr$^+$ + $^1$LiBr}             &       3.4530   &  ---       &    ---        &   3.508     &  ---       &      ---    &   ---    \\
  \hline         
\multicolumn{2}{c}{$^2${CHNaBr$_2^+ \quad  \rightarrow$}} & \multicolumn{5}{c}{} \\                                                
              \text{$^1$CHNaBr$^+$ + $^2$Br}             &       2.744   &  2.691     &     -53       &   2.971    &  2.959     &       -12   &   267    \\
              \text{$^2$CNaBr$^+$ + $^1$HBr}             &       2.646   &  2.723     &      77       &   2.799    &  2.791     &        -8   &    68    \\
              \text{$^2$CHBr$^+$ + $^1$NaBr}             &       3.715   &  ---       &     ---       &   3.563    &  ---       &      ---    &   ---    \\
  \hline         
\multicolumn{2}{c}{$^2${CHNaBrI$^+ \quad  \rightarrow$}} & \multicolumn{5}{c}{} \\                                               
              \text{$^1$CHNaBr$^+$ + $^2$I }             &       2.304   &  2.239     &     -65       &   2.589     &  2.536     &       -53   &    298   \\
              \text{$^1$CHNaI$^+$ + $^2$Br }             &       2.693   &  2.715     &      22       &   2.936     &  3.014     &        77   &    299   \\
              \text{$^2$CNaI$^+$ + $^1$HBr }             &       2.732   &  2.800     &      68       &   3.017     &  2.901     &      -116   &    101   \\
              \text{$^2$CNaBr$^+$ + $^1$HI }             &       2.785   &  2.867     &      82       &   2.980     &  2.945     &       -35   &     78   \\
              \text{$^2$CHBr$^+$ + $^1$NaI }             &       3.903   &  ---       &     ---       &   3.812     &  ---       &      ---    &    ---   \\
              \text{$^2$CHI$^+$ +  $^1$NaBr}             &       3.774   &  ---       &     ---       &   3.948     &  ---       &      ---    &    ---   \\
  \hline         
  \hline                         
\hline                         
\end{longtable*}                                                                                                                               
\pagebreak

 \begin{longtable*}{l|c|c|c|c|c|c|c} 	\caption{Dissociation energies. CCSD and CCSD(T) are calculated at the MP2 geometry with the same basis set. The EOM-EA-CCSD energies are calculated at the EOM-CCSD geometry with the same basis set. $\Delta^{XZ}=E^{XZ}_{CCSD(T)}-E^{XZ}_{EOMCCSD}$ (X=T, Q); $\Delta^{Trpl}=E^{TZ}_{CCSD}-E^{TZ}_{CCSD(T)}$; $\Delta^{TZ}_{QZ}=E^{TZ}_{EOMCCSD}-E^{QZ}_{EOMCCSD}$. Details regarding the absolute mean value $(|\overline{\Delta}|)$ and the absolute maximal energy difference value $(|{\Delta^{Max}}|)$ are given in Table... Energies are in eV, whereas the $\Delta$'s are in meV. } \\
    \cline{2-7}
    \cline{2-7}
   \label{tbl:dis3}
   \def\arraystretch{1.2}
      &   \multicolumn{3}{c|}{TZ, eV ($\Delta$, meV)}   &       \multicolumn{3}{c|}{QZ, eV ($\Delta$, meV)} \\
   \hline                                                                                    
	   \text{Molecule}               &  CCSD(T)     &  EA-CCSD   & $\Delta^{TZ}$  &  CCSD(T)  &  EA-CCSD    &  $\Delta^{QZ}$  &  $\Delta^{TZ}_{QZ}$ \\
  \hline
\multicolumn{2}{c}{$^1${CHCaBr$_2^+ \quad  \rightarrow$}} & \multicolumn{6}{c}{} \\                                                                          
	      \text{$^2$CHCaBr$^+$ + $^2$Br}            &       3.143    &    3.072     &   -71          &   3.260     &    3.252      &            -8  &  180   \\
	      \text{$^1$CCaBr$^+$ + $^1$HBr}            &       4.420    &    4.371     &   -48          &   4.470     &    4.418      &           -52  &   47   \\ 
	      \text{$^1$CBr$_2$ +  $^1$CaH$^+$}         &       4.745    &    4.742     &    -4          &   4.821     &    4.809      &           -13  &   67   \\ 
	      \text{$^1$CHBr$_2^+$ +  $^1$Ca}           &       4.766    &    4.766     &     0          &   4.977     &    4.965      &           -13  &  199   \\ 
  \hline         
\multicolumn{2}{c}{{$^1$CHCaBrI$^+ \quad  \rightarrow$}} & \multicolumn{6}{c}{} \\                                                         
              \text{$^2$CHCaBr$^+$ + $^2$I}             &       2.736    &    2.633     &  -103          &   2.863     &    2.821      &           -41  &  188   \\  
	      \text{$^2$CHCaI$^+$ + $^2$Br}             &       3.125    &    3.086     &   -39          &   3.246     &    3.267      &            21  &  181   \\
\multicolumn{2}{c}{{$^1$CHYbBr$_2^+ \quad  \rightarrow$}} & \multicolumn{6}{c}{} \\                                                      
              \text{$^2$CHYbBr$^+$ + $^2$Br}            &       3.135    &    ---       &                &   3.260     &    ---        &                &        \\
	      \text{$^1$CHBr$_2^+$ +  $^1$Yb}           &       4.491    &    4.593     &   102          &   4.639     &    4.700      &            61  &  107   \\ 
	      \text{$^1$CBr$_2$ +  $^1$YbH$^+$}         &       4.696    &    4.707     &    11          &    ---      &    4.731      &          ---   &   24   \\ 
	      \text{$^1$CYbBr$^+$ + $^1$HBr}            &       4.448    &    4.392     &  -56           &    ---      &    4.436      &          ---   &   44   \\

\hline
 \hline

 \end{longtable*}

 \begin{table}[]
        \caption{EEs w.r.t the size of the basis set. EOM-EE-CCSD for the singlet reference state and EOM-EA-CCSD for the doublet one}
\label{ee_basis}
   \begin{center}
   \begin{tabular}{|c|c|c|c|}
     \hline
     \hline
      \multicolumn{4}{|c|}{$^1$CHD$^{79}$Br$^{81}$Br}               \\
      \hline
     TZ    & aug-TZ &  QZ   &  aug-QZ   \\               
      \hline
     6.04  &  5.91  &  5.98  &  5.95     \\              
     6.12  &  5.99  &  6.05  &  6.01     \\
     6.38  &  6.25  &  6.31  &  6.28     \\
     6.56  &  6.42  &  6.49  &  6.45     \\
     7.82  &  7.39  &  7.69  &  7.47     \\
     7.99  &  7.43  &  7.84  &  7.49     \\
     8.33  &  7.48  &  8.16  &  7.56     \\
     8.54  &  7.53  &  8.30  &  7.59     \\

\hline
\hline
   \end{tabular}
   \end{center}
 \end{table}
                                                                                                                                                                                       
 \begin{table}[]
        \caption{The vibrational frequencies of neutral achiral CH$_2$BrI and CH$_2$Br$_2$ studied at the CCSD/TZ (CCSD) and  $\omega$B97M-V/TZ (DFT) are  compared with experimental measurements (Expt.).~\cite{CH2BrIvibFreq1942} The scaled frequencies for CH$_2$BrI  (CH$_2$Br$_2$) are presented using the vibrational scaling factors of 0.942 (0.947) for CCSD (scl-CCSD) and 0.946 (0.949) for  $\omega$B97M-V (scl-DFT). These scalings are inline with those presented in Refs.~\citenum{Irikura2005,NIST2022}.}

\label{freq_exprt}
   \begin{center}
   \begin{tabular}{|c|c|c|c|c||c|c|c|c|c|}
   \hline
   \multicolumn{5}{|c|}{CH$_2$BrI} & \multicolumn{5}{|c|}{CH$_2$Br$_2$}\\
     \hline
     \hline
       CCSD    &    DFT    &    Expt   & scl-CCSD  &    scl-DFT  &   CCSD  &    DFT    &   Expt   &  scl-CCSD  &    scl-DFT      \\
          147  &    147    &     144    &     138   &       139   &    174  &    173    &   173    &    165     &    164          \\
          548  &    555    &     517    &     516   &       525   &    604  &    604    &   576    &    571     &    572          \\
          673  &    671    &     616    &     633   &       634   &    694  &    688    &   638    &    657     &    653          \\
          781  &    775    &     754    &     735   &       733   &    832  &    824    &   806    &    787     &    781          \\
         1113  &   1104    &    1065    &    1048   &      1044   &   1136  &   1125    &  1090    &   1076     &   1067          \\
         1204  &   1192    &    1150    &    1134   &      1127   &   1241  &   1227    &  1183    &   1175     &   1164          \\
         1451  &   1430    &    1374    &    1366   &      1352   &   1463  &   1440    &  1390    &   1385     &   1366          \\
         3161  &   3149    &    2978    &    2977   &      2978   &   3162  &   3150    &  2988    &   2994     &   2989          \\
         3243  &   3233    &    3053    &    3054   &      3058   &   3243  &   3234    &  3054    &   3071     &   3069          \\
\hline
\hline
   \end{tabular}
   \end{center}
 \end{table}

\end{document}